\documentclass[lettersize,journal]{IEEEtran}

\usepackage{ifpdf}
 \ifpdf
 \else
 \fi

\ifCLASSINFOpdf
   \usepackage[pdftex]{graphicx}
\else
\fi

\usepackage{amsfonts}
\usepackage{multirow}
\usepackage{booktabs}
\usepackage{makecell}
\usepackage{graphicx}
\usepackage{subfigure}

\usepackage{color}
\definecolor{light_gray}{RGB}{255, 255, 204}
\usepackage{colortbl} 

\usepackage[pagebackref=false,breaklinks=false,letterpaper=true,colorlinks,citecolor=blue,linkcolor=blue, anchorcolor=blue, bookmarks=true]{hyperref}
\usepackage{amsmath}
\usepackage{amssymb}
\usepackage{array}
\usepackage{bm}
\usepackage{algorithm}
\usepackage{algorithmic}
\usepackage{pifont}
\newcommand{\cmark}{\ding{51}}%
\newcommand{\xmark}{\ding{55}}%
\newcommand{\et}{\emph{et al. }}
\newcommand{\etc}{\emph{etc}}

\usepackage{pdfsync}
\usepackage{float}

\usepackage{upgreek}

\usepackage{lettrine} 

\title{Revolution of Wireless Signal Recognition for 6G: Recent Advances, Challenges and Future Directions}
\author{Hao Zhang, \IEEEmembership{Graduate Student Member, IEEE}, 
Fuhui Zhou, \IEEEmembership{Senior Member, IEEE},\\
Hongyang Du, \IEEEmembership{Member, IEEE}, 
Qihui Wu, \IEEEmembership{Fellow, IEEE}, 
and Chau Yuen, \IEEEmembership{Fellow, IEEE}
\thanks{
H. Zhang, F. Zhou, and Q. Wu are with the College of Electronic and Information Engineering, Nanjing University of Aeronautics and Astronautics, Nanjing 211106 China. They are also with the Key Laboratory of Dynamic Cognitive System of Electromagnetic Spectrum Space (Nanjing University of Aeronautics and Astronautics) and with the Ministry of Industry and Information Technology, Nanjing, 211106, China (email: haozhangcn@nuaa.edu.cn; zhoufuhui@ieee.org; wuquhui2014@sina.com)

H. Du is with the Department of Electrical and Electronic Engineering, University of Hong Kong, Pok Fu Lam, Hong Kong (e-mail: duhy@eee.hku.hk).

C. Yuen is with the School of Electrical and Electronic Engineering, Nanyang Technological University, Singapore 639798 (email:chau.yuen@ntu.edu.sg)
}
}

\begin{document}
\bstctlcite{BSTcontrol}

\maketitle

\begin{abstract}
Wireless signal recognition (WSR) is a crucial technique for intelligent communications and spectrum sharing in the next six-generation (6G) wireless communication networks. It can be utilized to enhance network performance and efficiency, improve quality of service (QoS), and improve network security and reliability. Additionally, WSR can be applied for military applications such as signal interception, signal race, and signal abduction. In the past decades, great efforts have been made for the research of WSR. Earlier works mainly focus on model-based methods, including likelihood-based (LB) and feature-based (FB) methods, which have taken the leading position for many years. With the emergence of artificial intelligence (AI), intelligent methods including machine learning-based (ML-based) and deep learning-based (DL-based) methods have been developed to extract the features of the received signals and perform the classification. 
In this work, we provide a comprehensive review of WSR from the view of applications, main tasks, recent advances, datasets and evaluation metrics, challenges, and future directions. Specifically, intelligent WSR methods are introduced from the perspective of model, data, learning and implementation. Moreover, we analyze the challenges for WSR from the view of complex, dynamic, and open 6G wireless environments and discuss the future directions for WSR. This survey is expected to provide a comprehensive overview of the state-of-the-art WSR techniques and inspire new research directions for WSR in 6G networks. 
\end{abstract}


\section{Introduction}

\IEEEPARstart{T}{HE} fifth-generation (5G) network has been widely deployed in many areas, providing a high QoS and experience (QoE). Meanwhile, the next-generation 6G network is receiving increasing attention from researchers with characteristics of full-spectrum, full coverage, and intelligent communication \cite{tang2021survey,shi2023ris}. The 6G network is expected to provide a data rate of 1 Tbps, a latency of 1 ms, and a connection density of 10$^6$ devices per square kilometer. This will significantly increase the number of connected devices, which will be 100 times more than that of 5G. With the rapid increase in access equipment, the efficiency and security of 6G networks will be greatly challenged. To address these challenges, the 6G network will be equipped with intelligent communication capabilities, one of the most essential characteristics of 6G networks.

To achieve intelligent communication, wireless signal recognition (WSR) is one of the most important tasks. 
Firstly, WSR can be utilized to enhance network performance and efficiency. 
For example, by identifying different wireless signals, the 6G network can allocate spectrum resources more effectively and improve spectrum usage efficiency. 
Moreover, real-time identification of traffic patterns and demands in the network helps dynamically adjust resource allocation and improve the network's overall performance. 
Secondly, WSR can be utilized to improve QoS. 
For example, by identifying and managing various wireless signals, we can ensure that critical applications receive the necessary bandwidth and low latency, thereby improving user experience. 
Additionally, effectively identifying and managing data traffic during high-demand periods can reduce network congestion and ensure service quality. 
Thirdly, WSR can be utilized to improve network security and reliability. 
For example, identifying abnormal or malicious wireless signals helps respond to network attacks and threats promptly and enhance network security. 
Moreover, by monitoring and identifying signal anomalies, network faults can be quickly located and the reliability and stability of the network can be improved. 

WSR serves critical functions in both civilian and military domains. In civilian applications, WSR primarily enables three key capabilities through cognitive radio (CR) and spectrum management for efficiency enhancement, adaptive transmission for quality improvement, and anomaly detection across industrial, medical, and transportation sectors for security reinforcement. In military scenarios, WSR facilitates signal interception for capturing battlefield communications between transmitters and receivers, signal race for detecting and occupying communication channels to prevent legitimate transmissions, and signal deception for disrupting receiver operations through either jamming or misleading signals. These diverse applications demonstrate WSR's vital role in modern communication systems and military operations, where it continues to enhance spectrum efficiency, service quality, and operational capabilities.

WSR involves extracting signal descriptors, such as modulation type, signal types, and hardware-specific characteristics, to characterize a radio frequency (RF) waveform. 
It was motivated by its promising applications in military applications such as signal interception, signal race, and signal deception. 
In military communications applications, signal interception, jamming, and hijacking often require identifying the modulation types of adversarial signals. 
Nowadays, WSR has also been applied in many commercial and civil applications supporting the development of advanced technologies in communication systems, enhancing the efficiency, quality, and security of wireless communications \cite{zhang2024sswsrnet}. 
To improve network efficiency, WSR can be used to identify the modulation schemes of the users, which is essential for spectrum management and CR. 
For enhancing network quality, WSR is the basic task for the full awareness of the users, which is essential for adaptive transmission. 
Lastly, WSR can be used to identify abnormal signals, which is essential for network security \cite{zhu2015automatic}. 

\subsection{Scope}
WSR mainly includes four main tasks, including radio frequency fingerprint identification (RFFI), automatic modulation classification/recognition (AMC/AMR), wireless technology classification/recognition (WTC/WTR), and wireless interference identification (WII) \cite{li2019survey}. 
The RFFI of wireless devices mainly identifies wireless electronic devices by extracting the characteristics of radio frequency signals. 
AMC, the process of automatically identifying modulation formats, represents an intermediate stage between signal detection and demodulation. 
The goal of WTR is to classify wireless technologies in order to achieve better spectrum access and improve the security of wireless communication systems. 
WII seeks to classify interference signal types without any a priori information. 

Traditional WSR algorithms can be separated into two groups, namely likelihood-based (LB) and feature-based (FB) approaches \cite{dobre2007survey}. 
The LB methods, which consider WSR as a multiple hypothesis testing problem, can achieve optimal solutions \cite{dobre2003higher}. However, they suffer from high computation complexity. 
The DB methods identify the signal type by evaluating the equality of different signal distributions. 
The FB methods extract features from the received signals and can achieve near-optimal performance in classifying the signal types with lower complexity than LB methods. 
However, proper features and classifiers are needed for the classification. 
Different features are designed to represent the signal, such as time-domain, spectral-based, and statistics-based features. 
Then, threshold-based classifiers are usually utilized to classify these features. 
However, the simple classifiers cannot be adapted for all the scenarios of the features, which makes the traditional FB methods can only be applied for specific signal types under specific scenarios \cite{zhang2021noveltccn}. 
Recently, machine-learning (ML) based methods such as linear regression (LR), support vector machine (SVM), and decision tree are developed as the classifiers for the handcraft features. 

Nowadays, with the emergence of new technologies in computer science, especially the advanced development of hardware, deep learning (DL) has been widely applied across disciplines, such as computer vision \cite{zhang2019recent,xu2020automatic}, natural language processing, robotics, as well as wireless communications \cite{du2023beyond,du2023ai}. 
Great success has been made by DL models mainly due to their feature learning and decision-making ability in an end-to-end manner without the need for handcraft features. 
There is great potential for integrating DL in wireless communication systems since it can make the system more intelligent. 
The application of DL for WSR offers several advantages \cite{peng2018modulation}. 
First, the large datasets required for DL can be readily obtained in wireless communications systems. 
Second, manual feature selections are not required in DL, which can be a substantial issue in traditional modulation classification.
Third, the rapid advances in DL technologies present abundant opportunities to develop robust WSR solutions for complex communication scenarios.

DL-based WSR typically involves two sequential stages. First, the received signal is preprocessed into a suitable representation for subsequent processing. Second, deep neural network (DNN) models classify the modulation scheme based on the signal representations. This survey reviews DL-based WSR approaches, focusing on model architectures, data representations, learning techniques, and other key methods. Various models, including DNNs, 
deep belief networks (DBNs), convolutional neural networks (CNNs), recurrent neural networks (RNNs), and long short-term memory networks (LSTMs), can extract features from the input data. The received signal can be represented as images, sequences, or combinations before feeding into the models. Learning techniques like contrastive learning, multi-task learning, and transfer learning can enhance WSR classification performance. Additional AMC techniques include lightweight models, adversarial attacks, and transfer learning.

\begin{figure}[H]
	\centering
	\includegraphics[width=0.5\textwidth]{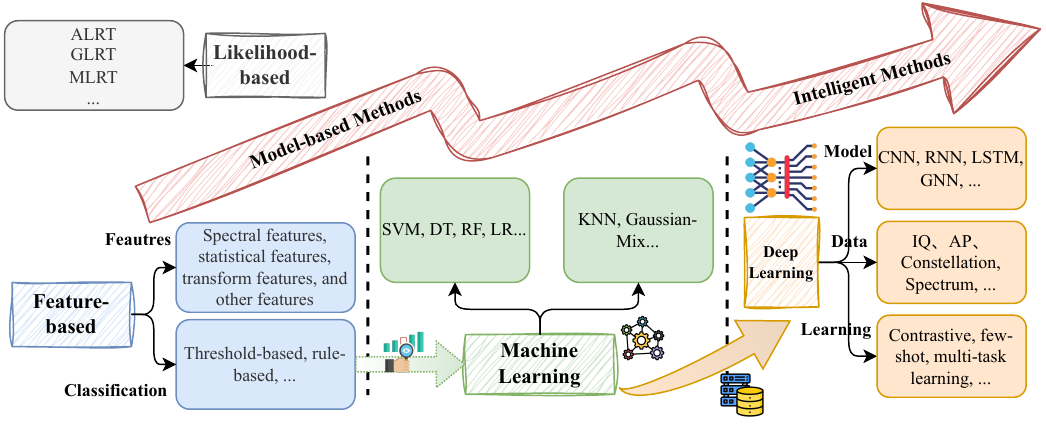}
	\caption{Historical development and evolution of WSR methods.}
	\label{fig:timeline}
  \end{figure}


\subsection{Existing Surveys and Reviews}
There exists numerous surveys, reviews, and technical reports related to WSR for wireless communications systems \cite{dobre2005blind,dobre2007survey,xu2010likelihood,hazza2013overview,zhu2015automatic,li2019survey,zhou2020deep,ghunaim2020deep,alshoubaki2021machine,abdel2021survey,jdid2021machine,peng2021survey}. 
Specifically, Dobre \et \cite{dobre2005blind} first systematically reviewed lots of existing techniques for digital modulation classification, providing helpful guidelines for selecting proper algorithms. 
Later, a more comprehensive version of various modulation classification techniques covering both digital and analog modulation types was presented in \cite{dobre2007survey}. 
The authors in \cite{xu2010distributed} and \cite{hazza2013overview} offered surveys of likelihood-ratio approaches and features-based methods for AMC, respectively. 
The book \cite{zhu2015automatic} presented a comprehensive survey for AMC, from the view of principles, algorithms and applications. 

\begin{table*}[htbp]
	\scriptsize 
	\centering
	\caption{Existing Surveys and Reviews Related to Wireless Signal Recognition}
	  \begin{tabular}{m{0.5cm}<{\centering}m{3cm}<{\centering}m{9cm}<{\centering}m{0.5cm}<{\centering}m{0.5cm}<{\centering}m{0.5cm}<{\centering}}
	  \toprule
	  Year &  Publication & Summary & App. & LB/FB & ML/DL \\
	  \midrule
	  2005 &  Dobre \et \cite{dobre2005blind}   & a review of techniques for digital MR systematically, which provides useful guidelines for choosing appropriate classification algorithms for different modulations.  & \xmark & \cmark & \xmark \\
	  \midrule
	  2007 & Dobre \et \cite{dobre2007survey} & a comprehensive survey of different MR techniques systematically, including likelihood-based methods and feature-based methods.  & \xmark & \cmark & \xmark \\
	  \midrule
	  2010 & Xu \et \cite{xu2010likelihood}   & a review of AMC methods based on likelihood functions, studies various classification solutions derived from the likelihood ratio test, and discusses the detailed characteristics. & \xmark & \cmark & \xmark \\
	  \midrule
	  2013 & Hazza \et \cite{hazza2013overview}  & an overview of FB methods developed for AMC, including features and ML classifiers.  & \xmark & \cmark & \xmark \\
	  \midrule
	  2015 & Zhu \et \cite{zhu2015automatic}  & a comprehensive survey for AMC, from the principles, algorithms and applications. & \cmark & \cmark & \xmark \\
	  \midrule
	  2019 & Li \et  \cite{li2019survey}   & a brief overview of signal recognition approaches including classical methods, emerging machine learning, and deep learning schemes. & \xmark & \xmark & \cmark \\
	  \midrule
	  2020 & Zhou \et \cite{zhou2020deep}  & a brief review of the most widely used DL techniques for recognizing a wireless signal in terms of modulation schemes & \xmark & \xmark & \cmark \\
	  \midrule
	  2020  & Ghunaim \et \cite{ghunaim2020deep} & a review studying the implementation of DL algorithms in AMC, which mainly from four perspectives including DL techniques/models, performance metrics used, DL models adopted, and types of AMC. & \xmark & \xmark & \cmark \\
	  \midrule
	  2021  & Alshoubaki \et \cite{alshoubaki2021machine} &  a survey of the DL neural network models and the techniques used in recognizing different modulation types of intercepted radar waveform.  & \xmark & \xmark & \cmark \\
	  \midrule
	  2021  & Abdel \et \cite{abdel2021survey} &  summarizing the AMC methods from the view of traditional methods and the advanced methods, comparing them, and presenting the commercial software packages for AMC. & \xmark & \cmark & \cmark \\
	  \midrule
	  2021  & Jdid \et \cite{jdid2021machine}  & a comprehensive SOTA review of the most recent ML-based AMR methods for SISO and MIMO systems. & \xmark & \xmark & \cmark \\
	  \midrule
	  2021  & Peng \et \cite{peng2021survey}  & a comprehensive survey of the SOTA DL-based MC algorithms, especially the techniques of signal representation and data preprocessing utilized in these algorithms.  & \xmark & \xmark & \cmark \\
	  \midrule
	  2022  & Jagannath \et \cite{jagannath2022comprehensive} & a comprehensive survey of the SOTA DL-based modulation classification algorithms, especially the techniques of signal representation and data preprocessing utilized in these algorithms.  & \xmark & \xmark & \cmark \\
	  \midrule
	  2022  & Zhang \et \cite{zhang2022deep} &  a review of current research on DL approaches for both SISO and MIMO systems from both accuracy and complexity perspectives. & \xmark & \xmark & \cmark \\
	  \midrule
	  \cellcolor{light_gray}\textbf{2024}  & \cellcolor{light_gray}\textbf{This work} & \cellcolor{light_gray}\textbf{a comprehensive survey of traditional and advanced DL-based WSR methods for AMC, WTC, WII and RFFI, where DL methods are concluded from the view of model, data, learning, and implementation aspects}. & \cellcolor{light_gray}\cmark & \cellcolor{light_gray}\cmark & \cellcolor{light_gray} \cmark \\
	  \bottomrule
	  \end{tabular}
	\label{tab:survey}
  \end{table*}

After the emergence of DL-based models for WSR, recent surveys and reviews have put their attention on ML and DL-based methods. 
For example, the authors in \cite{li2019survey} give a brief overview of signal recognition approaches including classical methods, emerging machine learning, and deep learning schemes. In addition, the opening problems and new challenges are discussed.
Peng \et \cite{peng2021survey} reviewed multiple papers in the view of signal representation and data preprocessing. 
Apart from the mentioned surveys, and reviews for AMC, there are several works considering the performance evaluation and comparison of AMC methods. 
The work \cite{ghasemzadeh2018performance} executed a performance analysis of the performance of AMC applied for civilian modulations with various features under practical scenarios. 
Mouton \et \cite{mouton2020comparison} compared clustering algorithms on the basis of classification accuracy and execution time for estimating modulation order, determining centroid locations, or both. 
The authors in \cite{cai2021performance} constructed a large dataset comprising various signal types at different median SNRs and utilized this big data to train a deep-learning model for automatic modulation classification. 
The main contributions of the existing WSR-related survey and tutorial papers are contrasted in Table \ref{tab:survey} to this survey. 
Earlier surveys and reviews only focus on the traditional model-based methods, or the performance evaluation and comparison of AMC methods, without considering the advanced DL-based methods. 
Recently, several surveys and reviews have focused on the DL-based methods for AMC, while not considering the traditional model-based methods and the methods for RFFI, WTC, and WII. 
However, there is no comprehensive survey that introduces the model-based and intelligent WSR methods for RFFI, AMC, WTC, and WII. 
Moreover, the challenges and future directions for WSR in the complex, dynamic, and open 6G wireless environments are not discussed in the existing surveys and reviews.

\begin{figure*}[t]
  \centering
	\includegraphics[width=0.99\linewidth]{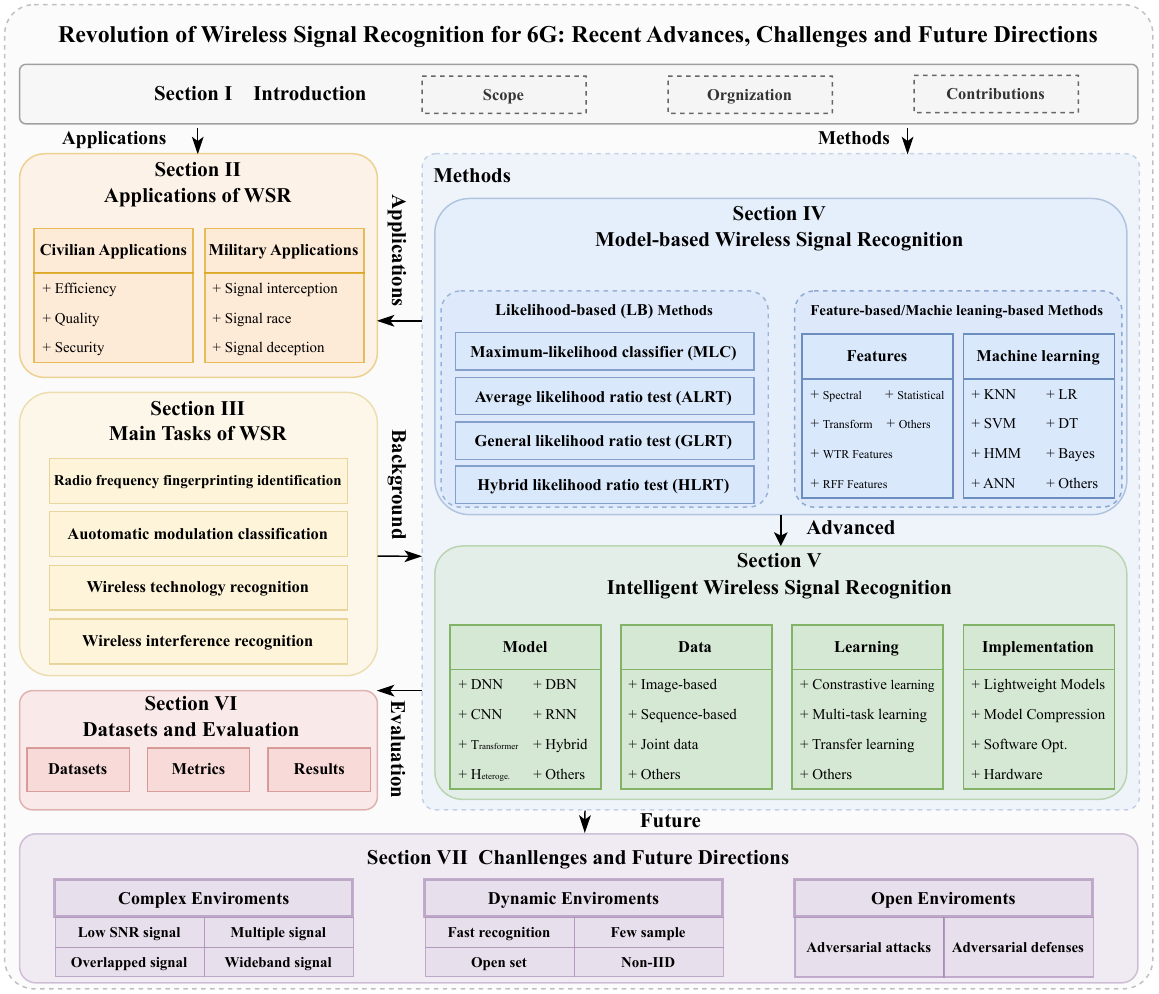}
	\caption{Organization and contents of this survey and its sections.}
	\label{fig:overview}
\end{figure*}

\subsection{Contributions and Organization}
Recent studies in the field have predominantly focused on employing ML and DL for WSR, whereas earlier research primarily centered on FB and LB models. 
This survey sets itself apart by offering a comprehensive review of works related to WSR, encompassing LB, FB, ML, and DL methods. The main contributions of this survey can be summarized as follows
\begin{enumerate}
  \item This survey provides a comprehensive survey on the paradigm of model-based WSR and intelligent WSR from the scope of applications, tasks, techniques, datasets, evaluation metrics, and challenges and future directions on WSR including tasks of radio frequency fingerprinting identification (RFFI), automatic modulation classification (AMC), wireless technique classification (WTC), and wireless interference identification (WII). 
  \item This survey categorizes the existing WSR methods into model-based WSR and intelligent WSR. Model-based WSR include likelihood-based (LB) methods, feature-based (FB) methods, and machine learning-based methods. Intelligent WSR methods are divided into four aspects, namely, model, data, learning, and others. 
  \item This survey discusses both the advantages and disadvantages of each WSR method, which are essential for designing future studies on intelligent WSR. 
  \item This survey analyzes the challenges for WSR from the view of complex, dynamic, and open 6G wireless environments, which are essential for designing future studies on intelligent WSR. 
\end{enumerate}

The rest of this survey is organized as follows, as shown in Fig. \ref{fig:overview}. Section \ref{sec:app} introduces the applications of WSR in terms of civilian and military aspects. Section \ref{sec:main_elements} presents the main tasks of WSR including the concepts of different tasks. Section \ref{sec:traditional_methods} explains model-based WSR methods involving likelihood-based methods, feature-based methods, and machine learning-based methods. Section \ref{sec:advanced_methods} introduces intelligent WSR methods mainly deep learning-based methods from the view of model, data, and learning. Section \ref{sec:data_evaluation} shows the datasets and evaluation metrics. Section \ref{sec:challenges_issues} draws the future challenges and open issues for WSR and Section \ref{sec:conclusion} concludes this survey. The list of abbreviations is presented in Table \ref{tab:abbrv}.

\begin{table}[h]
	\scriptsize 
	\caption{List of Abbreviations.}
	\begin{center}
	\begin{tabular}{m{2cm}m{5cm}}
	\toprule
	Acronym & Definition \\
	\midrule
	ALRT & Average likelihood ratio test\\
	AP & Amplitude and phase\\ 
	AM & Amplitude modulation\\
	AMC & Automatic modulation classification\\
	AWGN & Additive white Gaussian noise\\
	BT & Bluetooth\\
	CR & Cognitive radio\\
	CNN & Convolutional neural network \\
	CWT & Continuous wavelet transform\\
	DBN & Deep belief network\\
	DFT & Discrete Fourier transform\\
	DL & Deep learning\\
	DNN & Deep neural network\\ 
	DSA & Dynamic spectrum access\\
	DRSN & Deep residual shrinkage network\\
	DWT & Discrete wavelet transform\\
	FB & Feature-based\\
	FC & Fully connected\\
	FFT & Fast Fourier transform\\
	FM & Frequency modulation\\
	GAN & Generative adversarial network\\
	GAP & Global average pooling\\
	GLRT & Generalized likelihood ratio test\\
	HLRT & Hybrid likelihood ratio test\\
	HOC & Higher-order cumulant\\
	IQ &  In-phase and quadrature\\
	LB & Likelihood-based\\
	LSTM & Long short-term memory\\
	MC & Modulation classification\\
	ML & Machine learning\\
	MLC & Maximum likelihood classification\\
	MIMO & Multi-input-multi-output \\
	NOMA & Non-orthogonal multiple access\\
	PSK & Phase shift keying\\
	OSR & Open set recognition\\
	QAM	& Quadrature amplitude modulation\\
	QPSK & Quadrature phase shift keying\\
	QoS & Quality of service\\
	QoE & Quality of experience\\
	SNR & Signal-to-noise ratio\\
	SISO & Single-input-single-output \\
	SOTA & State-of-the-art \\
	WLAN & Wireless local area network\\
	WSR & Wireless signal recognition\\
	WTC & Wireless technology classification \\
	\bottomrule
	\end{tabular}
	\label{tab:abbrv}
	\end{center}
\end{table}

\section{Applications of Wireless Signal Recognition}\label{sec:app}
This section introduces the applications of WSR in terms of civilian and military aspects. 
For civilian applications, WSR can be utilized to enhance network performance and efficiency, improve QoS, and improve network security and reliability. 
For military applications, WSR can be applied for signal interception, signal race, and signal deception. 

\begin{figure*}[t]
	\centering
	  \includegraphics[width=0.95\linewidth]{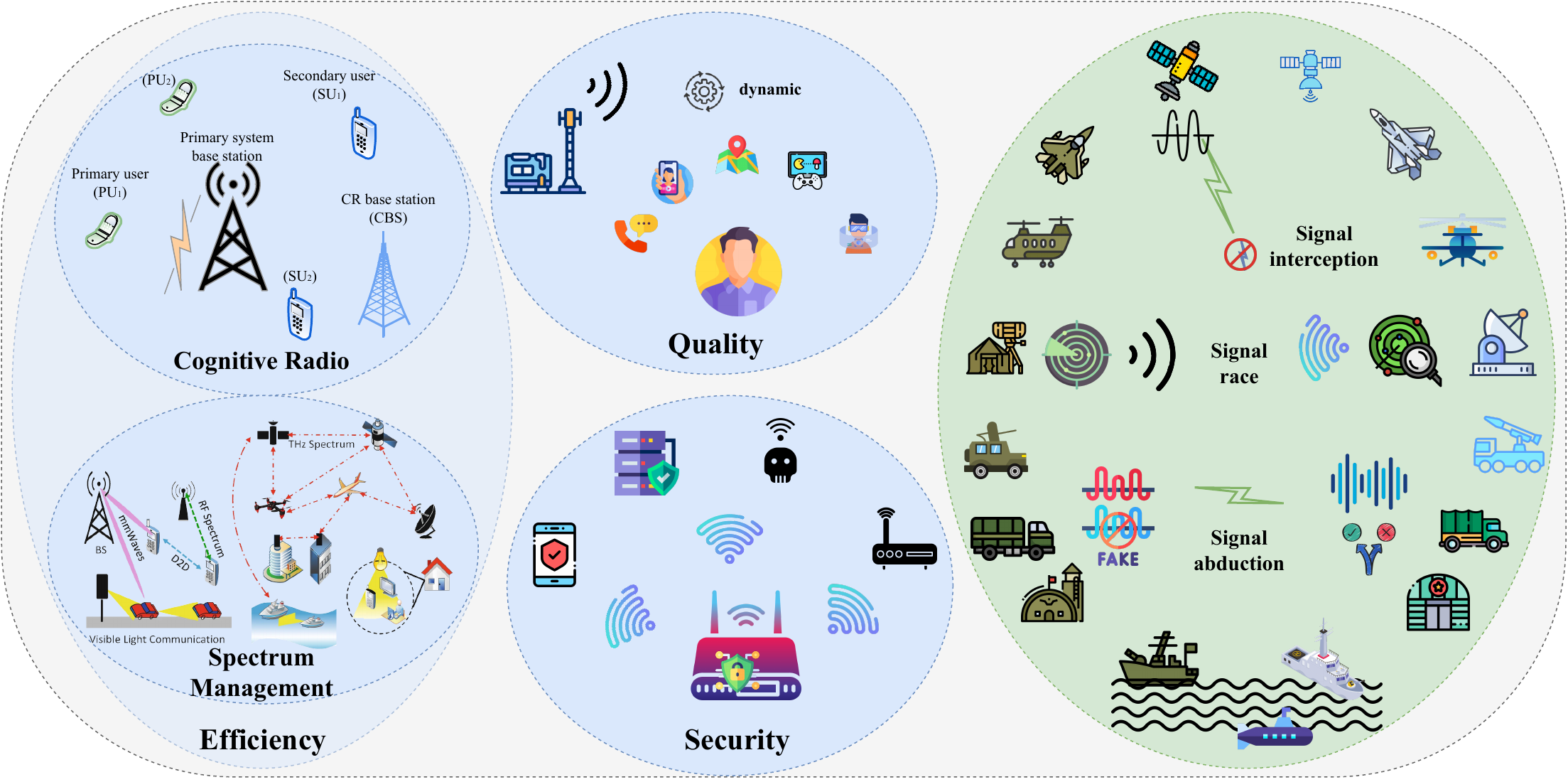}
	  \caption{Applications of wireless signal recognition, including civilian applications (blue circles) and military applications (green circle). }
	  \label{fig:app}
  \end{figure*}

\subsection{Civilian Applications}

For civilian applications, WSR has three primary use cases, including efficiency, quality, and security, as shown in Fig. \ref{fig:app}. The first use case is efficiency, which can be achieved through CR and spectrum management. The second use case is quality, which can be achieved through adaptive transmission to improve the quality of service. The third is security, which can be achieved through abnormal signal detection to enhance network security \cite{sun2020micro,fu2018low}.  

\subsubsection{Efficiency}
With the rapid growth of wireless communications and increasing demand for wireless services, available spectrum resources are quickly becoming exhausted \cite{ali2016advances,awin2018blind}. Smarter spectrum utilization techniques are therefore needed. CR has emerged as a promising solution, significantly improving spectrum efficiency by allowing licensed and unlicensed users to share licensed bands. A typical CR network comprises a primary system base station, a CR base station, several primary users, and several secondary users. In CR networks, WSR can detect nearby radio sources and optimize spectral efficiency without a priori signal knowledge, across diverse unknown channels. 
Moreover, WSR can be used to identify the modulation schemes of the users, which is essential for spectrum management. 

\subsubsection{Quality} 
Another example of civilian applications is enhancing the QoS through sensing the status of the users including the modulation schemes. Then, adaptive transmission can be utilized to improve the QoS. Adaptive modulation schemes aim to achieve the maximum possible transmission rates in a wireless system. These schemes fully utilize time-varying channels to achieve this objective. Typically, the transmitter must share information on the modulation scheme with the receiver via network protocol overhead. However, this overhead could be eliminated if the receiver automatically recognizes modulation schemes. Thus, WSR enables receivers to identify modulation formats and select modulation schemes adaptively.

\subsubsection{Security} 
Detecting abnormal signals is critical across industry, medicine, and transportation domains. However, the vast quantity and diversity of signals make anomaly detection challenging under all conditions. Abnormal signal detection remains an intractable problem in many fields, as manual detection is cost-prohibitive and automated methods lack sufficient accuracy. Most existing anomaly detection techniques comprise two stages: feature extraction and anomaly classification, formulating abnormal signal detection as a multi-class categorization task. In wireless communications, WSR enables direct anomaly detection by identifying modulation schemes of received signals. Furthermore, WSR models can be transferred to anomaly detection in a multi-class framework. Thus, WSR provides a fundamental technique for abnormal signal detection across problem contexts.

\subsection{Military Applications}
Military applications for WSR can be classified into three main aspects, namely, signal interception, signal race, and signal deception. 

\subsubsection{Signal interception}
Signal interception \cite{gardner1988signal} aims at intercepting the communication signals between two devices, such as transmitters and receivers on the battlefield. 
Intercepting wireless signals is of great interest to law enforcement and military entities, as it provides critical intelligence on illicit activities or enemy operations. Signal interception first necessitates detecting the signal presence. Spread spectrum communications further complicate detection by submerging signals below the noise floor to evade monitoring. Effective signal detection systems must therefore operate at low SNR to uncover hidden transmissions. Moreover, in unintended reception scenarios, receivers lack a priori knowledge of signal activity or the intercept environment. The detection system must thus function absent any priors on signal presence in the targeted band or signal statistics.

\subsubsection{Signal race}
The goal of the signal race is to occupy a communication channel between two or more devices. Thus, the true transmitters and receivers cannot communicate through their channels. To achieve this goal, the existence of the signal must first be detected by using spectrum sensing, AMC or WTR. Then, the attackers should generate interference, noise or disruptive signals to block or degrade other parties' ability to communicate through a shared channel or frequency. 

\subsubsection{Signal deception}
Signal jamming and deception refer to electronic countermeasures designed to disrupt receiver operations by emitting deliberate frequency signals. These signals aim to overwhelm the receiver with either noise or misleading information. The act of inundating the receiver to render its display unreadable is typically referred to as jamming. On the other hand, systems that generate contradictory or confusing signals are known as deception. However, it is common to use the term “jamming” to encompass all such countermeasure systems.

\section{Main Tasks of WSR}\label{sec:main_elements}
In this section, we first present a basic signal model of WSR. Then we introduce four typical tasks including RFFI, AMC, WTC, and WII. 

\subsection{Signal Model}
Suppose that the received signal is presented as 
\begin{equation}
r(t) = s(t)\exp{j(2\pi f_c t+\Theta)}+w(t),
\end{equation}
where $s(t)$ is the send signal. $f_c$ and $\Theta$ refer to the frequency and the phase of the carrier, respectively. $w(t)$ stands for additive Gaussian noise with mean $0$ and variance $\sigma^2$. 
For AMC, WTR, WII, and RFFI tasks, $s(t)$ is the signal with the different modulation schemes, or different types of wireless technologies, or the different types of interference signals, or sent by different devices.

For practical implementation of WSR algorithms, the continuous-time received signal $r(t)$ is sampled at a rate satisfying the Nyquist criterion to obtain discrete-time samples. The discrete received signal can be represented as:
\begin{equation}
r[n] = r(t)|_{t=nT_s} = s[n] \exp j(2\pi f_c nT_s + \Theta) + w[n], \tag{1a}
\end{equation}
where $T_s$ is the sampling period, $n$ is the sample index ($n = 1, 2, \ldots, N$), $N$ is the total number of samples used for classification, and $s[n]$, $n[n]$ are the discrete-time versions of the transmitted signal and noise, respectively.

\subsection{Radio Frequency Fingerprint Identification (RFFI)}

In wireless device identification, traditional methods rely on unique attributes such as public identifiers or secret keys. However, beyond these explicit credentials, an additional layer of identification can be tapped into: the distinctive characteristic fingerprints of the devices. These fingerprints arise from observable traits related to various components such as operating systems, drivers, clocks, and radio circuitry. Analyzing these components to extract identifiable features, akin to the concept of biometric fingerprints, is referred to as RFFI. This approach, as highlighted by Danev \et \cite{danev2012physical}, leverages the subtle but unique signatures inherent in the hardware and software configurations of wireless devices for more nuanced identification. 
RFFI involves analyzing these signal characteristics, including frequency, phase, amplitude, and signal shape. The goal is to create fingerprints that are unique to each device, similar to biometric fingerprints.

\subsection{Automatic Modulation Classification (AMC)}
Modulation is the process of transferring the properties of a carrier signal to a modulation signal containing the information to be transmitted. The purpose of modulation is to impress the information onto the carrier wave for conveyance from transmitter to receiver. Modulation enables multiple channels of information transmission through a single communication medium 
, for example, using frequency division multiplexing (FDM), though other techniques such as time division multiplexing (TDM) or code division multiplexing (CDM) can also apply.
There are two primary forms of modulation: analog modulation and digital modulation. Analog modulation schemes include amplitude modulation (AM), frequency modulation (FM), and Gaussian minimum shift keying (GMSK), among others, while M-ary amplitude shift keying (M-ASK), M-ary frequency shift keying (M-FSK), M-ary phase shift keying (M-PSK), and M-ary quadrature amplitude modulation (M-QAM) are common digital modulations. The modulated signal carries characteristics of the modulation scheme, which can be recognized from the received signal. AMC aims to blindly identify the modulation type of incoming signal at the receiver in wireless systems without a priori knowledge of the transmitted data or other unknown parameters like signal power, carrier frequency/phase, timing, \etc.

\subsection{Wireless Technology Classification (WTC)} 

Heterogeneous wireless networks—comprising diverse coexisting wireless technologies such as Wi-Fi, Bluetooth, Zigbee, LTE, and GSM sharing spectrum—are a promising solution for improving spectrum sharing. A key enabler for developing coexistence protocols is correctly identifying wireless technologies anticipated to share spectrum, and shifting users between available wireless options to optimize usage and minimize interference. This problem is known as WTC, referring to the identification of wireless signals without requiring preprocessing like channel estimation or timing/frequency synchronization \cite{yucek2009survey}.

\subsection{Wireless Interference Identification (WII)}

Interference signals \cite{zhou2013large,zhou2020joint} can be categorized as deception or suppression. Deception interference intentionally generates signals mimicking target communications to confuse receivers. The aim is to induce misunderstanding or incorrect use of the obtained data, including techniques like interrupted sampling repeaters and dense false targets. Suppression interference overpowers signals using high power to disrupt communications, such as single-tone and linear frequency-modulated interferences. WII is a critical process in wireless communication systems, especially as the airwaves become more congested with various signals. In WII, the goal is to classify and identify the type of interference affecting the wireless signal. This is challenging because it often needs to be done without any prior information about the nature of the interference.

\section{Model-based Wireless Signal Recognition}\label{sec:traditional_methods}
In this section, we introduce the model-based WSR methods including likelihood-based methods, feature-based methods, and machine learning-based methods. 
Moreover, both the advantages and disadvantages of each WSR method are discussed. 

\subsection{Likelihood-based (LB) Methods}

\begin{figure}[t]
\centering
\includegraphics[width=0.99\linewidth]{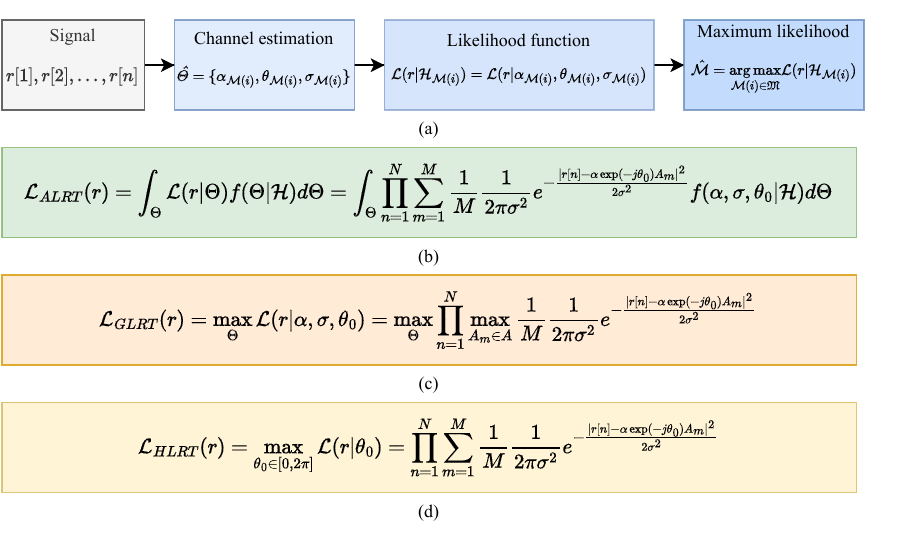}
\caption{Illustration of the Likelihood-based methods for WSR, (a) MLC in AWGN channel, (b) ALRT, (c) GLRT, and (d) HLRT.}
\label{fig:mlc}
\end{figure}

LB methods are the most prevalent techniques for WSR, especially for AMC, owing to the complexity of the other three tasks. For WII, Zhao \et \cite{zhao2017discrimination} applied a generalized likelihood ratio test discriminator to recognize deception interference. These approaches leverage probability density functions (PDFs) of the observed waveform. First, the likelihood is estimated for the observed signal samples under each signal hypothesis. The likelihood functions are then computed and updated using the selected signal model to reduce complexity or enable non-cooperative settings. Subsequently, the likelihood ratio test discerns the signal by comparing the signal fit to the candidates via a threshold. This survey covers several likelihood-based methods, including maximum-likelihood classifier (MLC), average likelihood ratio test (ALRT), general likelihood ratio test (GLRT), and hybrid likelihood ratio test (HLRT) techniques, as shown in Fig. \ref{fig:mlc}.

\subsubsection{Maximum-likelihood classifier (MLC)}

In an MLC with perfect channel state information \cite{wei2000maximum}, computing likelihoods equates to calculating the probabilities of the received signal sample conditioned on all parameters except the signal type. Thus, classification amounts to MLC estimation of the signal type from a finite candidate set. 
For an AWGN channel, a signal sample $r[n]$ corresponds to a signal type $\mathcal{M}$. The likelihood of a particular type is given by
\begin{equation}
\mathcal{L}(r[n])=p(r[n]|\mathcal{M},\sigma).
\end{equation}

Given the complex PDF of the observed sample over the AWGN channel, the likelihood function can be expressed as
\begin{equation}
\mathcal{L}(r[n]|M,\sigma)=\sum\limits_{m=1}^M\frac{1}{M}\frac{1}{2\pi\sigma^2}\exp\left(-\frac{|r[n]-A_m|^2}{2\sigma^2}\right).
\end{equation}

The likelihood is computed by taking the average likelihood value between the observed signal sample and each signal type $A_m$. The joint likelihood for multiple received samples is calculated as the product of the individual sample likelihoods
\begin{equation}
\mathcal{L}(r|M,\sigma)=\int\limits_{n=1}^N\sum\limits_{m=1}^M\frac{1}{M}\frac{1}{2\pi\sigma^2}\exp\left(-\frac{|r[n]-A_m|^2}{2\sigma^2}\right).
\end{equation}

The maximum likelihood classification (MLC) decision is made assuming a finite pool $\mathfrak{M}$ of signal candidates. The hypothesis $\mathcal{H}_{\mathcal{M}_{(i)}}$ for each modulation type $\mathcal{M}_{(i)}$ is computed by estimating the channel parameters $\hat{\Theta}_{\mathcal{M}_{(i)}}$ comprising the channel gain $\alpha$, noise variance $\sigma^2$, and phase offset $\theta_0$, along with an appropriate likelihood function $\mathcal{L}(r|\mathcal{H}_{\mathcal{M}_{(i)}})$. The decision is obtained by selecting the hypothesis with the maximum likelihood as follows

\begin{equation}
	\hat{\mathcal{M}}=\underset{\mathcal{M}(i)\in \mathfrak{M} }{\arg\max } \mathcal{L}(r|\mathcal{H}_{\mathcal{M}(i)}).
\end{equation}

Initial work on MLC for WSR was conducted by Sills \cite{sills1999maximum}, who proposed an MLC algorithm for coherent detection of PSK and QAM signals. Wei and Mendel  \cite{wei2000maximum} then applied MLC to classify digital quadrature modulations, proving its ability to correctly classify any finite constellation set given sufficient observed symbols. Su \et \cite{su2008real} subsequently developed a real-time MLC-based AMC method for software-defined radios without pilot symbols. Shi \et \cite{shi2012automatic} explored a phase-based MLC approach for linearly modulated signals. Fki \et \cite{fki2015blind} leveraged MLC functions on the real and imaginary components of the equalized signal. MLC techniques have also been applied to AMC in multiple-input multiple-output systems \cite{salam2015automatic}. Motivated by the simplicity of using signal moment statistics, Abu-Alshaeer \et \cite{abu2018automatic} proposed a hybrid MLC-features classifier. For wireless technology recognition, Alayaoui \et \cite{alyaoui2011fourth} utilized Gaussian MLC to identify LTE signals. A key assumption in MLC is known channel state information. However, in practice, these parameters are often unknown.

\subsubsection{Average likelihood ratio test (ALRT)}

The ALRT was proposed to address the limitations of MLC. The ALRT likelihood function marginalizes uncertain parameters by integrating over all possible values weighted by their probabilities. Since the classifier lacks knowledge of the channel parameter set $\Theta$  comprising gain $\alpha$, noise variance $\sigma$, and phase offset $\theta_0$, 
the ALRT likelihood is given as in Fig. \ref{fig:mlc} (b). 

The application of ALRT to modulation classification was pioneered by Polydoros and Kim \cite{polydoros1990detection} and expanded upon by \cite{huan1995likelihood}, \cite{sills1999maximum}, and \cite{hong2000bpsk} for BPSK detection. Huan \et \cite{huan1995likelihood} specifically proposed an ALRT algorithm to classify M-PSK signals in additive white Gaussian noise. Beidas \et \cite{beidas1995higher,beidas1998asynchronous} developed ALRT classifiers to lower-bound optimal likelihood ratio test performance for synchronous and asynchronous modulation classification. Hong \et \cite{hong2003classification} leveraged a Bayesian ALRT approach to distinguish BPSK from QPSK without a priori signal power knowledge. Shah \et \cite{shah2019effective} proposed a reduced complexity ALRT technique using zero-forcing equalization for orthogonal space-time block-coded MIMO systems. As Zhu \et \cite{zhu2014automatic,zhu2015automatic} discussed, the ALRT likelihood function becomes much more complex with unknown parameters, success depends on accurate models for these variables. Thus, ALRT degrades to a suboptimal approximation of the optimal classifier when accurate channel models are unavailable. Further inaccuracy stems from estimating additional hyperparameters. Moreover, integration introduces additional complexity to the likelihood function.

\subsubsection{General likelihood ratio test (GLRT)}

To address the limitations of ALRT, Panagiotou \et \cite{panagiotou2000likelihood} proposed the GLRT as an alternative. GLRT integrates maximum likelihood estimation with maximum likelihood classification. The likelihood function substitutes integration over unknown parameters with a maximization of the likelihood over a feasible interval for the unknowns. The GLRT likelihood is expressed as in Fig. \ref{fig:mlc} (c).

While reducing complexity, the GLRT classifier becomes biased at low and high SNR regimes when classifying nested modulations. As an example, consider GLRT classification between 4-QAM and 16-QAM. At low SNR, when signals are dispersed, a 4-QAM modulated signal will likely produce a higher 16-QAM likelihood since the denser 16-QAM constellation has more possible symbols. Conversely, at high SNR with tightly clustered signals, maximizing the likelihood via channel gain scaling can cause overlap between the 4-QAM alphabet and a subset of the 16-QAM symbols. This phenomenon seen with nested modulations produces equal likelihoods between low and high-order modulations when classifying the lower-order scheme. Therefore, GLRT exhibits inherent bias favoring higher-order modulations.

\subsubsection{Hybrid likelihood ratio test (HLRT)}
To address the limitations of the GLRT, Panagiotou \et \cite{panagiotou2000likelihood} proposed an alternative likelihood ratio approach called HLRT. The HLRT likelihood is computed by averaging over the transmitted symbols and maximizing the resulting function with respect to the carrier phase. Thus, the HLRT probability density function can be expressed as in Fig. \ref{fig:mlc}
 (d). 

The HLRT method follows a similar procedure to the ALRT and GLRT methods but include a threshold parameter $\gamma_H$ to optimize classification accuracy.
Setting the threshold to one results in the MLC method being used.
The likelihood ratio used in HLRT can be computed using the equation below.
This likelihood function evaluates the probability of each signal sample belonging to each alphabet symbol, eliminating the possibility of biased classification due to a nested constellation.
Additionally, the maximization process replaces the integration of unknown parameters and their probability density functions with a much simpler analytical and computational approach.

Ozdemir \et \cite{ozdemir2015asynchronous} introduced a novel hybrid maximum likelihood classification scheme that utilizes a generalized expectation maximization algorithm. Their proposed algorithm's efficiency and effectiveness were demonstrated through simulation results. Likewise, Wimalajeewa and co-authors \cite{wimalajeewa2015distributed} presented a distributed hybrid maximum likelihood algorithm that accounts for unknown time offset, phase offset, and channel gain. Simulation and experimental results were used to showcase the efficacy of their proposed algorithm. In another study, Zheng and collaborators \cite{zheng2018likelihood} proposed an HLRT-based blind AMC method for scenarios with unknown CSI. They presented an efficient implementation of the expectation-maximization algorithm to estimate channel fading coefficients and noise variance. The effectiveness of their proposed AMC algorithms was confirmed through computer simulations.

LB methods have been widely used in WSR, especially for AMC. However, these methods are limited by several factors, including the need for accurate channel state information, the complexity of the likelihood function, and the bias of the likelihood function. Thus, they are not adaptive to the dynamic and complex wireless environments. 
Moreover, with the increasing complexity of wireless signals, the LB methods are not suitable for future 6G. 

\begin{figure}
	\centering
	\includegraphics[width=0.99\linewidth]{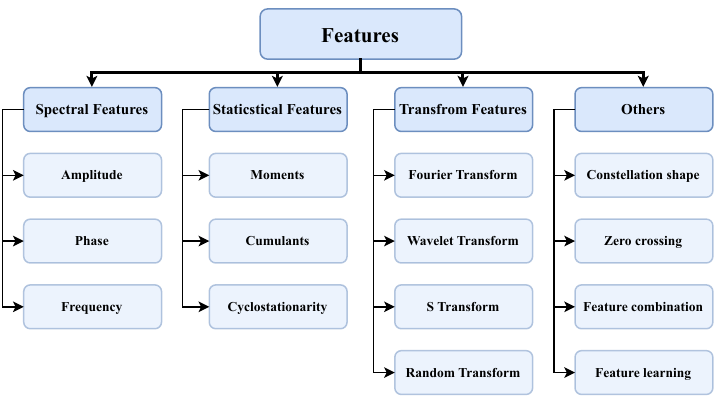}
	\caption{Features for wireless signals, including spectral features, statistical features, transform features and others.}
	\label{fig:fea}
\end{figure}
  
\subsection{Feature-based (FB) Methods}
LB methods are decision theoretic-based classifiers, which depend on the signal distribution. 
However, in the specific of certain modulations, such as AM and FM, it is evident that signal distribution alone is inadequate for accurately classifying modulation types. 
Thus, FB methods are proposed to extract the key features from the signal and then classify the signal based on these features. 
The received signal can be recognized by identifying the key features including spectral features, statistical features, transform features, and other features, as shown in Fig. \ref{fig:fea}.

\subsubsection{Spectral features}
Spectral features can be used to describe various aspects of the signal, such as the frequency content, power spectrum, and phase information.
These parameters and the variation of them are developed to WSR. 

In the 1990s, Nandi and Azzouz \cite{nandi1995automatic, azzouz1995automatic, azzouz1996procedure} introduced essential spectral-based features for the automated classification of fundamental analog and digital modulations. Their proposed features built upon and enhanced the feature extraction techniques suggested by Fabrizi \cite{fabrizi1986receiver} and Dobre et al. \cite{dobre2003higher}. Table \ref{tab:fea} offers nine key features for AMC, including $\gamma_{max}$, $\sigma_{ap}$, \etc.

Various features extracted from the signal have been employed for AMC. In \cite{nandi1998algorithms}, three key features derived from the instantaneous amplitude, phase, and frequency of the intercepted signal are used and an overall success rate of over 96\% is achieved at an SNR of 15 dB. Teng \et \cite{teng2008modulation} developed a classification method based on spectral correlation and support vector machine (SVM). The experimental results demonstrate that the algorithm is robust with high accuracy even at low SNR. The authors in \cite{shen2014automatic} selected seven feature parameters with fine classification information based on analyzing the signal characteristics in time and frequency domains. Combined with a 1-nearest neighbor pattern classifier, the method achieved better performance compared to the method based on principal component analysis, which is widely used. Mendis \et \cite{mendis2016deep} proposed a DBN-based AMC method that employs spectral correlation function. Simulation results illustrate the efficiency of the proposed method in classifying four FSK, 16-QAM, BPSK, QPSK, and OFDM modulations in various environments. These existing works have applied various spectral features and classified these features using ML classifiers. While achieving good classification accuracy, these methods rely heavily on the choice and number of features extracted from the signal.

For WTC, spectral features contain valuable characteristics such as bandwidth, center frequency, and power spectral density, which are critical for WTC. Spectral-domain data provide effective frequency domain information that can significantly mitigate the problems of data transmission and storage in the implementation process. In \cite{ahmad2010fuzzy}, a novel fuzzy logic (FL) method was proposed for recognizing wireless local area network (WLAN), Bluetooth (BT), and FSK signals. Spectral features, namely the power spectral density (PSD) information, were utilized to extract the bandwidth and center frequency of the signals. These spectral features were then used to label the signals according to the respective wireless standards. The authors demonstrate that the proposed FL approach effectively extracts explicit signal features. Ahmad \et \cite{ahmad2010neuro} present a neuro-fuzzy signal classifier (NFSC) for recognizing various wireless signals based on their PSD profiles. An adaptive neuro-fuzzy inference system is employed for the classification task. Experimental results demonstrate that the NFSC performance is improved by using wideband PSD data acquisition in real-time coexistence environments. This suggests that capturing the full PSD shape, which is possible in wideband mode, provides more discriminative features for better signal recognition.

\subsubsection{Statistical features} 
Statistical features are used to describe the properties of a signal and can be used to extract useful information from the signal. Some common statistical features of signals include Moments, cumulants, and cyclostationarity. 

\paragraph{Moments} 
Mathematically, the $k$th-order moment calculation of the signal phase is performed as 
\begin{equation}
\mu_k(r)=\frac{1}{N}\sum\limits_{n=1}{N}\phi^k(n),
\end{equation}
where $\phi(n)$ is the phase of the $n$th sample of the signal.

The moment for a received complex-valued signal $r = [r[1], r[2], \cdots, r[N]]$ is estimated as
\begin{equation}
\mu_{xy}(r)=\frac{1}{N}\sum\limits_{n=1}{N}r^x[n]\cdot r^{*y}[n],
\end{equation}
where $x+y=k$ and $r*[n]$ is the complex conjugate of $r[n]$.

Previous works have explored using moment-based features for AMC. Soliman and Dominguez \cite{soliman1992signal} developed an AMC algorithm that utilizes moments of the signal phase to classify general M-PSK modulation types. Lopatka et al. \cite{lopatka2000automatic} extracted moments of the signal amplitude, phase, and frequency as inputs to a fuzzy logic classifier for AMC. They found this approach performed well even at low SNRs. More recently, Moser \et \cite{moser2015automatic} proposed using instantaneous signal features like amplitude, phase, and frequency for AMC algorithms. The effectiveness of instantaneous features suggests combining them with moment-based features could lead to more accurate and robust modulation identification. Future AMC research can build on these works by fusing statistical and instantaneous signal features to improve classification performance.

\paragraph{Cumulants}
Cumulants are the most prevalent features used to classify modulation schemes. 
The $n$th-order cumulant with $m$ conjugate variables is given in Table \ref{tab:cum}. 
Other higher-order cumulants (HOCs), such as $C_{80}$, $C_{81}$, $C_{82}$, $C_{83}$, $C_{84}$, $C_{10,0}$, $C_{10,1}$, and $C_{10,2}$, can be similarly obtained with estimated moments as in \cite{aslam2012automatic} and \cite{lee2017deep}.

Several works have explored using HOCs for AMC, especially under multipath fading conditions. Swami and Sadler \cite{swami2000hierarchical} proposed a simple AMC method based on 4th-order cumulants that demonstrated comparable performance. Dai \et \cite{dai2002joint} introduced a joint power estimation and modulation classification approach. Other studies \cite{xi2006robust,wu2008novel, orlic2009automatic, chang2015cumulants} have investigated AMC using HOCs under multipath fading channels. Lee \et \cite{lee2019effective} analyzed the impacts of different cumulants for AMC and found that pruning meaningless features and only using effective features can reduce computation time for potential real-time applications. For example, they showed $C_{40}$ is most effective while $C_{20}$ is least effective in classifying BPSK, QPSK, 8PSK, 16QAM, and 64QAM using the cumulant set \{$C_{20}$,$C_{21}$,$C_{40}$,$C_{41}$,$C_{42}$\}. Using a DNN trained on only the effective cumulants provided up to $8\%$ classification accuracy improvement. These works demonstrate the promise of using HOCs and feature selection for robust AMC, especially in fading conditions. Future research can build on these findings to develop real-time AMC solutions.

Previous works have explored using signal features like amplitude modulation, HOCs, and bispectrum for wireless interference rejection (WIR) to detect jamming signals. Xiaoyan \et \cite{xiaoyan2005pattern} analyzed and compared three different features-amplitude modulation, higher order cumulants, and bispectrum-for jamming signal detection. The authors in  \cite{jian2011signal} analyzed the features of deception interference signals and studied filter bank methods based on HOCs and bispectrum. These works demonstrate analyzing moments and spectral features can help identify interference and jamming signals for WIR. Future research can build on these findings to develop improved WIR techniques, potentially combining analysis of multiple signal features like modulation, cumulants, and bispectrum to better detect jamming signals in wireless systems.

\paragraph{Cyclostationary features}
Gardner and Spooner \cite{gardner1988cyclic} first implemented cyclostationary analysis for WSR  problems, exploiting the distinct differences between the cyclic spectrum patterns of different signal types. 
The statistical characteristics of the cyclostationary signals are changed with time \cite{gardner2006cyclostationarity}. 
Cyclostationarity is a unique feature of many communication signals that can be leveraged to make them robust to interference and noise. Spectral correlation function (SCF) analysis is commonly used to test and analyze cyclostationary signal properties, which is defined as
\begin{equation}
\footnotesize
S_x^\alpha(f)=\lim_{\Delta f\to\infty} \lim_{\Delta f\to\infty} \frac{1}{\Delta t}\int_{-\frac{\Delta t}{2}}^{\frac{\Delta t}{2}} \Delta f X_{\frac{1}{\Delta f}}(t,f+\frac{\alpha}{2}) X_{\frac{1}{\Delta f}}^*(t,f-\frac{\alpha}{2})dt,
\end{equation}
where
\begin{equation}
X_{\frac{1}{\Delta f}}(t,v)=\int_{t-\frac{1}{2\Delta f}}^{t+\frac{1}{2\Delta f}}x(u)\exp(-i2\pi vu)du,
\end{equation}
denotes the complex envelope, $\alpha$ is the cyclic frequency, $\Delta f$ represents the bandwidth, and $\Delta$ denotes the measurement interval. 
Different modulation types can be classified because they have different SCF values.

Ramkumar \cite{ramkumar2009combined} proposed combining cyclostationary feature detection with a predictor-based recursive blind equalizer for AMC. This approach exploits the cyclostationarity property of communication signals to improve classification robustness. Simulations illustrated the concept and yielded promising results. The study demonstrates the potential of leveraging cyclostationary feature detection to enhance AMC performance. 
Satija \cite{satija2014performance} investigated the efficacy of digital modulation classification techniques that employ cyclostationary features in conjunction with various classifiers, including NN, SVM, k-NN, Naive Bayes, Linear Discriminant Analysis, and Neuro-Fuzzy classifiers. The study revealed that the integration of cyclostationary features with Naive Bayes and Linear Discriminant Analysis classifiers resulted in superior classification accuracy while requiring less computational complexity.
Ma \et \cite{ma2018automatic} proposed a new MC method by using the cyclic correntropy spectrum (CCES), effectively suppressing impulsive noise. Simulations demonstrate that the proposed method outperforms other existing schemes in impulsive noise cases, especially under a low SNR condition.

For WTR, Oner \cite{oner2004cyclostationarity} first used the second-order cyclostationay-based feature to identify the global system for mobile communication (GSM) signals. 
Later in \cite{kim2008specific}, the second-order cyclostationarity-based features are employed to identify IEEE 802.11 signals. Such features are also exploited to identify long-term evolution downlink (LTE-DL) versus WiMAX signals in \cite{dobre2011second}, GSM versus LTE-DL signals in \cite{karami2015identification}. 
The authors in \cite{7969734} presented an algorithm that relies on the signal cumulative distribution function as an identification feature and on the Kolmogorov-Smirnov test as the decision criteria.

\paragraph{Other statistical features}
The authors in \cite{brik2008wireless} proposed a radiometric signature-based device identification called PARADIS (Passive RAdiometic Device Identification System), which quantifies the transmitter’s radiometric identity by comparing the signal with an ideal signal in the modulation domain on a frame-by-frame basis. 
The paper \cite{patel2015non} suggested identifying ZigBee devices using non-parametric features like mean, median, mode, and linear model coefficients. These features, extracted from each signal's Region of Interest, were classified with a random forest of 1000 trees. This method achieved over 97\% accuracy for signals with SNR above 10dB and showed up to 9\% improvement at 8dB SNR compared to parametric features. 
Lukacs \et \cite{lukacs2015classification} employed RF-DNA fingerprinting, normalized power spectral density, and discrete Gabor transform to identify ultra-wideband noise radar devices. They used multiple discriminant analysis with maximum likelihood and generalized relevance learning vector quantization-improved classifiers for effective device identification.

\subsubsection{Transform features}

Transform features are used to convert a signal from its original domain to a new domain, where it may be analyzed more effectively. Some common transform features of signals include Fourier transform features, wavelet transform features, $\mathcal{S}$ transform features, and random transform features. 

\begin{table}
\centering
\scriptsize  
\caption{Transform Features}
\label{tab:trans}
\begin{tabular}{m{1.5cm}<{\centering}m{6.5cm}<{\centering}}
\toprule
Transform  & Definition \\
\midrule
Fourier transform & $\displaystyle S_{x}(m  \Delta(f))=\sum_{n} x(n  \Delta(t))  \exp(-j  2  \pi  m  \Delta(t)  n  \Delta(t))$ \\
\midrule
Wavelet transform & $\begin{array}{cc} CWT(a, \tau)=\int_{-\infty}^{\infty} r(t) \psi_{a, \tau}^{*}(t) d t, \\ \psi_{a, \tau}(t)=\frac{1}{\sqrt{a}} \psi(\frac{t-\tau}{a})\end{array}$.\\
\midrule
Haar wavelet transform & $\psi(t)=\left\{\begin{array}{cc}
	1, & \text { if } 0 \leq t<T/2 \\
	-1, & \text { if } T/2 \leq t<T \\
	0, & \text { otherwise }
	\end{array}\right.$\\
\midrule
$\mathcal{S}$ transform & $S(\tau, v)=\int_{-\infty}^{+\infty} g(t) \frac{|v|}{\sqrt{2 \pi}} \exp(-\frac{(t-t)^{2} v^{2}}{2}) \exp(-i 2 \pi v t) d t$\\
\midrule
Radon transform & $\begin{array}{cc}F_{R}(\rho, \theta)=\int f(x, y)dl_{\rho,\theta}\\=\iint f(x, y) \delta(\rho-x \cos \theta-y \sin \theta)dxdy\end{array}$\\
\bottomrule
\end{tabular}
\end{table}

\paragraph{Fourier transform features}
The Fourier transform is an important tool for signal processing and analysis, providing a more efficient representation compared to the time domain. The fast Fourier transform (FFT) has been widely adopted for AMC. The discrete Fourier transform (DFT) is used to obtain the amplitude and phase (AP) in the frequency domain, which are then utilized to classify the modulation type of the observed signal. With the DFT, the carrier frequency can be estimated by evaluating different DFT frequencies $m\Delta(f)$. Proper sampling parameters including the sampling frequency $f$, number of samples $n$, and symbol time $n\Delta(t)$ must be set for the AMC task. Ultimately, the DFT provides the AP of the constellation frequencies present in the received signal $S_x$, enabling effective feature extraction for robust AMC. Further research can build on these Fourier transform-based techniques to advance AMC algorithms, leveraging frequency domain features to reliably classify modulation types in wireless channels. 

Several works have investigated using Fourier transform-based techniques for AMC. Yu \et \cite{yu2003m} developed a fast Fourier transform classifier (FFTC) algorithm that can effectively classify M-FSK signals with only reasonable knowledge of the received signal. They showed the FFTC performs well in classifying 2-FSK to 32-FSK signals at SNRs above 0dB. Liu \et \cite{liu2012novel} proposed using higher-order statistical moments of the wavelet and fractional Fourier transforms as features for an AMC algorithm. Their results demonstrated superior classification performance compared to other existing classifiers. More recently, Zhou \et \cite{zhou2017learning} proposed building a deep learning network to learn the short-time Fourier transform (STFT), a core component of traditional spectrum sensing algorithms, for modulation classification. Zeng \et \cite{zeng2019spectrum} transformed 1D signals into spectrogram images using the short-time discrete Fourier transform and applied CNNs for recognition. Their approach achieved higher accuracy than other deep learning methods. These works highlight the effectiveness of Fourier-based transforms combined with statistical, deep learning, and image processing techniques for robust AMC. Further research can build on these techniques to develop real-time AMC solutions under dynamic wireless environments.

Danev \et \cite{danev2009transient} proposed using FFT-based Fisher features for wireless node identification. Their approach extracts RF fingerprints by first detecting signal transients using a variance-based threshold algorithm \cite{rasmussen2007implications}. The relative difference between adjacent FFT spectra is then determined by applying a 1D Fourier transform on the extracted transients. Fisher feature vectors are extracted from the FFT spectra difference using linear discriminant analysis (LDA) \cite{martinez2001pca}. The LDA matrix is derived from scatter matrices using standard procedures. Finally, fingerprints are matched by calculating the Mahalanobis distance between the reference and test signal templates. This approach demonstrates the use of Fourier transform and statistical methods to extract robust device fingerprints from transient signals. Further research can enhance these techniques using advanced feature extraction and matching methods for improved wireless node identification.

\paragraph{Wavelet transform features}
The continuous wavelet transform (CWT) of a signal $r(t)$ is defined as the inner product of $r(t)$ with scaled and shifted versions of a mother wavelet $\psi(t)$

Among different mother wavelet functions, the Haar wavelet is commonly used for wavelet analysis of wireless signals due to its simple form and low computational complexity. The Haar wavelet is defined as in Table \ref{tab:trans}. 

Several works have proposed AMC techniques using wavelet transform for feature extraction combined with pattern recognition or machine learning methods for classification. Yuan \et \cite{yuan2004modulation} developed an AMC algorithm using wavelet transform and pattern recognition that achieved efficient performance at 15dB SNR. Dan \et \cite{dan2005new} proposed using wavelet analysis with wavelet support vector machine (W-SVM), improving accuracy and efficiency for AMC. Park \et \cite{park2008automatic} presented a wavelet transform and SVM-based method that succeeded with $95\%$ accuracy at 10dB SNR without requiring a priori knowledge. Avci \cite{avci2009selecting} extracted features using discrete wavelet transform (DWT) and adaptive wavelet entropy, then classified signals with a hybrid genetic algorithm-SVM approach outperforming standard SVM. Hassan \et \cite{hassan2010automatic} used HOCs of continuous wavelet transform (CWT) as features and an NN classifier, achieving highly accurate AMC over a wide SNR range. These works demonstrate wavelet transform is effective for feature extraction in AMC.

For WTC, Liu \cite{liu1999identification} and Ho \cite{ho1999improving} proposed a wavelet-based algorithm to identify GSM and UMTS signals. The algorithm uses wavelet transform to extract transient behaviors arising from modulation differences. It then performs template matching in the wavelet domain to identify the signals.

For WII, the authors in \cite{greco2008radar} proposed an adaptive wavelet estimator to detect and classify radar interference signals. Zhao \et \cite{zhao2017discrimination} adopted a likelihood ratio test discriminator to recognize deception jamming. However, maximum likelihood approaches require perfect channel knowledge and have high computational complexity. In contrast, wavelet transform can extract distinguishing features for WIR without needing perfect channel information. By transforming the signal to the time-frequency domain, wavelet analysis can characterize the interference signals based on their spectral correlations and temporal patterns. Therefore, wavelet transform provides an effective and efficient technique for WIR that does not rely on channel estimations or complex optimizations.

For RFFI, wavelet analysis has been used to extract unique device signatures from transient signals \cite{toonstra1995transient,toonstra1996radio}, preambles \cite{klein2009application} and wireless signal responses \cite{choe1995novel,bertoncini2011wavelet}. Toonstra \et \cite{toonstra1995transient,toonstra1996radio} identified FM transmitters by characterizing features in transients using multi-resolution wavelet analysis. Choe \et \cite{choe1995novel} proposed an identifier using Daubechies wavelet transform and neural networks. Klein \cite{klein2009application} achieved 80\% accuracy fingerprinting 802.11a devices from preamble features extracted by discrete wavelet transform. Bertoncini \et \cite{bertoncini2011wavelet} presented a dynamic wavelet fingerprinting technique to identify RFID tags. These works demonstrate wavelet transform can effectively extract device-specific features from wireless signal transients, preambles, and responses for fingerprinting and identification. By analyzing signals in the joint time-frequency domain, wavelet-based techniques can robustly characterize hardware imperfections and transmit signatures for RFFI.

\paragraph{$\mathcal{S}$ transform features}
The Stockwell transform ($\mathcal{S}$-transform) is a time-frequency analysis technique that combines elements of the short-time Fourier transform (STFT) and continuous wavelet transform (CWT). The $\mathcal{S}$-transform of a time domain signal $g(t)$ is defined as in Table \ref{tab:trans}, where $v$ and $\tau$ are the frequency and time coordinates. The Fourier basis functions are modulated by a Gaussian window centered at time $\tau$. This adds time localization to the frequency analysis, similar to the STFT. However, the Gaussian window has a width that adapts with the frequency $f$, making the time-frequency resolution variable like in the CWT. At lower frequencies, the Gaussian window is wider, providing better frequency resolution. At higher frequencies, the window narrows to improve time resolution. Therefore, the S-transform provides a frequency-dependent resolution that balances time and frequency localization. This makes it well-suited for analysis of non-stationary signals with time-varying frequency content.

The Stockwell transform has been explored for AMC due to its time-frequency analysis capabilities. Satija \et \cite{satija2015automatic} extracted features using the S-transform and classified different digital modulation schemes using NN, SVM, LDA, Naive Bayes (NB), and K-nearest neighbors (KNN). They found that S-transform features achieved better classification accuracy and lower complexity compared to wavelet features. Zhao \et \cite{zhao2016anovel} proposed an S-transform-based method for underwater acoustic modulation classification. These works demonstrate the Stockwell transform can effectively extract discriminative features for AMC while providing lower complexity than the wavelet transform. The variable time-frequency resolution of the S-transform allows it to analyze transient, non-stationary characteristics useful for classifying wireless signals.

\paragraph{{\color{blue}Radon} transform features}
The Radon transform is defined as the line integral of a two-dimensional function along all possible straight lines \cite{wood1988performance}. It transforms the function into the Radon domain where each point corresponds to a set of lines characterized by two parameters - $\rho$ and $\theta$. $\rho$ represents the perpendicular distance of the line from the origin, while $\theta$ defines the angle of the normal vector to the line. By integrating the function along lines of all possible orientations, the Radon transform provides a complete characterization of the two-dimensional function. It transforms spatial information into a domain organized by line parameters, enabling analysis of patterns and shapes. 
So, the Radon transform $F_R(\rho, \theta)$ of $f(x,y)$ is given as in Table \ref{tab:trans}. 

The Radon transform essentially converts points in the spatial domain into lines in the radom domain. A single point $f(x,y)=\delta (x-x_o,y-y_o)$ gets transformed into a cosine curve $\rho=\rho_o\cos(\theta-\theta_o)$ where $\rho_o$ and $\theta_o$ parameterize the line passing through $(x_o,y_o)$. Multiple points lying along the same straight line in the spatial domain will intersect at the same $\rho, \theta$ values in the Radon domain. This intersection arises because integrating along the line passing through those points yields identical line integral values. Therefore, the Radon transform maps points on a line to a single point on the corresponding cosine curve. This is how the Radon transform converts spatial information into the line parameter domain.

\subsubsection{Other features} Instead of using the features mentioned above, combined features, feature selection, and similarity-based features can also be utilized as features for WSR. 

\paragraph{Combined features}
Deep neural networks (DNNs) have recently been explored for AMC to leverage their powerful feature learning capabilities. The authors in \cite{lee2017robust} extracted 28 cumulant and spectral features to train a DNN, achieving excellent classification of various modulations in fading channels. Shi \et \cite{shi2019particle} used 12 cumulants and spectral features as input to a DNN optimized by particle swarm optimization, improving convergence and accuracy. The work \cite{huang2016automatic} proposed an AMC framework that first separates overlapped signals using blind channel estimation, then classifies each signal with a maximum likelihood-based multi-cumulant DNN. These works show deep learning can effectively exploit cumulant, spectral, and spatial features for robust AMC across various wireless propagation environments and interference conditions. The nonlinear modeling capacity of DNNs makes them well-suited for distilling discriminative signatures from the complex wireless channel.

\paragraph{Feature selection}
Feature selection techniques have been applied in radio frequency fingerprinting and AMC to improve model robustness and generalization. Bihl \et  \cite{bihl2016feature} proposed a dimensional reduction analysis using eigen-based fusion of FDA loadings to select relevant RF-DNA fingerprint features. This improved device identification accuracy and robustness to noise. Wu \et \cite{wu2017robust} selected noise-insensitive features from a large set to enable robust AMC under varying SNRs. By removing redundant and noisy features, these works show that feature selection can enhance model performance across changing real-world conditions. Selecting the most discriminative features makes RF fingerprinting and modulation classification more reliable without overfitting noise or irrelevant signatures.

\paragraph{Similarity-based features}
Information theoretic learning (ITL) has been explored for AMC to exploit its robust statistical similarity measures. Fontes \et \cite{fontes2012automatic,fontes2015performance} proposed an AMC method using the correntropy coefficient from ITL to compare extracted features. This provided robustness to non-Gaussian noise and outliers. Hakimi \et \cite{hakimi2017optimized} also utilized correntropy as a local similarity measure for distributed AMC across wireless sensor networks. By replacing conventional metrics with ITL-based ones like correntropy, these works achieved more reliable AMC, especially in impulsive noise environments. ITL enables modulation classifiers to overcome complications like low SNRs and interference that violate assumptions of traditional estimation techniques.

\paragraph{Other types of feature extraction}
There are several other types of features to be extracted, such as zero-crossing rate \cite{hsue1989automatic,hsue1990automatic,grimaldi2001automatic}, Mel frequency cepstral coefficients (MFCCs) \cite{al2012automatic}, and phase features \cite{yang1998asymptotic}.

FB methods for WSR, while useful, have distinct disadvantages. 
First, they require precise feature selection and extraction, which can be complex and may not always capture all relevant information about the signal, potentially leading to incomplete or inaccurate classification. Additionally, these methods can be sensitive to variations in signal conditions such as noise, channel effects, and interference, which can degrade performance if the features are not robust enough. Furthermore, the effectiveness of feature-based methods depends heavily on the training data; if the training data isn't comprehensive or representative of all possible signal scenarios, the model may not generalize well to new, unseen conditions. Lastly, these methods can also be computationally expensive, especially when extracting complex features, which can be a limitation in real-time applications.

\subsection{Machine Learning-based Methods}

Feature-based methods for WSR typically rely on multi-stage decision trees with hand-designed thresholds. However, ML techniques can overcome these limitations in two ways. First, they provide easier-to-implement classification using methods like neural networks that avoid complex decision trees. Second, they enable more optimal feature sets through automated feature selection and generation. This allows considering a broader feature space to improve discrimination while maintaining efficiency. Overall, ML complements feature-based WSR by replacing manual decision processes with adaptive classification and identifying the most salient features through dimensionality reduction. This enhances performance and robustness without the burden of designing and optimizing decision thresholds across multiple stages. The combination of in-depth domain knowledge in feature extraction and the learning capacity of ML techniques offers a promising direction for advancing WSR research.

\begin{figure}[t]
	\centering
	  \includegraphics[width=0.99\linewidth]{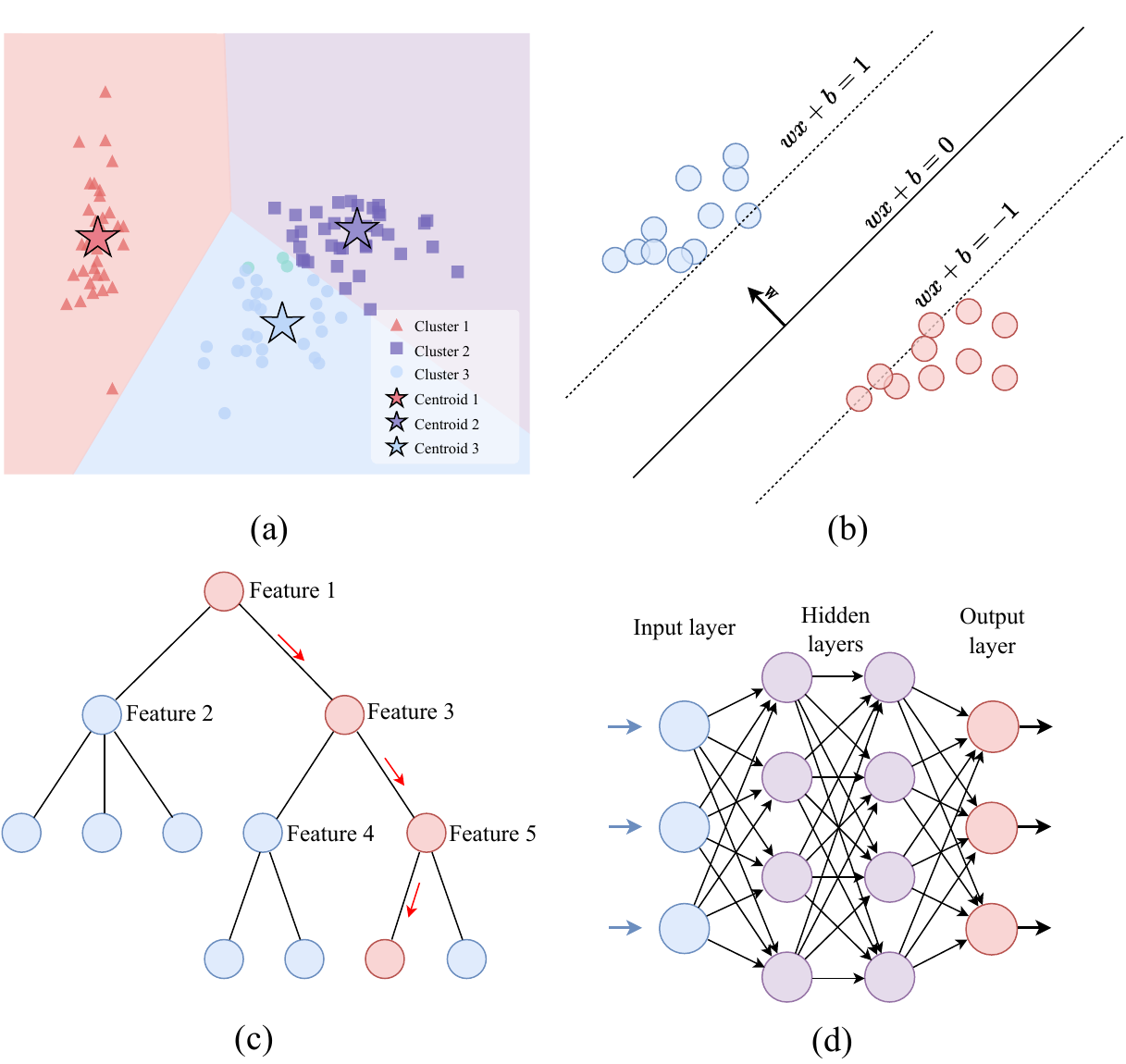}
	  \caption{Different kinds of machine learning algorithms, (a) K-nearest neighbors (KNN), (b) support vector machine (SVM), (c) decision tree (DT), and (d) Artificial neural network (ANN).}
	  \label{fig:ml}
  \end{figure}

\subsubsection{K-nearest neighbors (KNN)} 
KNN is a non-parametric classification method. It assigns a class to a test observation by looking at the $k$ closest training samples in the feature space, as illustrated in Fig. \ref{fig:ml} (a). There are three main steps in KNN classification:
\begin{itemize}
	\item \emph{Reference feature space}: 
	To enable accurate KNN classification, the feature space must contain sufficient reference samples from each modulation class. Typically, $M$ reference feature vectors per class are needed to characterize the distribution of possible test signals. The reference samples should span the diverse range of expected variations within each class, such as different channel effects, noise levels, and interference. With a robust set of reference features covering intra-class variabilities, the KNN algorithm can effectively interpolate to classify unknown test samples based on their local neighborhood relationships. Careful design of the training dataset is crucial so that proximity patterns in the feature space align with the modulation labels. This allows KNN to exploit local feature similarities to generalize well to new test data.

	\item \emph{Distance definition}: 
	The KNN classifier assigns a data point to the class that is most common among its K nearest neighbors, as determined by a distance function. One of the most important distance metrics used in the KNN classifier is the Euclidean distance. 
	Given a set of features $\mathbb{F}=\{\mathbb{F}_1,\mathbb{F}_2,\cdots,\mathbb{F}_L\}$, the Euclidean distance is calculated by using the characteristic sets of signals $A$ and $B$ according to $L$  characteristics as $D(\mathbb{F}(A),\mathbb{F}(B))=\sqrt{\sum_{l=1}^L[\mathbb{F}(A)_l-\mathbb{F}(B)_l]^2}$.
	\item \emph{KNN decision}: 
	Once distances to all training samples are computed, the k-closest samples are selected as the nearest neighbors. The value of $k$ is usually chosen to be an odd number to avoid tied votes. A larger k reduces noise effects but makes decision boundaries less distinct. The test sample is assigned to the class that appears most frequently among the k nearest neighbors. With enough reference samples per class and a properly sized $k$, the majority class of the local neighborhood provides a robust classification outcome even with some outliers. The key aspects are having a sufficiently large and representative training dataset, and selecting a suitable $k$ through validation. This allows the local proximity structure in the feature space to generalize well to new test data.
\end{itemize}

Zhu \et \cite{zhu2010augmented} were the first to propose utilizing genetic programming (GP) with a KNN classifier for automatic classification of digital modulation types, including BPSK, QPSK, 16QAM and 64QAM. Their simulation experiments demonstrated that this proposed method was able to successfully identify the above modulations at SNRs of both 10dB and 20dB. Building on this work, Aslam \et \cite{aslam2012automatic} developed a more generalized digital AMC algorithm, still employing GP with KNN, but using HOCs as the input features. More recently, Yang \et \cite{yang2019adaptive} proposed an adaptive spatial modulation MIMO (SM-MIMO) system based on machine learning, using KNN and SVM algorithms to achieve statistically consistent solutions with superior performance and lower complexity compared to conventional optimization-driven designs.

For RFFI, Kennedy \et \cite{kennedy2008radio} utilized KNN for radio transmitter identification based on frequency domain characteristics in their work on RFFI. Their results demonstrated that this approach achieved $97\%$ accuracy in identifying eight identical USRP transmitters at a SNR of 30dB, and 66\% accuracy at an SNR of 0 dB.

\subsubsection{Logistic regression (LR)} 
LR is a method used to construct a linear combination of features to separate two categories or classes. Suppose there are $k$ existing features that are merged into a new feature. The linear combination for logistic regression can be evaluated using the following equation
\begin{equation}
\mathbb{F}_{new}=\omega_0+\sum\limits_{k=1},^K\omega_k\mathbb{F}_k
\end{equation}
where $\omega_0$ is a constant value, $\omega_k$ is the weight of the $k$th feature $\mathbb{F}_k$, and $K$ is the total number of the combined features. LR is the process of optimizing these weights, with the aim of maximizing the difference in the new feature values between the various classes. It has been adopted by Zhu \et \cite{zhu2013robustness} for dimension reduction when using distribution-based features. There are two common algorithms for LR: binomial logistic regression and multinomial logistic regression. LR projects the signal using a logistic function $p(\cdot)$ where $p(\cdots)=1$ when $\mathcal{M}(i)$ is used and $p(\cdots)=0$ when $\mathcal{M}(j)$ is applied. 
\begin{equation}
p(\mathbb{F})=\frac{1}{1+exp(-g(\mathbb{F}))}=\left\{
\begin{aligned}
&~1, \mathrm{~when~} \mathcal{M}(i) \\
&~0, \mathrm{~when~} \mathcal{M}(j) ,
\end{aligned}
\right.
\end{equation}
{\color{blue}
where $\mathbb{F}$ denotes the collection of existing features and the inverse of the logistic function $p(\cdot)$, $g(\mathbb{F})$ is the logit function, which is given by}
\begin{equation}
g(\mathbb{F})=B_0+\sum\limits_{k=1}^K B_k\mathcal{F}_k.
\end{equation}

Using iterative methods, the parameters $B_0$ and $B_k$ can often be estimated. LR provides an important tool for selecting and combining features. However, multi-class classification is not always optimal when using LR for feature selection and combination.

Jiang \et \cite{jiang2018feature} proposed a novel modulation classification method called LRGP for overlapped sources, which utilizes multifaceted genetic group programming-based feature engineering to transform received signal cumulants into highly discriminative features. It also uses a LR-based classifier trained to identify combinations of modulation formats instead of signal separation. Extensive simulations demonstrated superior performance for LRGP compared to existing methods.

\subsubsection{Support vector machine (SVM)} 
SVM provides an alternative approach to classification in the present feature space with multiple dimensions. SVM achieves classification by identifying a hyperplane that separates data points belonging to distinct categories. The optimal hyperplane is obtained by maximizing its distance to the nearest training samples on either side. SVM classifiers can be divided into linear and nonlinear versions, depending on the type of data being classified. Linear SVM classifiers make use of linear kernel functions. A linear kernel can be defined as
\begin{equation}
k(\mathbf{x},\mathbf{w})=\mathbf{x}^T\mathbf{w},
\end{equation}
where $\mathbf{w} = [w_1,w_2,\cdots,w_K]$ is the weight vector to be optimized and $\mathbf{x} = [x_1,x_2,\cdots,x_K]$ is the input feature vector $\mathbb{F}=\{\mathbb{F}_1,\mathbb{F}_2,\cdots,\mathbb{F}_k\}$. 
The kernel defines a linear separation hyperplane as
\begin{equation}
g(\mathbf{x})=\mathbf{x}^T\mathbf{w}+w_0,
\end{equation}
where $w_0$ is a constant.

Sengur \et \cite{sengur2009multiclass} presented the application of multiclass least-squares SVM (MC-LS-SVM) for classifying analog modulated signals, achieving $100\%$ correct classification in simulations using 10-fold cross-validation. The authors in \cite{wang2009algorithm} proposed a new algorithm based on HOCs and SVM for recognizing six digital modulations, including 2ASK, 4ASK, 8ASK, 4PSK, 8PSK and 16QAM. The algorithm utilized fourth and sixth-order cumulants as features. The work \cite{zhang2014automatic} extracted several statistical features to represent signals, then applied SVMs to classify unknown modulation schemes. Their results showed the proposed algorithm had high robustness over a wide SNR range. Li \et \cite{li2017automatic} combined SVM and statistical and spectral features for digital MC, achieving higher efficiency than other methods at low SNRs. Müller \et \cite{muller2011front} presented a novel discriminative learning-based SVM method for AMC. It utilized the ordered magnitude and phase of received symbols at the matched filter output as features.

\subsubsection{Decision tree (DT)}
DT is a non-parametric supervised learning method that provides an interpretable and computationally efficient approach for classification. DTs have lower complexity compared to other classifiers, requiring less memory, but their accuracy can suffer in noisy conditions or when insufficient training features are available. DT classifiers are binary, recursively partitioning the feature space into regions associated with class labels. They can be categorized into fine-tree, coarse-tree, and medium-tree types based on the number of leaves or terminal nodes. Fine-tree DTs have a large number of leaves, enabling high classification accuracy suitable for problems with many classes. Coarse-tree DTs minimize the number of leaves to improve robustness and interpretability for problems with few classes, although at the cost of reduced accuracy. Medium-tree DTs aim for a balance, using a moderate number of leaves to achieve better accuracy than coarse trees.

Grimaldi \et \cite{grimaldi2007automatic} and De Rore \et \cite{de2008improved} presented decision tree approaches for digital MC without requiring a priori knowledge of the signal parameters. The work \cite{kharbech2016classifiers} designed and compared four commonly used classifiers for feature-based AMC including decision trees, K-nearest neighbors, artificial neural networks, and support vector machines. Their comparison showed artificial neural networks provided the best performance-complexity tradeoff. Venkata \et \cite{venkata2018automatic} proposed new supervised learning algorithms for MC using decision trees. The work \cite{triantafyllakis2017phasma} incorporated a random forest classifier for AMC in their proposed architecture, achieving the classification of various digital and analog modulations under different SNRs. The authors in \cite{clark2019developing} compared two expert system architectures to traditional image processing architectures. The first utilized decision fusion on binary classifiers while the second employed a hierarchical decision tree relying on expert knowledge. Although performance gains were minor compared to standard approaches, the expert architectures provided greater adaptability, interpretability, and future-proofing.

\subsubsection{Hidden Markov model (HMM)}
An HMM is a Markov model in which the system is represented as a statistical process with hidden states. The underlying states are not directly observable, and instead, the observations are based on the emission properties of the states. In mathematical terms, an HMM can be defined as the pair ${X_t,Y_t;t\in N}$ in a probability space.
$Y_t$ denotes the observation sequence, and $(X, Y)$ is finite and can be regarded as a stationary finite-state system (SFSS), given by
\begin{align}
\Pr(Y_{t+1}=y_{t+1},X_{t+1}=x_{t+1}|Y_1^t=y_1^t,X_1^t=x_1^t) \nonumber\\
=\Pr(Y_{t+1}=y_{t+1},X_{t+1}=x_{t+1}|X_t=x_t),
\end{align}
where $X_t$ and $Y_t$ denote the state and output processes, respectively, of a state-space system. The model for a hidden Markov process (HMP) is referred to as a hidden Markov model (HMM). Consider a discrete HMM with $N$ states and $M$ symbols characterized by a state transition probability matrix $\mathbf{P}\in \mathbb{R}^{N\times N}$, an output symbol probability matrix $\mathbf{B}\in \mathbb{R}^{N\times M}$, and an initial state probability vector $\mathbf{S}\in \mathbb{R}^N$. This HMM can be defined as  $\zeta=\{\mathbf{P},\mathbf{B},\pi\}$, with its model parameters optimized through the Baum-Welch algorithm (BWA). The BWA is an expectation-maximization (EM) algorithm specialized for HMMs. 
The probability of generating the observation sequence from the model can be given by
\begin{equation}
\Pr(y_1^T|\zeta)=\pi \mathbf{B}(y_1) \mathbf{P} \mathbf{B}(y_2) \mathbf{P} \cdots \mathbf{P} \mathbf{B}(y_T)\mathbf{1}^t
\end{equation}
where $\mathbf{B}(y_k),k=1,2,\cdots,T$ denotes the generated symbol probability from different states. 
$\Pr(y_1^T|\zeta)$ is usually treated as a log-likelihood logarithm.

In \cite{kim2007cyclostationary}, the authors employed the cycle frequency domain profile (CDP) technique for signal detection and preprocessing, followed by a threshold-test method to extract signal features for classification. To leverage the robust pattern-matching ability of the HMM, the extracted signal features were processed using the HMM. The results showed that the HMM-based classifier was effective in classifying signals with low SNRs.
Zhang \et \cite{zhang2018continuous} investigated the classification of continuous phase modulation (CPM) signals under unknown fading channels. The time-varying phases of CPM were first formulated as an HMM by observing the memorable properties of CPM. A likelihood-based classifier was then proposed using the Baum-Welch algorithm, which is able to estimate the unknown parameters in the HMM. Simulation results demonstrated that the proposed algorithm outperformed the existing scheme utilizing approximate entropy in terms of classification accuracy.

\subsubsection{Naive Bayes classifiers}
As a kind of classic statistical algorithm, naive Bayes classifiers depend on Bayes' theorem by assuming that each feature is statistically independent of all other features \cite{wong2008naive}. 
Providing the feature set $f_1,f_2,\cdots,f_N$, the modulation type with the largest probability is chosen by the naive Bayes classifiers, given by
\begin{equation}
m_i=\underset{m_j\in M}{\arg\max} \Pr(m_j) \prod\limits_{i\in 1,2,\cdots,N} \Pr(f_i|m_j)
\end{equation}
where $m_i$ denotes the candidate signal types and the priors $\Pr(m_i)$ are derived using the prior knowledge from the training set. The conditional probabilities $\Pr(f_i|m_i)$ are then generated.

Wong \et \cite{wong2008naive} evaluated the use of higher order statistical measures coupled with a classical Naive Bayes classifier for fast identification of adaptive modulation schemes. Their approach achieved better performance compared to maximum likelihood classifiers and SVM-based classifiers. In related work, Satija \et \cite{satija2014performance} studied the performance of digital MC based on cyclostationary features and different classifiers. They found that combining cyclostationary features with Naive Bayes and Linear Discriminant Analysis classifiers led to improved classification accuracy with lower computational complexity. Further, Mughal \et \cite{mughal2018signal} proposed a new technique for signal classification and jamming detection in wide-band radios using key spectral features and the naïve Bayes classifier. Their proposed algorithm demonstrated better performance compared to a recently proposed feature-based jamming detection algorithm.

\subsubsection{Artificial neural network (ANN)} 
ANN classifiers are supervised ML algorithms that require training on labeled data. To integrate new features with lower dimensions and improved performance, ANN classifiers can be utilized to combine existing features and learn non-linear transformations of these features. For example, in a single-layer perceptron network, the trained network performs a linear combination of the input features. This can be represented mathematically as
\begin{equation}
\mathbb{F}_o=w_0+\sum\limits_{k=1}^K w_k\mathbb{F}_i(k).
\end{equation}

Multi-layer perceptron (MLP) is one of the most widely used neural network architectures due to its simple design and efficient hardware implementation. The MLP is a feedforward network topology consisting of single nonlinear processing units, termed neurons, arranged in layers. Inputs are propagated layer-by-layer across the network, enabling the MLP to learn nonlinear mappings from the input to output space. Specifically, the MLP performs a nonlinear combination of the input features, which can be expressed mathematically as
\begin{equation}
y_k=\phi(\sum\limits_{i=1}^q w_{ki}\phi(\sum\limits_{j=1}^p w_{ij} x_j)),
\end{equation}
where $w_{ij}$ is the weight value from neuron $j$ to neuron $i$, and $\phi$ is the activation function. 
$x_j$ is the $j$th input feature $\mathbb{F}_{i}(j)$ from the feature set. 
$y_k$ is the output of the MLP network and the combination of the features $\mathbb{F}_{o}(j)$ being optimized at the $k$th output node of the MLP network illustrated in Fig. \ref{fig:ml} (d).

where $w_{ij}$ is the weight value connecting the $j$th input neuron to the $i$th output neuron, $\phi$ denotes the activation function, $x_j$ is the $j$th input feature $\mathbb{F}_i(j)$ from the full input feature set, and $b_k$ is the bias term for the $k$th output neuron. The output $y_k$ represents the nonlinear combination of the input features $\mathbb{F}o(j)$ being optimized at the $k$th output node of the MLP network architecture shown in Fig. \ref{fig:ml}(d). Through iterative training, the MLP learns the optimal set of weights $w{kj}$ and biases $b_k$ that minimize the error between the target outputs and the network's predictions. This enables the MLP to approximate complex nonlinear functions for classification and pattern recognition tasks.

Zhao \et \cite{yaqin2003automatic} proposed a modified architecture and learning algorithm for ANNs to recognize baseband signal modulation types under additive white Gaussian noise. Specifically, they developed an ANN classifier with a straightforward structure and an optimized learning algorithm tailored for noisy environments. Simulation experiments demonstrated that their approach was effective at low signal-to-noise ratios, achieving high overall classification success rates. 
Wong \et \cite{wong2004automatic} investigated the use of backpropagation with momentum and an adaptive learning rate to accelerate the training of ANNs for AMC. Through genetic algorithm-based feature selection, they identified an optimal subset of just six features that enabled their ANN classifier to achieve 99\% recognition accuracy across a wide range of signal-to-noise ratios. 
The work \cite{popoola2011automatic} outlined a three-step approach for developing an AMC system, which involves extracting statistical feature keys, developing an ANN-based classifier, and evaluating its performance. The results demonstrate that the developed classifier has a high success rate of over $99.0\%$ in accurately classifying nine modulation schemes.
The authors in \cite{jagannath2018artificial} proposed statistical features for AMC signals and designed an ANN-based classifier that performs well over a wide range of SNRs. The authors implemented the proposed architecture on a software-defined radio (SDR) testbed, demonstrating its practical feasibility and superior performance compared to a hybrid hierarchical AMC (HH-AMC) system.

\subsubsection{Other classifiers} classifiers such as fuzzy classifiers \cite{liu2011automatic,zhang2018novel}, polynomial classifiers \cite{abdelmutalab2014automatic,abdelmutalab2016automatic,abdelmutalab2016automaticieee}, auto-encoder \cite{dai2016automatic,ali2017automatic,ali2017automaticelse,ali2017unsupervised,xu2019deep} have also been applied for WSR. 
Liu \et \cite{liu2011automatic} proposed a new MC method combining clustering and neural network techniques, in which they introduced a novel algorithm for feature extraction. Simulation experiments demonstrated that their combined clustering and neural network approach achieved significantly higher classification accuracy compared to using clustering alone. 
Zhang \et \cite{zhang2018novel} presented an AMC method for M-QAM signals based on an adaptive fuzzy clustering model. Through Monte Carlo simulations and theoretical analysis, they demonstrated that their proposed fuzzy clustering-based approach provided promising AMC performance for M-QAM signals. Compared to existing AMC methods, the model developed by Zhang \et \cite{zhang2018novel} offered robustness, flexibility, and strong classification capabilities for the modulation types of interest. Their work highlights the potential of fuzzy clustering techniques for MC in modern communication systems.

Ahmad \cite{ahmad2010fuzzy} presented a new fuzzy logic (FL) approach for classifying WLAN, Bluetooth (BT), and FSK signals. They utilized power spectral density (PSD) information to extract key signal features like bandwidth and center frequency, which were used for labeling signals to their corresponding standards. Their results demonstrated that the proposed FL strategy efficiently extracted explicit discriminative features for wireless signal classification. 
In follow-on work, Ahmad \et \cite{ahmad2010neuro} developed a neuro-fuzzy signal classifier (NFSC) to recognize nanoNET, WLAN, Atmel, and BT signals, again leveraging measured PSD data. Through real-time coexistence experiments, they showed improved classification performance by integrating both wideband and narrowband data acquisition modes into their NFSC design. Their work highlighted the capabilities of neuro-fuzzy approaches for automated wireless signal classification using accessible spectral measurements.

ML-based methods have been widely applied in WSR, providing a more flexible and adaptive approach to WSR compared to FB methods.
However, they still face several limitations. 
Firstly, they still rely on FB methods for feature extraction, which can be complex and may not always capture all relevant information about the signal, potentially leading to incomplete or inaccurate classification. 
Additionally, they require large amounts of labeled training data to perform effectively, which can be difficult and expensive to obtain. 
Furthermore, these methods may struggle to adapt to new or changing signal conditions unless continuously retrained or updated, which can be resource-intensive.

\subsection{Summary of Model-based Methods}
Through comprehensive review of model-based WSR methods, several key lessons emerge. Likelihood-based methods excel in theoretical optimality but suffer from high computational complexity and sensitivity to model assumptions. While feature-based approaches offer better computational efficiency and robustness, their performance heavily depends on expert feature selection and may not generalize well to new scenarios. Machine learning-based methods strike a balance between performance and complexity, but require significant training data and careful parameter tuning. The progression from likelihood-based to machine learning-based approaches reflects a shift from theoretical optimality to practical applicability. For real-world deployments, hybrid approaches combining multiple model-based methods often achieve the best results by leveraging their complementary strengths. Furthermore, the choice of method should be guided by specific application requirements - likelihood-based methods for scenarios demanding theoretical guarantees, feature-based methods for resource-constrained systems, and machine learning-based methods for complex environments with abundant training data.

\section{Intelligent Wireless Signal Recognition}\label{sec:advanced_methods}

ML-based models can learn to separate the characteristics defined by the expert features from a data perspective. 
However, the generated characteristics from the FB models are not accurate for WSR. 
Thus, DL-based methods are proposed to execute the feature extraction and classification simultaneously in a unified framework. 
In this part, we introduce the advanced intelligent WSR methods from different aspects, namely, model, data, learning and others. 
From the model perspective, models such as DNN, DBN, convolutional neural network (CNN), recurrent neural network (RNN), and some hybrid models are given. 
Moreover, model fusion and model compression are also introduced. 
For data, image-based data such as constellation images and eye diagrams, and sequence-based data such as IQ samples, and pole features are presented. 
Moreover, DL models with different types of data can \lq\lq learn\rq\rq from these data in various ways, such as multi-task learning, curriculum learning, and transfer learning. 

\subsection{Model} 
In the field of deep learning, the model refers to the design of different kinds of DNNs. For instance, DBNs, convolutional neural networks (CNNs), recurrent neural networks (RNNs), transformers, hybrid models, and heterogeneous models are given. Moreover, model compression and other techniques are also introduced.

\begin{figure*}[t]
	\centering
	\includegraphics[width=0.85\linewidth]{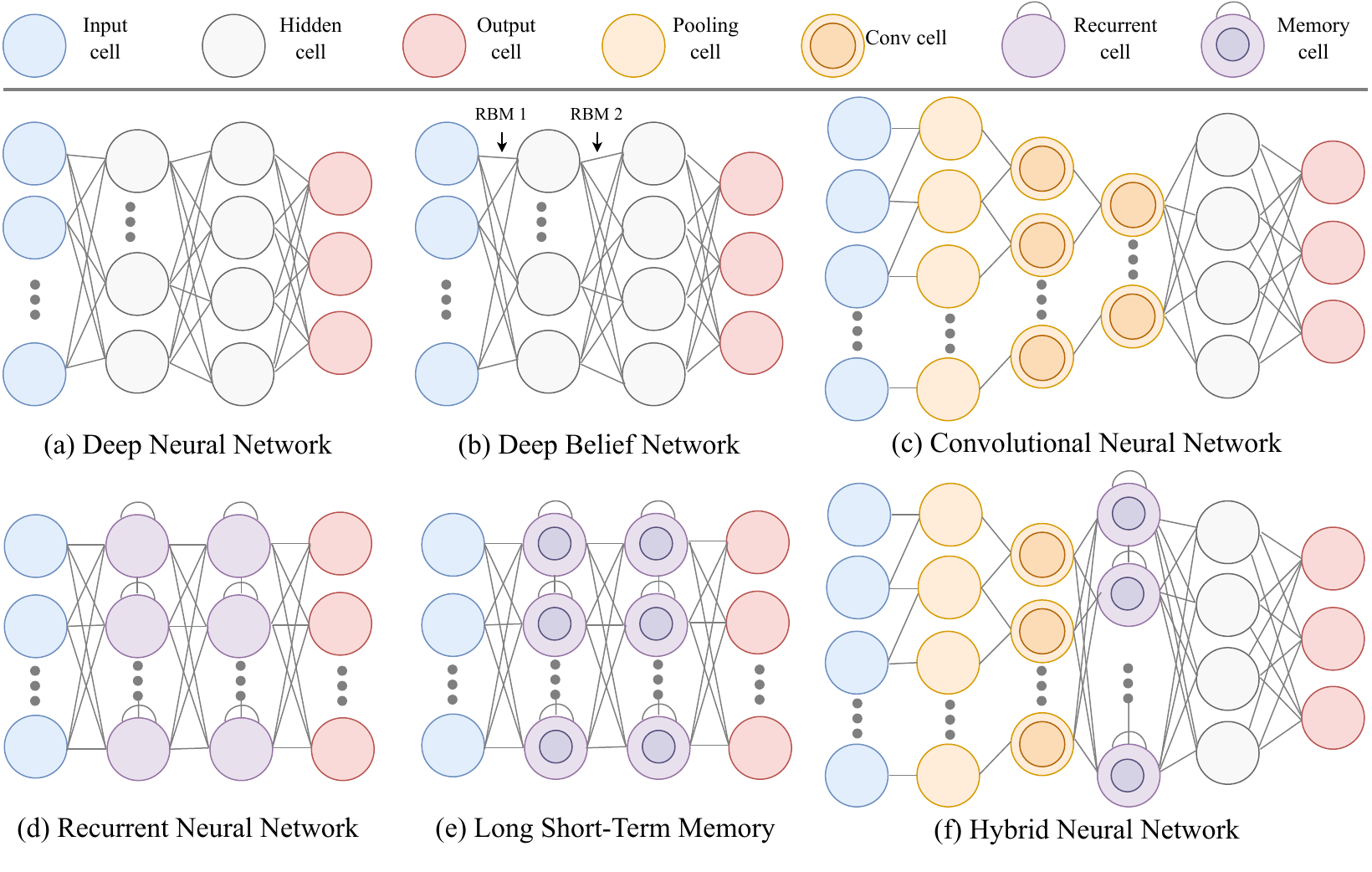}
	\caption{Basic structures of typical deep learning models, (a) deep neural network, (b) deep belief network, (c) convolutional neural network, (d) recurrent neural network, (e) long short-term memory, (f) hybrid neural network.}
	\label{fig:nn}
\end{figure*}

\subsubsection{Deep neural networks (DNNs)} 
DNNs can be treated as a deeper version of an artificial neural network (ANN) with multiple hidden layers, as shown in Fig \ref{fig:nn} (a). 
The authors in \cite{kim2016deep} explored the use of DNN for AMC, selecting 21 statistical features and implementing a fully connected DNN with three hidden layers. The results demonstrated significant performance improvements over existing classifiers, particularly in high Doppler fading channels.

\subsubsection{Deep belief networks (DBNs)} 
DBN is a kind of probabilistic generation model that can constitute a universal distribution between labels and observed data. 
As illustrated in Figure \ref{fig:nn}(b), a DBN comprises multiple layers of restricted Boltzmann machines (RBMs), which are energy-based models capable of capturing complex data distributions. Each RBM contains two layers, including a visible layer and an upper hidden layer. The hidden layer of the first RBM (RBM1) learns to encode features from the input layer $i$, with the data then serving as input to the second RBM (RBM2). For supervised training with labeled data, the visible layer of the final RBM includes both the hidden layer of the previous RBM, as well as the labeled output unit. Through this multi-layer stacked RBM architecture, DBNs can learn robust feature representations that capture complex statistical relationships in the input data for classification and pattern recognition tasks.

DBNs have been explored for WSR in recent years. Mendis \et \cite{mendis2016deep,mendis2017deep} utilized DBNs for AMC, taking the spectral correlation function (SCF) as a data pre-processing technique. The 3D SCF patterns of received modulation signals were transformed into 2D SCF patterns. The resulting 2D SCF gray-scale images were then used to train the DBN in a semi-supervised manner. A similar scheme was adopted in \cite{mendis2016deep1} to detect and identify micro unmanned aerial systems. Ma \et \cite{ma2016dbn} used amplitude information and spectrum of receiving signals as training data for their DBN-based classifier. A limitation of these approaches is that the classification accuracy of noisy PSK signals tends to be lower, as phase information is more obscured in the training data. More commonly, DBNs are employed for feature extraction, which allows their use in various classification frameworks. Wang \et \cite{wang2019deep} proposed combining DBN feature extraction with a SVM classifier, using the DBN to extract features that are classified by the SVM. Ghasemzadeh et al. \cite{ghasemzadeh2020new,ghasemzadeh2020novel} investigated individual filter bank-based AMC methods, and proposed an adaptive framework that intelligently switches between these FB classifiers to optimize the balance between accuracy and execution time. Simulation results showed DBNs can achieve substantially higher classification accuracy compared to other techniques.

\subsubsection{Convolutional neural networks (CNNs)}  
CNN is a kind of NN that depends on convolution operation instead of general matrix multiplication. 
CNNs consist of three types of layers, namely, convolutional layers, pooling layers, and fully-connected (FC) layers, with the convolutional layers carrying out most of the calculations. The convolutional operation is characterized by sparse connectivity, parameter sharing, and pooling operation, which increase statistical efficiency and reduce memory requirements compared to traditional neural networks like ANN/DNN. The pooling layer performs down-sampling and prevents over-fitting by replacing the output of a statistical operation at a specific spatial position, with max-pooling and average-pooling being the two main pooling methods.
LeCun \et \cite{lecun1989backpropagation} first proposed a convolutional neural network, LeNet, for handwritten number image classification, and it defines the basic structure of modern convolutional neural networks. 
After that, CNN grows deeper in structure and various learning and optimization methods are developed. 
The representative CNN architectures are AlexNet \cite{krizhevsky2012imagenet}, VGGNet \cite{simonyan2014very}, GoogLeNet \cite{szegedy2015going},and ResNet \cite{he2016deep}. 
CNNs have achieved unprecedented performance compared to traditional methods including FB and ML methods. 
However, it is still hard to design an efficient architecture for WSR. 

CNNs are widely applied for WSR, especially for AMC \cite{OShea2016,zhang2017modulation,OShea2018,liu2018method,meng2018automatic,sun2018automatic,nie2019deep,wu2019convolutional,zhou2019robust,wang2019data,wang2020deep,li2020deep,hermawan2020cnn,zhang2020autocorrelation}. 
For example, the authors in \cite{OShea2016} first proposed a three-layer CNN for AMC and achieved comparable performance compared to a relatively well-expert-regarded approach. 
Then, different architectures of CNN have been applied for AMC, such as ResNet-based CNN \cite{OShea2018}, VGG-based CNN \cite{liu2018method}, and other deep CNNs. 

For WTC, CNNs have been explored using time-domain features for training. Kulin \et \cite{kulin2018end} used IQ vectors and amplitude/phase (AP) vectors to train CNN classifiers. Their results demonstrated that the proposed scheme was well-suited for recognizing ZigBee, WiFi, and Bluetooth signals. Selim \et \cite{selim2017spectrum} also employed AP representations to train a CNN, showing successful recognition of radar signals even with coexisting LTE and WLAN signals. These studies indicate that time domain features and CNN classifiers can achieve effective WTC, robust to interference from other wireless signals.

CNNs have also been explored for wireless interference identification (WII). CNNs were proposed for barrage jamming detection and classification by Junfei et al. \cite{junfei2018barrage}. Wang \et \cite{wang2019recognition} applied CNNs to recognize active jamming, demonstrating their effectiveness in distinguishing active jammers. The authors in \cite{schmidt2017wireless} proposed a CNN-based WII approach that outperformed state-of-the-art interference identification methods. Zhang \et \cite{zhang2019deep} investigated various CNN architectures for interference recognition, presenting a generalization analysis. Qu \et \cite{qu2020jrnet} proposed the JRNet architecture, using an asymmetric convolutional structure to enhance the recognition ability for WII. Their simulations showed effective classification of interference even at high noise levels. Wang \et \cite{wang2022low} addressed training quantized CNNs with low-precision weights and activations for WII. Optimizing for low-bit widths is challenging due to the non-differentiable quantization function, but can improve efficiency. Overall, CNNs have proven effective for robust WII under various conditions.

The authors \cite{sankhe2019oracle} proposed a CNN framework for RFFI named ORACLE (Optimized Radio clAssification through Convolutional neuraL nEtworks) by utilizing IQ samples to separate a unique radio from a large amount of similar devices. Simulation results demonstrate that ORACLE achieves $99\%$ classification accuracy while balancing computational time and accuracy. 
Shen \et \cite{shen2021radio} proposed an RFFI scheme for Long Range (LoRa) systems based on the spectrogram and CNN, achieving a performance of 97.61\% for 20 LoRa devices under test.

\subsubsection{Recurrent neural networks (RNNs)}
RNNs have been proposed for modeling time-series data to overcome the limitations of DNNs and CNNs. RNNs have been widely applied in natural language processing, speech recognition, handwriting recognition and other sequential modeling tasks. The special memory function of RNNs allows them to process data that is not independent across time steps. Theoretically, there is no limit on the sequence length RNNs can process. However, in practice, long sequences cannot be handled due to the problem of vanishing or exploding gradients. 

Hong \et \cite{hong2017automatic} proposed a novel AMC method based on RNNs, which exploits the temporal sequence characteristics of received communication signals without needing manual feature extraction. Their proposed RNN-based method outperformed a CNN-based approach, particularly for SNRs above -4dB. This demonstrates that RNNs can effectively leverage temporal dependencies for robust AMC.

To address the vanishing and exploding gradient problems of conventional RNNs, a variant called long short-term memory (LSTM) was developed \cite{hochreiter1997long}. LSTM has an input gate, memory cell, forget gate, and output gate. The forget gate determines which data to exclude from the cell state. The input gate decides what values to update in the cell state. The output gate controls the extent to which the cell state value is used to compute the LSTM unit's activation. The final output depends on both the output gate and cell state. 
Chen \et \cite{chen2020automatic} proposed a single-layer LSTM model with an attention mechanism for AMC. Their model uses signal embedding to enhance modulation information in the input and applies weighting to the LSTM hidden state outputs in an attention module to capture temporal features of modulated signals. This resulted in faster convergence and better classification performance compared to a model without attention. Ke \et \cite{ke2022realtime} presented an LSTM denoising autoencoder framework to extract robust features from noisy radio signals for modulation and technology type inference. Their algorithm achieved state-of-the-art accuracy with a compact architecture suitable for low-cost hardware, outperforming current methods on both synthetic and over-the-air data. These studies demonstrate that LSTM can effectively exploit long-term dependencies in signals for AMC.

Gated recurrent units (GRUs) are a type of RNN similar to LSTM but with a simpler architecture. A GRU contains an update gate and a reset gate. The update gate controls how much previous information is retained, while the reset gate determines how much of the previous state is discarded. Compared to LSTM, GRU has fewer parameters and lower computational complexity, making it more efficient for resource-constrained applications. Utrilla \et \cite{utrilla2020gated} proposed a GRU neural network solution tailored for AMC on resource-limited IoT devices. Huang \et \cite{huang2020automatic} developed a gated recurrent residual neural network (GrrNet) for feature-based AMC using the AP of received signals as input. GrrNet employs a ResNet module to extract representative features, and a GRU to capture temporal information. Simulations showed it outperforms other recent DL-based AMC methods. These works demonstrate GRU can achieve efficient and accurate AMC while requiring less computation than LSTM. Their simplified architecture makes them suitable for embedded and real-time AMC implementations.

\subsubsection{Transformers} 
\begin{figure}
\centering
\includegraphics[width=0.8\linewidth]{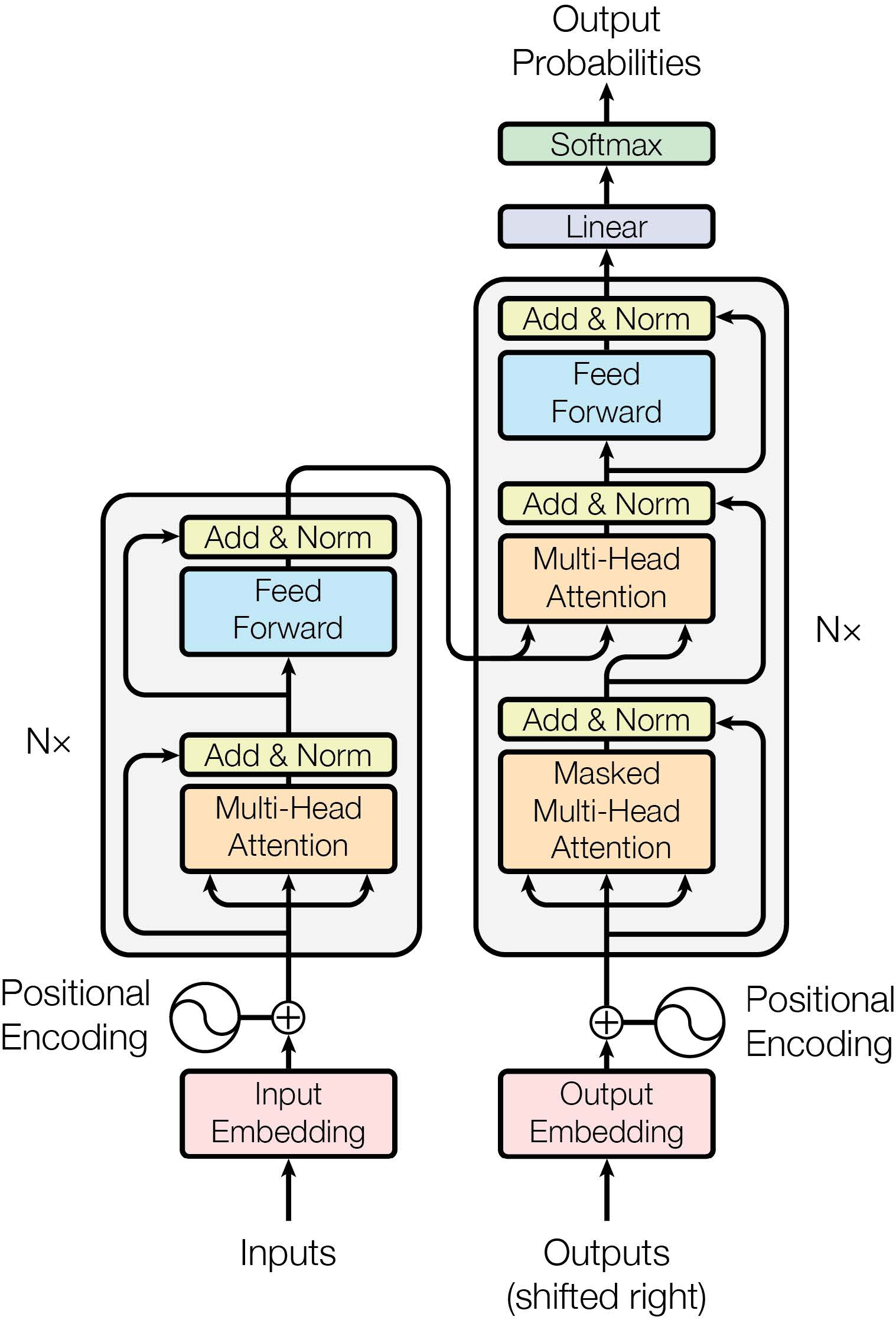}
\caption{Transformer architecture \cite{vaswani2017attention}.}
\label{fig:transformer}
\end{figure}

Transformers \cite{vaswani2017attention} are a type of neural network architecture that has revolutionized natural language processing and various other domains since their introduction in 2017. Originally designed for machine translation tasks, transformers utilize a mechanism called self-attention, which allows the model to weigh the importance of different parts of the input data when processing each element. This approach enables transformers to capture long-range dependencies and context more effectively than previous architectures such as RNNs. The key innovation of transformers is their ability to process entire sequences in parallel, leading to faster training and improved performance on many tasks. Their success has led to the development of powerful language models such as BERT \cite{devlin2018bert} and GPT \cite{radford2018improving}, which have achieved SOTA results in numerous NLP tasks and have been adapted for use in other fields such as computer vision and audio processing. The architectural of transformers is shown in Fig. \ref{fig:transformer}. 

The authors in \cite{hamidi2021mcformer} proposed a transformer-based architectural, MCformer by combining  convolutional and self-attention layers to achieve state-of-the-art accuracy in AMC of radio signals, while using fewer parameters for efficient operation. 
Hu \et \cite{hu2023feature} investigated FCAformer, a novel approach for AMC that combines Markov Transformation Field with I/Q sequences, uses a new patchify module for efficient feature extraction, and employs a convolution-aided encoder with improved attention mechanism, resulting in higher accuracy and lower parameter count compared to other methods on the RadioRML2016.10a dataset. 
Similarly, works \cite{cai2022signal,ying2023convolutional,dao2023vt} also explored the use of transformer-based models for AMC, demonstrating their effectiveness in improving classification accuracy. 

While transformer-based large models such as BERT and GPT have been applied to various aspects of wireless communications, including spectrum sensing \cite{shao2024wirelessllm} and channel estimation \cite{liu2024llm4cp}, their use in WSR remains unexplored.

\subsubsection{Hybrid models} 
Hybrid neural network models refer to combining different architectures, typically CNNs and RNNs/LSTMs, to leverage their complementary modeling capabilities. CNNs are effective at capturing spatial variations while RNNs/LSTMs excel at modeling temporal dynamics. 
CNN-LSTM hybrids have been widely explored for WSR. As shown in Fig. \ref{fig:nn} (f), an LSTM layer is added to the CNN architecture. The CNN extracts implicit time-domain information and passes higher-level features to the LSTM layer \cite{zhang2020automatichybrid}. Liu \et \cite{liu2017deep} found a convolutional LSTM DNN (CLDNN) achieved the best performance among tested architectures. The work \cite{lin2020hybrid} proposed HybridNet with a bidirectional GRU (Bi-GRU) after the CNN to explicitly capture temporal dependencies. Zhou \et \cite{zhou2020lstm} added an LSTM module to a CNN (Incept-LSTM) to learn time-related signal features, improving classification accuracy, especially at low SNRs.
The long-term memory of LSTMs suits the temporal causality of time-domain radio signals. Studies have shown fusion models significantly outperform individual CNNs or RNNs/LSTMs \cite{zhang2018automatic} and \cite{sang2018application}, with lower complexity. Beyond CNN-LSTM, other combinations like DNN-LSTM, and CNN-LSTM-DNN have been explored, with architectural analysis in \cite{sainath2015convolutional}. Overall, hybrid neural networks demonstrate superior WSR by merging spatial and temporal modeling.

Hybrid architectures combining multiple neural network models have been explored for further improving AMC performance. Jagannath \et \cite{jagannath2017design} designed a hierarchical hybrid AMC (HH-AMC) using both feature-based and likelihood-based classifiers to reduce complexity while improving accuracy. 
The authors in \cite{zhang2022modulation} combined RNN and CNN to jointly exploit automatic feature extraction along with temporal modeling for underwater acoustic modulation recognition. Their model achieved higher accuracy with lower latency compared to conventional deep learning approaches.
Shen \et \cite{shen2021radio,shen2021radiojsac} incorporated estimated carrier frequency offset into a hybrid classifier to enhance deep learning accuracy in classifying LoRa devices. Experiments showed a spectrogram-CNN model achieved $96.40\%$ accuracy with the lowest complexity and training time among evaluated hybrid architectures.

\subsubsection{Heterogeneous models}  

A heterogeneous model is the model fusion in serial or parallel ways. Advanced heterogeneous neural networks have been developed to further boost WSR performance. Huang \et \cite{huang2020identifying} proposed a multi-module fusion neural network using novel pixel-coloring constellation image features, outperforming other deep learning-based AMC methods and demonstrating superior classification accuracy. 
Lyu \et \cite{lyu2020robust} developed a robust AMC method with a convolutional and recurrent fusion network combining CNNs and simple recurrent units to effectively suppress Doppler shift. Their model classified eight modulation types under AWGN and varying Doppler, merging CNN spatial modeling with recurrent temporal modeling. 
Ali \et \cite{ali2021automatic} designed a feature extraction module to select optimal HOC combinations up to sixth order, leveraging their logarithmic properties to improve cumulant distributions.

\subsubsection{Other models}
Model compression \cite{huang2019automatic,ramjee2019fast,huynh2020exploiting,huynh2020mcnet,kim2020lightweight,lin2020improved,ma2020cross,wang2020lightamc} is adopted to accelerate the inference time of deep learning models to meet the requirements of resource-limited devices, such as sensors and edge computers. 
Huang \et \cite{huang2019automatic} proposed a compressive CNN that takes two different constellation image types as inputs - regular constellation images and contrast-enhanced grid constellation images. These are generated from the received signals and fed into the CNN architecture.
Additionally, dictionary learning \cite{zhang2018data,zhang2018dictionary}, complex-valued networks \cite{tu2020complex,wang2021efficient,xiao2023complex}, zero-forcing (ZF) equalization \cite{wang2020automatic,wang2020transfer}, distributed learning \cite{wang2020distributed} have also been utilized for WSR. 
We present the advanced methods from the model perspective in Table \ref{tab:model}.

\begin{table*}[]
	\scriptsize
	\centering
	\caption{Performance comparison of DL-based methods in terms of Model}
	\label{tab:model}
	\begin{tabular}{|m{1.25cm}<{\centering}|m{2.5cm}<{\centering}|m{2.5cm}<{\centering}|m{2cm}<{\centering}|m{2.5cm}<{\centering}|m{0.75cm}<{\centering}|m{1cm}<{\centering}|m{0.5cm}<{\centering}|m{0.5cm}<{\centering}|}
	\hline
	Model & Ref  & Method & Feature types   & Modulation type  & Symbols & SNR   & Acc. L & Acc. H \\
	\hline
	\multirow{2}{*}{DNN} & Lee \et \cite{lee2017deep} & DNN & 28 features & BPSK, QPSK, 8PSK, 16QAM, 64QAM & 2000 & [5] & 70.64 & 99.95 \\
	\cline{2-9} &
	O'Shea \et  \cite{OShea2016}    & DNN                & IQ               & RadioML 2016                                                                & 128  & [-20:2:18] & 15 & 87.4  \\
	\hline
	\multirow{15}{*}{CNN} &
	Kulin \et  \cite{kulin2018end}     & CNN                & IQ               & RadioML 2016                                                               & 128  & [-20:2:18] & 9  & 82    \\
	\cline{2-9} & Yashashwi  \et \cite{yashashwi2019} & CM+CNN& IQ & RadioML2016& 128  & [-20:2:18] & 10 & 90    \\
	\cline{2-9} & Tekbiyik \et \cite{tekbiyik2020robust} & CNN & IQ & HisarMod2019.1 & 1024  & [-20:2:18] & 26 & 94    \\
	\cline{2-9} & Shi \et \cite{shi2022combining} & Multiscale+SE block & IQ & RadioML2018.01A & 1024  & [-20:2:30] & 0 & 100    \\
	\cline{2-9} & Thien \et \cite{huynh2020mcnet} & MCNet & IQ & RadioML2018.01A & 1024  & [-20:2:30] & 5 & 93.59    \\
	\cline{2-9} & Zhang \et \cite{zhang2021real} & TRNN & IQ & OFDM Modulation Classification Dataset  & 1024  & [-10:2:10] & 64 & 100    \\
	\cline{2-9} & Hermawan \et \cite{hermawan2020cnn} & IC-AMCNet & IQ & RadioML2016  & 128  & [-20:2:18] & 10 & 91.7    \\
	\cline{2-9} & Liu \et \cite{liu2017deep} & ResNet, DenseNet & IQ & RadioML2016.10b  & 128  & [-20:2:18] & 10 & 88.5    \\
	\cline{2-9} & Yashashwi \et \cite{yashashwi2018learnable} & CM+CNN & IQ & RadioML2016  & 128  & [-20:2:18] & 10 & 90    \\
	\cline{2-9} & Chen \et \cite{chen2020signet} & SigNet & IQ & RML2016.10a   & 128  & [-20:2:18] & 10 & 91    \\
	\cline{2-9} & Zeng \et \cite{zeng2019spectrum} & SCNN & Spectrum & RML2016.10a   & -  & [-20:2:18] & 11 & 92    \\
	\hline
	\multirow{4}{*}{RNN} &
	Hong \et  \cite{hong2017automatic} & GRU & IQ & RadioML 2016& 128  & [-20:2:18] & 10  & 91.9    \\
	\cline{2-9} & Rajendran \et  \cite{rajendran2018deep} & LSTM & AP & RadioML2016.10a & 128  & [-10:2:18] & 10  & 90    \\
	\cline{2-9} & Hu \et  \cite{hu2018robust} & LSTM & AP & Non-public dataset & 128  & [0:2:20] & 60  & 98    \\
	\cline{2-9} & Ke \et  \cite{ke2021real} & LSTM + DNN & IQ & RadioML2016.10a & 128  & [0:2:20] & 9.3  & 92.75    \\
	\hline
	\multirow{4}{*}{Transformer} & Hamidi-Rad \et  \cite{hamidi2021mcformer} & Transformer & IQ & RadioML2016.10b & 128  &  [-20:2:18] & 10  & 93   \\
	\cline{2-9} & Cai \et  \cite{cai2022signal} & Transformer & IQ & RadioML2016.10c & 128  & [0:2:20] & 29  & 92.5    \\
	\cline{2-9} & Ying \et  \cite{ying2023convolutional} & Transformer & IQ & RadioML2016.10a & 128  & [0:2:20] & 10  & 93.5    \\
	\cline{2-9} & Dao \et  \cite{dao2023vt} & Transformer & IQ & RadioML2018.01A & 1024  & [-20:2:30] & 5  & 99.24    \\
	\hline

	\multirow{12}{*}{Hybrid} &
	West \et  \cite{west2017deep} & LSTM+CNN & IQ & RadioML 2016& 128  & [-20:2:18] & 11  & 86    \\
	\cline{2-9} & Liu \et \cite{liu2017deep} & CLDNN & IQ & RadioML2016.10b  & 128  & [-20:2:18] & 10 & 88.5    \\
	\cline{2-9} & Njoku \et \cite{njoku2021cgdnet} & CGDNet & IQ & RadioML2016.10b  & 128  & [-20:2:18] & 20 & 93.5    \\
	\cline{2-9} & Zhang \et \cite{zhang2021efficient} & PET-CGGDNN & IQ & RadioML2016  & 128  & [-20:2:18] & 10 & 93.41    \\
	\cline{2-9} & Xu \et \cite{xu2020spatiotemporal} & MCLDNN & IQ & RadioML2016  & 128  & [-20:2:18] & 10 & 92    \\
	\cline{2-9} & Ghasemzadeh \et \cite{ghasemzadeh2020novel} & DBN + SNN & I/Q + Amplitude/Phase + High-order & RadioML 2018.01   & -  & [-10:2:30] & 18 & 100    \\
	\cline{2-9} & Ghasemzadeh \et \cite{ghasemzadeh2022gs} & GS-QRNN & I/Q & RadioML 2018.01   & -  & [-10:2:30] & 55 & 100    \\
	\cline{2-9} & Zhang \et \cite{zhang2018automaticmodulation} & CNN-LSTM-IQFOC & I/Q & RadioML2016.10a    & 128  & [-20:2:18] & 12 & 88    \\
	\cline{2-9} & Zhang \et \cite{zhang2020automatic} & CNN+LSTM & I/Q+A/P & RadioML2016.10a    & 128  & [-6:2:18] & 55 & 91    \\
	\cline{2-9} & Wang \et \cite{wang2021deeplearning} & MLDNN & I/Q+A/P & RadioML2016.10a    & 128  & [-20:2:18] & 20 & 97.8    \\
	
	\hline
	
\end{tabular}
\end{table*}

\subsection{Data} 

Data plays an important role in the DL area because DL is driven by data and the performance heavily depends on the number of the training data. 
For WSR, the data can be classified into two main categories, namely image-based data, sequence-based data and combined data. 
Image-based data includes the constellation diagram, eye diagram, feature point image, ambiguity function image, spectral correlation function image, cyclic correntropy spectrum graph, and bispectrum graph. 
Sequence-based data consists of in-phase and quadrature (IQ) sequences, AP sequences, fast Fourier transformation (FFT) sequences, and amplitude histogram (AH) sequences. 
Moreover, different types of data can be combined to further improve classification performance. 
A summarisation of advanced methods from the data perspective is presented in Table \ref{tab:data}.

\subsubsection{Image-based data}
\begin{figure*}
	\centering
	\includegraphics[width=0.99\linewidth]{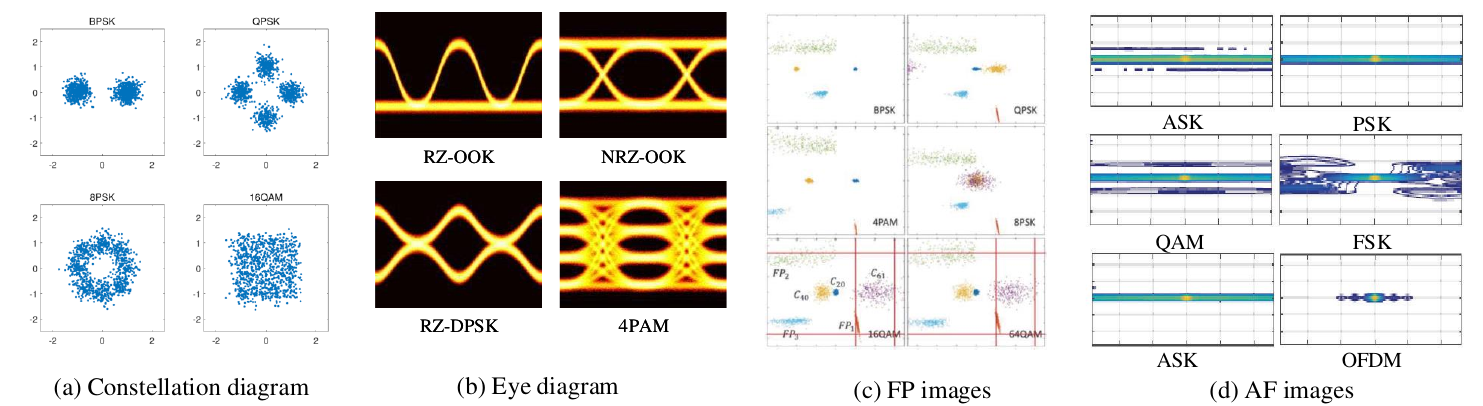}
	\caption{Different types of Image-based data, (a) constellation diagram, (b) eye diagram \cite{wang2017modulation}, (c) {\color{blue}FP (Feature Point) images \cite{lee2019feature}}, (d) AF images \cite{dai2016automatic}.}
	\label{fig:imagedata}
\end{figure*}

: Deep learning is a powerful tool first applied for image classification in the field of computer vision because image-based data and DNNs can be implemented parallelly accelerating with GPUs. 
From the perspective of computer vision, signal recognition can be executed by converting the received signal into an image and utilizing DNNs to recognize the transferred image, such as the constellation diagram, eye diagram, \etc, as shown in Fig. \ref{fig:imagedata}.

\paragraph{Constellation diagram} 
Constellation diagram \cite{kumar2020automatic,liu2021new} is a 2D scatter plot mapping signal samples to complex coordinate points. Peng \et \cite{peng2017modulation} demonstrated the feasibility of using constellation diagrams and DNN for MC. They converted the received signal into a $227 \times 227$ binary constellation image and used an AlexNet model to classify QPSK, 8PSK, 16QAM and 64QAM modulations. Lee \et \cite{lee2019feature} also used a CNN and transform-based 2D constellation image features for improved AMC performance. Kumar \et \cite{kumar2020automatic} designed a new algorithm based on constellation density matrices, forming color images that are classified by ResNet-50 and Inception ResNet V2 models. Mao \et \cite{mao2021attentive} proposed attentive Siamese networks operating on multi-timing constellation diagrams for more robust AMC. For RFFI, the authors in \cite{wang2020radio} used a deep complex residual network and statistical constellation analysis. Peng \et \cite{peng2022radio} introduced heat constellation trace figures and slice integration cooperation to improve fingerprinting accuracy.
    
\paragraph{Eye diagram}
An eye diagram is a type of display on an oscilloscope that utilizes the received signal for the vertical input and applies the data rate to trigger the horizontal sweep repetitively.
As shown in Fig. \ref{fig:imagedata} (b), Wang \et \cite{wang2017modulation} collected high-resolution ($900 \times 1200$) colored eye diagram images from an oscilloscope. To reduce computation, the colored images were transformed to grayscale and down-sampled to $28 \times 28$ pixels before feeding them into a CNN architecture for classification.

\paragraph{Feature point image}
Feature point image is generated from signal features, such as HOCs, PAR, PRR, maximum value of power spectral density, and skewness. 
In \cite{lee2019feature}, these features are plotted on a complex plane within a square area of $[-3.5, 3.5]$. The intrinsically complex feature values, including $C_{20}$, $C_{40}$, and $C_{61}$, are directly plotted without additional transformation. The real-valued features, including $C_{21}$, $C_{42}$, RPA, RPR, $\gamma_{max}$, and Sk, are transformed into complex Fingerprints (FPs) before plotting. Spectral correlation function image \cite{mendis2016deep}, cyclic correntropy spectrum graph \cite{ma2019automatic}, bispectrum \cite{li2019automatic} which are derived from the modulated signals can also be utilized for AMC. For RFFI, the authors of \cite{shen2021radio,shen2021radiojsac} used spectrogram to represent the time-frequency characteristics of LoRa signals and found that the drifting carrier frequency offset (CFO) can compromise system stability and cause misclassification. To mitigate this issue, they proposed a hybrid classifier that incorporates estimated CFO to adjust the CNN outputs, improving the classification accuracy.
	
\paragraph{Ambiguity function image}
Ambiguity function (AF), a bivariate function with respect to time delay $\tau$ and frequency offset $
\omega$, is defined by 
\begin{equation}
AF_s(\omega, \tau)=\int_{-\infty}^{\infty} y(t+\frac{\tau}{2}) y^*(t-\frac{\tau}{2})\exp{-j\omega t}dt,
\end{equation}
where $y(t)$ represents the continuous received signal and $*$ denotes the complex conjugate, a 2D matrix of $AF_s(\omega, \tau)$ values can be obtained for varying $\omega$ and $\tau$. This matrix comprises the ambiguity function (AF) image. In prior work \cite{dai2016automatic}, central $28 \times 28$ normalized AF images were extracted and input to sparse autoencoders for AMC of the received signal. As illustrated in Fig. \ref{fig:imagedata} (d), different modulation types (e.g. ASK, PSK, QAM, FSK, MSK, linear frequency modulation (LFM), and orthogonal frequency-division multiplexing (OFDM)) yield distinguishable AF image patterns.

\begin{figure*}[t]
	\centering
	\includegraphics[width=0.99\linewidth]{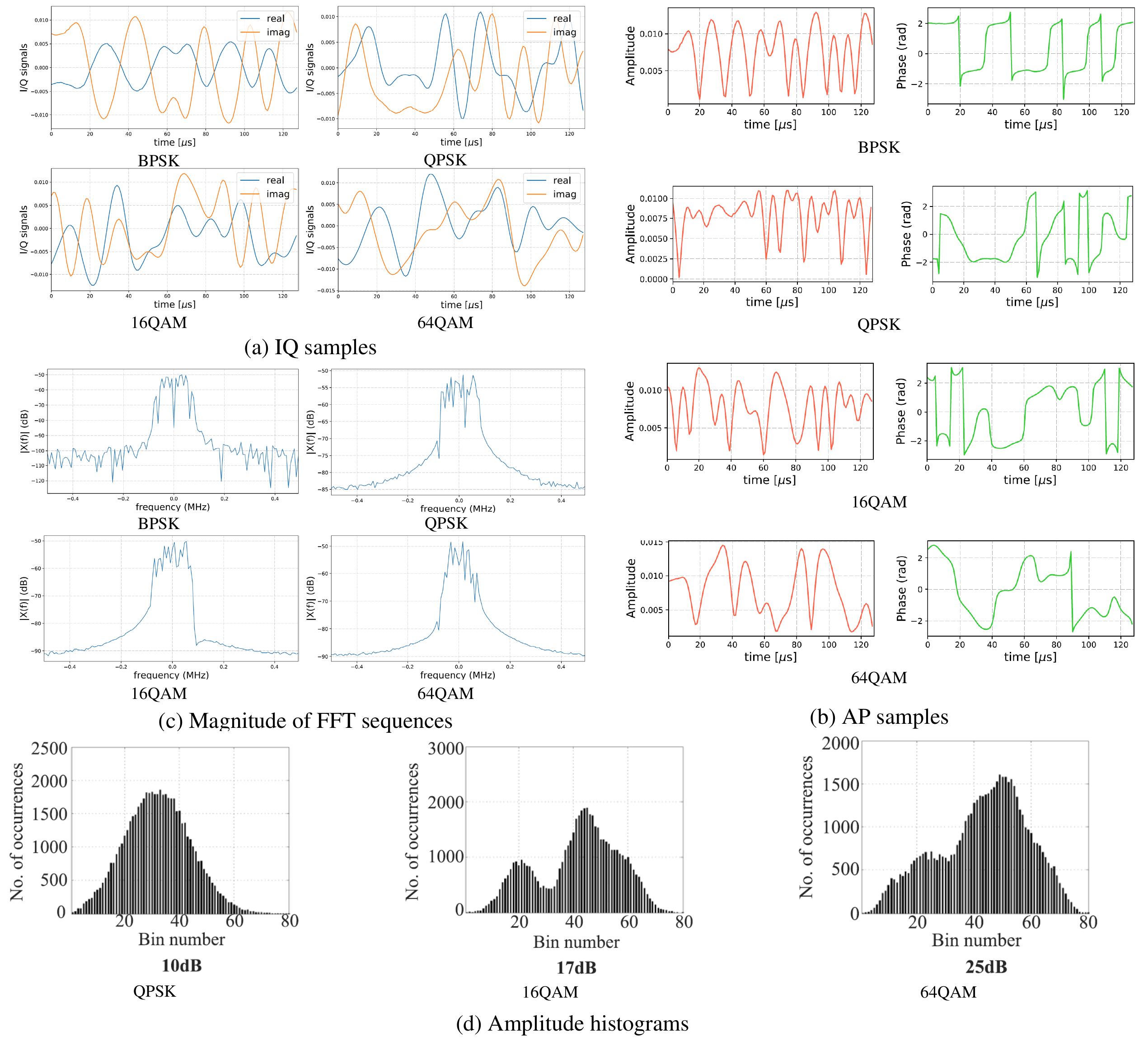}
	\caption{Different types of sequence-based data, (a) IQ samples \cite{kulin2018end}, (b) AP samples \cite{kulin2018end}, (c) Magnitude of FFT sequences \cite{kulin2018end}, (d) Amplitude histograms \cite{khan2016modulation}.}
	\label{fig:seqdata}
\end{figure*}

\paragraph{Other image data}
Apart from the image data mentioned above, spectral correlation function (SCF) image \cite{mendis2016deep,mendis2019deep}, logarithmic cyclic correntropy spectrum (Log-CCES) graphs \cite{ma2019automatic}, and amplitude spectrum of bispectrum (ASB) graphs \cite{li2019automatic} can also be utilized as image data for DL-based methods. 

\begin{table*}[]
	\scriptsize
	\centering
	\caption{Performance comparison of DL-based methods in terms of Data}
	\label{tab:data}
	\begin{tabular}{|m{1cm}<{\centering}|m{2cm}<{\centering}|m{1.5cm}<{\centering}|m{3cm}<{\centering}|m{4cm}<{\centering}|m{0.75cm}<{\centering}|m{1cm}<{\centering}|m{0.5cm}<{\centering}|m{0.5cm}<{\centering}|}
	\hline
	Data & Ref  & Method & Feature types   & Modulation type  & Symbols & SNR   & Acc. L & Acc. H \\
	\hline
	\multirow{6}{*}{Features} & Lee \et \cite{lee2017deep} & DNN & 28 features & BPSK, QPSK, 8PSK, 16QAM, 64QAM & $2000$ & [5] & $70.64$ & $99.95$ \\
	\cline{2-9}
	&Ali and Fan \cite{ali2017automatic} & SAE-DNN &Cumulants& BPSK, QPSK, 8PSK, 16QAM, 64QAM & 512  & [0:5:15] & 72  &100\\
	\cline{2-9}
	 &Xie \et \cite{xie2019deeplearning} & DNN & Cumulants &2FSK, 4FSK, BPSK, QPSK, 2ASK, 4ASK & / &   [-5,-2] & 87.53 & 100 \\
	\cline{2-9}
	 &Lee \et \cite{lee2019effective} & DNN &Cumulants &BPSK, QPSK, 8PSK, 16QAM, 64QAM& / &    [-5:5:10] &85.61 &99.69\\
	\cline{2-9}
	 &Shah \et \cite{shah2019classification} & RBFN, SAE-DNN &Cumulants, statics, spectral  &GMSK, CPFSK, GFSK, OQPSK& / &    [0:5:15] & 50 & 95\\
	\cline{2-9}
	 &Shi \et \cite{shi2019particle} &PSO-DNN &12 features &BPSK, QPSK, 8PSK, 16QAM, 64QAM, 256QAM&&    [0:2:12]&  87 & 100\\
	\hline
	\multirow{20}{*}{Sequences}& O'Shea \et  \cite{OShea2016}    & DNN                & IQ               & RadioML 2016                                                                & 128  & [-20:2:18] & 15 & 87.4  \\
	\cline{2-9}
	  & Liu \et    \cite{liu2017deep}     & CNN, CLDNN, ResNet & IQ               & BPSK, QPSK, 8PSK, 16QAM, 64QAM,   BFSK, CPFSK, 4PAM, WB-FM, AM-SSB, AM-DSB  &      & [-25:5:30] & 10 & 88.5  \\
	\cline{2-9}
	  & Ma \et    \cite{ma2020cross}      & HDNN, LRN          & IQ               & BPSK, QPSK, 8PSK, 16QAM, 64QAM,   BFSK, CPFSK, 4PAM, WB-FM, AM-SSB, AM-DSB  & 128  & [-20:2:18] & 10 & 92    \\
	\cline{2-9}
	\cline{2-9}
	  & Ali and   Fan \cite{ali2017ksparse}  & UDNN               & IQ               & QPSK, 16PSK, 16QAM, 128QAM                                                  & 128  & [0:5:15]   & 71 & 100   \\
	\cline{2-9}
	  & Hong \et  \cite{hong2019deep}      & CNN                & IQ, AP, Cumulant & BPSK, QPSK, 8PSK,16QAM, 64QAM                                               &      & [0:5:30]   & 40 & 92    \\
	\cline{2-9}
	  & shi \et    \cite{shi2020deep}     & CNN                & IQ               & BPSK, QPSK, 8PSK,16QAM                                                      &      & [-10:5:20] & 48 & 100   \\
	\cline{2-9}
	  & Meng \et   \cite{meng2018automatic}     & CNN                & IQ               & BPSK, QPSK, 8PSK, 16APSK,   32APSK, 16QAM, 64QAM                            &      & [-6:2:10]  & 30 & 96    \\
	\cline{2-9}
	  & Zheng \et  \cite{zheng2019fusion}     & CNN, ResNet        & IQ               & BPSK, QPSK, 8PSK, OQPSK,16QAM,   32QAM, 64QAM, 2FSK, 4FSK, 8FSK, 4PAM, 8PAM &      & [-20:2:30] & 10 & 100   \\
	\cline{2-9}
	  & Gu \et   \cite{gu2019blind}       & CNN                & IQ               & 2FSK, DQPSK, 16QAM, 4PAM, MSK,   GMSK                                       &      & [0:2:12]   & 83 & 98    \\
	\cline{2-9}
	  & Hu \et    \cite{hu2019deep}      & RNN                & IQ               & BPSK, QPSK, 8PSK, 16QAM                                                     &      & [0:2:20]   & 61 & 98    \\
	\cline{2-9}
	  & Wang \et   \cite{wang2020deeptvt}     & CNN                & IQ               & BPSK, QPSK, 8PSK, 16QAM                                                     &      & [-10:5:10] & 49 & 100   \\
	\cline{2-9}
	  & Mossad \et  \cite{mossad2019deep}    & CNN                & FFT, IQ          & 10 in RadioML2016                                                           & 128  & [-20:2:18] & 11 & 86.97 \\
	\cline{2-9}
	  & Khan \et  \cite{khan2016modulation,khan2017joint}      & DNN                & AH               & QPSK, 16QAM, 64QAM                                                          &      & [0:5:20]   & 99 & 100  \\
	\hline
	\multirow{20}{*}{Images} & Peng \et  \cite{peng2017modulation,peng2018modulation}   & AlexNet, GoogLeNet & Constellation Diagram                 & BPSK, QPSK, 8PSK, OQPSK,16QAM,   32QAM, 64QAM, 4ASK       & 1000  & [0:1:10]      & 74.1  & 100  \\ 
	\cline{2-9}
	 & Lee \et  \cite{lee2019feature}    & CNN                & Feature point image                   & BPSK, QPSK, 8PSK, 4PAM, 16QAM,   64QAM                    & 60000 & [0:2:10]      & 95.41 & 100  \\ 
	\cline{2-9}
	 & Wang \et  \cite{wang2017graphic}   & DBN                & Constellation Diagram                 & BPSK, QPSK, 8PSK, 16QAM                                   & 1024 & [-6:2:14]     & 83.5  & 100  \\
	\cline{2-9}
	 & Tang \et  \cite{tang2018digital}    & AlexNet, ACGAN     & Contour Stellar image                 & BPSK, 4ASK, QPSK, 8PSK,   OQPSK,16QAM, 32QAM, 64QAM       & / & [-5:2:15]     & 46    & 100  \\ 
	\cline{2-9}
	 & Huang \et \cite{huang2019automatic}   & CCNN               & Constellation Diagram                 & BPSK, QPSK, 8PSK,  16QAM, 64QAM                           & 1024 & [-5:2:15]     & 46    & 100  \\ 
	\cline{2-9}
	 & Huang \et  \cite{huangjiang2019automatic}  & CFCNN              & Constellation Diagram                 & BPSK, QPSK, 8PSK,  16QAM, 64QAM                           & 128 & [-5:5:15]     & 20    & 100  \\ 
	\cline{2-9}
	 & Xie \et  \cite{xie2019deep}    & CNN                & Constellation Diagram                 & BPSK, QPSK, 8PSK                                          & 1000 & [-5:1:3]      & 48    & 100  \\ 
	\cline{2-9}
	 & Tu \et   \cite{tu2019deep}    & AlexNet            & Contour Stellar image                 & 4ASK, BPSK, QPSK, OQPSK, 8PSK,   16QAM, 32QAM, 64QAM      & / & [-6:2:6]      & 82    & 100  \\ 
	\cline{2-9}
	 & Kumar \et \cite{kumar2020automatic}   & ResNet, Inception  & Constellation Diagram                 & 2ASK, 4ASK, BPSK, QPSK, 8PSK,   8QAM, 16QAM, 32QAM, 64QAM & 512 & [-5:5:15]     & 95.2  & 100  \\ 
	\cline{2-9}
	 & Wang \et  \cite{wang2017modulation}   & CNN                & Eye Diagram                           & RZ-OOK, NRZ-OOK, RZ-DPSK, 4PAM                            & / & [10:1:25]     & 95.2  & 100  \\ 
	\cline{2-9}
	 & Dai \et  \cite{dai2016automatic}    & SAE-DNN            & Ambiguity Function Image              & ASK, PSK, QAM, FSK, MSK, LFM,   OFDM                      & 28x28 & [-10:1:25]    & 90.4  & 99.8 \\ 
	\cline{2-9}
	 & Mendis   \et \cite{mendis2016deep} & DBN                & Spectral correlation function   image & 4FSK, 16QAM, BPSK, QPSK, OFDM                             & 128 & [-2:1:5]      & 89    & 100  \\ 
	\cline{2-9}
	 & Ma \et  \cite{ma2019automatic}     & ResNet             & Cyclic correntropy spectrum graph     & BPSK, 2ASK, 8ASK, 4QAM, 16QAM                             & 200x200 & [-5:5:10]     & 39    & 100  \\ 
	\cline{2-9}
	 & Li \et  \cite{li2019automatic}     & AlexNet            & Bispectrum Graph                      & BPSK, 2ASK, 2FSK, 4FSK, 8FSK,   LFM, OFDM                 & 227x227 & [-3,0,3,5,10] & 15    & 97.7 \\ 
	\hline
	\multirow{6}{*}{Combined} & Zhang \et  \cite{zhang2018automaticmodulation}    & CNN, LSTM  & Cumulants, IQ                                                     & RadioML2016                                                           & 128 & [-20:2:18] & 12   & 87   \\ 
	\cline{2-9}
	 & Wang \et   \cite{wang2019data}    & DrCNN, CNN & IQ, Constellation Diagram                                         & BPSK, QPSK, 8PSK, GFSK, CPFSK,   4PAM, 16QAM, 64QAM                   &  128   & [-8:2:18]  & 43   & 98   \\
	\cline{2-9}
	 & Wu \et   \cite{wu2019convolutional}      & CNN        & Cyclic Spectra Image,   Constellation Diagram                     & RadioML 2016                                                          & 128 & [-20:2:18] & 16   & 92   \\ 
	\cline{2-9}
	 & Wang \et  \cite{wang2019automatic}     & CNN        & Joint Time-Frequency Image,  Instantaneous Autocorrelation Image & LFMM SF, BPSK, QPSK, 2FSK, 4FSK                                       &  750  & [6:2:14]   & 97.1 & 99.8 \\ 
	\cline{2-9}
	 & Zhang \et  \cite{zhang2019automatic}    & ResNet     & SPWVD, BJD Images                                                 & BPSK, QPSK, 2FSK, 4FSK, 2ASK,   4ASK, 16QAM, 64QAM, OFDM              &   512  & [-4:2:8]   & 90   & 98.5 \\ 
	\cline{2-9}
	 & Hiremath   \et \cite{hiremath2019deep} & CNN        & IQ, DOST Sequences                                                & RadioML 2018.01A &  1024   & [-8:2:8]   & 50   & 97.3 \\ 
	\hline
\end{tabular}
\end{table*} 

\begin{figure*}[!h]
	\centering
	\includegraphics[width=0.9\linewidth]{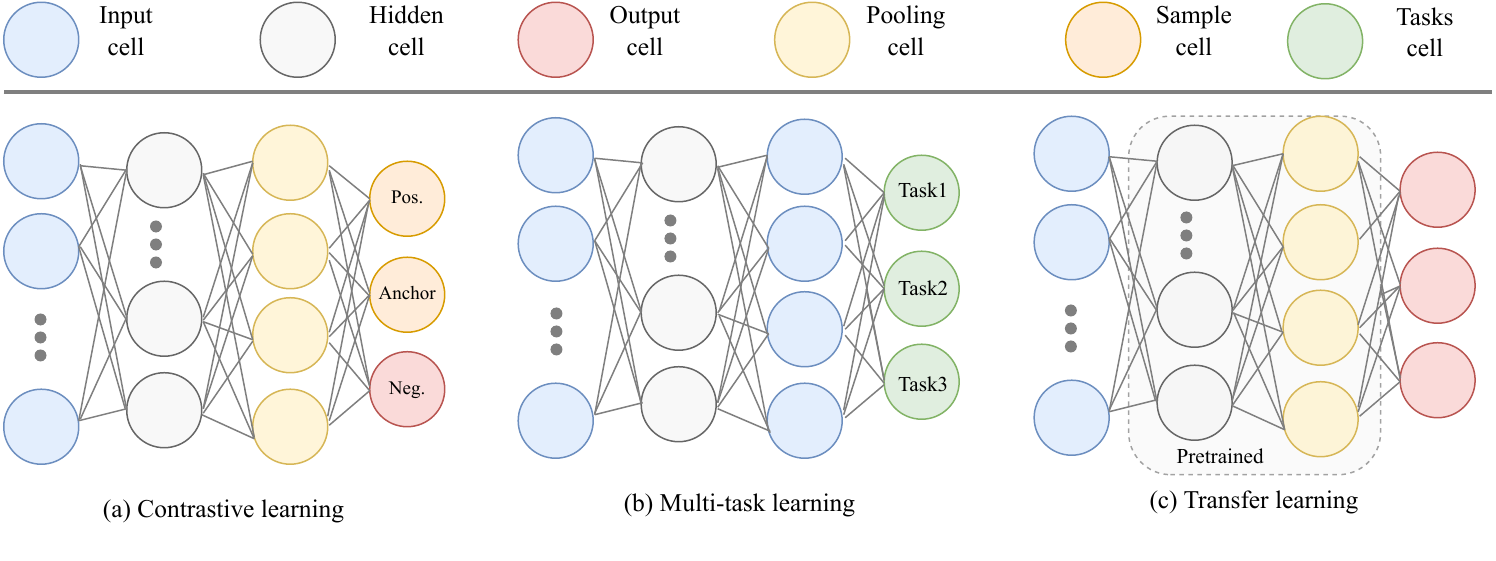}
	\caption{Different learning types, (a) contrastive learning, (b) multi-task learning, and (c) transfer learning.}
	\label{fig:learning}
\end{figure*}

\subsubsection{Sequence-based data} Sequence-based data is the basic representation way in communications systems because the received signals are organized in a sequence. Typical sequence-based data include in-phase and quadrature (IQ) data, AP data, and FFT sequences, as shown in Fig. \ref{fig:seqdata}.

\paragraph{IQ data}
O'Shea \et \cite{OShea2016} pioneered the use of in-phase and quadrature (IQ) data for deep learning-based AMC, classifying received signals from 8 digital and 3 analog modulation schemes \cite{o2016radio}. In their work, received signals were segmented into 128-point samples using a rectangular window. A subsequent study by O’Shea \et \cite{OShea2018} expanded the IQ dataset using advanced software-defined radio, generating more single-carrier modulation types. The new dataset comprises compositions with 11 and 24 modulations and explores varying wireless channel scenarios, including over-the-air transmissions. Moreover, IQ sequence lengths were increased from 128 to 1024 points. The authors implemented VGGNet and ResNet models to effectively learn from these sequences and achieve AMC. Building on these datasets, additional techniques and networks have since been developed for deep learning-based AMC. Huang \et \cite{huang2019data} analyzed three data augmentation methods to enlarge the datasets and mitigate overfitting. For WTC, Kulin \et \cite{kulin2018end} used time domain features like IQ vectors and amplitude/phase vectors to train CNN classifiers, successfully recognizing ZigBee, WiFi, and Bluetooth signals. For RFFI, Riyaz \et \cite{riyaz2018deep} proposed a CNN-based method using IQ datasets that learned the inherent signatures of different transmitters. Their technique outperformed conventional machine learning approaches in identifying 5 identical hardware devices.

\paragraph{Amplitude and phase (AP) data}
Similar to IQ-sequence representation, AP sequences can also be used to characterize the received signals. 
Building on the IQ dataset from \cite{o2016radio}, Kulin \et \cite{kulin2018end} mapped the IQ signals into 128-point AP sequences. To enable fair comparison with \cite{OShea2016}, they utilized a similar convolutional neural network (CNN) architecture to classify the AP sequences. Their simulations demonstrated that the AP-sequence data led to performance gains up to 2\% at medium to high SNRs compared to the IQ-sequence data. In other work, Selim \et \cite{selim2017spectrum} adopted AP difference representations to train a CNN for radar signal recognition. Their method successfully classified radar signals even with coexisting LTE and WLAN signals.

\paragraph{FFT sequences}
The FFT representation maps the time-domain IQ signal to frequency-domain complex data comprising real and imaginary FFT sequence pairs. Kulin \et \cite{kulin2018end} performed 128-point FFTs on the IQ dataset from \cite{o2016radio} and fed the FFT magnitude sequences for eight digital modulations into a CNN, confirming the feasibility of classification using this representation. To highlight the advantages of the FFT sequences, Mossad \et \cite{mossad2019deep} compared against IQ sequences. For higher-order modulations like QPSK, 8PSK, 16QAM and 64QAM, the FFT representation yielded improved classification accuracy. As noted in \cite{mossad2019deep}, FFT sequences can be short with small FFT sizes, enabling low-complexity MC suitable for resource-constrained systems like the electrosense network \cite{rajendran2017electrosense}. Histograms present another accurate representation constructed by dividing the data range into intervals and tallying occupants per interval. Fig. \ref{fig:imagedata} (d) illustrates amplitude histograms of QPSK, 16QAM and 64QAM signals at varying SNRs \cite{khan2016modulation,khan2017joint}, exhibiting unique, modulation-dependent shapes potentially recognizable by DNNs. Khan \et \cite{khan2016modulation} collected 56,000 equalized signal samples to construct 80-bin amplitude histograms as 80-point sequences. Feeding many histogram sequences into a two-layer DNN enabled accurate classification of QPSK, 16QAM and 64QAM. In summary, FFT and histogram representations provide useful alternatives to IQ sequences as inputs for DL-based AMC. FFT sequences enable lower complexity classification while histograms leverage unique data distributions. Further research can elucidate trade-offs between these approaches.

\subsubsection{Joint features} The received signal can be represented by either features, images, or sequences. 
Thus, joint features such as features and sequences \cite{zhang2018automaticmodulation}, images and sequences \cite{wang2019data}, multiple images \cite{wu2019convolutional,wang2019automatic}, and multiple sequences can be utilized for WSR. 

\paragraph{Features and sequences}
Zhang \et \cite{zhang2018automaticmodulation} enhanced IQ signal classification by adding fourth-order cumulant features as a third component to the traditional RGB three-channel format, resulting in a $3 \times 128$ signal representation. This approach, using CNN and LSTM models, improved AMC accuracy by about 8\% and increased noise robustness. It shows that combining handcrafted features with IQ data enhances deep learning performance, highlighting the benefits of merging data-driven and model-driven methods in WSR. Future research could investigate additional handcrafted features and fusion architectures to maximize this synergy.

\paragraph{Images and sequences}
Wang \et \cite{wang2019data} proposed a novel algorithm leveraging both constellation diagrams and IQ sequences to address classification confusion between 16QAM and 64QAM that can occur at low SNRs when using IQ data alone. The authors noted that 16QAM and 64QAM exhibit distinguishable constellation patterns that can complement the IQ sequences. By combining IQ and constellation representations, their proposed approach could accurately classify all eight modulation types considered. The results demonstrate the potential for improved classification accuracy by fusing multiple representational views of the wireless signal. 

\paragraph{Multiple images}
While cyclic spectra images provide noise-resilient representations, high-order MC using only cyclic spectra can suffer performance degradation. To address this, Wu \et \cite{wu2019convolutional} developed a two-branch convolutional neural network (CNN) that extracts high-level features from both cyclic spectra and constellation diagrams. By fusing the features through multi-feature integration, the combined representation improved classification accuracy and reduced complexity compared to state-of-the-art approaches. 
Similarly, the authors in \cite{wang2019automatic} proposed joining time-frequency and instantaneous auto-correlation images for unified classification, since neither image alone was effective for all signal types. The received signal was converted into both image representations and combined into a single binary image input to a seven-layer CNN. Simulations showed the joint image outperformed the individual images in accuracy. Similarly, Zhang \et \cite{zhang2019automatic} represented signals as two $224 \times 224$ images using smooth pseudo-Wigner-Ville distribution (SPWVD) and Born-Jordan distribution (BJD), and fed them into fine-tuned CNNs for feature extraction. By fusing the SPWVD and BJD features through multi-modality fusion, accurate classification was achieved. Furthermore, handcrafted features like $\gamma_{max}$, $\delta_{aa}$, $\delta_{ap}$, $\delta_{dp}$, and cumulants could be simultaneously integrated, improving accuracy further. 

\paragraph{Multiple sequences}
Hiremath \et \cite{hiremath2019deep} proposed combining IQ sequences and discrete orthogonal stockwell transform (DOST) \cite{stockwell2007basis} representations of IQ data as inputs to a CNN model for AMC. The motivation was to jointly leverage time domain features from the IQ sequences and time-frequency domain features from the DOST of IQ for improved performance. By extracting and fusing information from both the original and transformed IQ data using DOST, their approach achieved gains in overall classification accuracy compared to using either representation alone.

\paragraph{Feature fusion \& selection}
The method proposed by \cite{qi2020automatic} utilizes a fusion technique called waveform-spectrum multimodal fusion (WSMF) to achieve AMC using deep residual networks. This approach demonstrates strong performance, even when dealing with complex digital modulation types such as 256QAM and 1024QAM. 
The authors \cite{shah2019novel} proposed a novel method to select the most various $2^m$ features from a larger feature set by using the Bhattacharyya distance metric. The study evaluated the proposed method using three NN-based classifiers under AWGN and frequency-selective fading channels. The approach achieves a significant reduction in computational complexity while maintaining an acceptable level of classification performance.

\subsubsection{Data processing} data processing plays an important role in the implementation of deep learning models for WSR because processing the signal data properly before inputting would improve the performance of a DL-based model in the WSR problem. 
Data processing includes data preprocessing, data augmentation, and \etc.

\paragraph{Data preprocessing}
In \cite{zhang2020data}, the authors proposed a novel data preprocessing method (DPM) to address the problems of the raw input of the signals. The performance improvements resulting from the use of DPM were demonstrated in the experimental results.

\paragraph{Data augmentation}
To address limited training data for AMC, Patel \et \cite{patel2020data} proposed enhancing AMC training with limited data through a conditional generative adversarial network (CGAN) augmentation, generating synthetic samples that mimic real data distributions, thus boosting CNN model accuracy. Another approach by \cite{zheng2020spectrum} used a two-level spectrum interference technique, applying bidirectional noise masks in the frequency domain to diversify training data, proven effective on the RadioML 2016.10a dataset. Additionally, the authors in \cite{dong2023signal} introduced segment-wise and signal-wise generation methods for new signal creation, using techniques like cyclic segment shift and multiple signal concatenation, showing superior performance compared to previous methods.

\subsection{Learning}
Deep learning is strongly based on data and network structures, as well as learning methods. 
Different kinds of learning methods such as contrastive learning, multi-task learning, transfer learning, and curriculum learning can be utilized for AMC, as shown in Fig. \ref{fig:learning}.

\subsubsection{Contrastive learning}
Contrastive learning is an unsupervised machine learning approach that learns general features of unlabeled data by training models to recognize which data points are similar or different. 
It can allow the model to learn the similarity between similar or different kinds of data by enlarging the distance between data from different classes and reducing the distance between data from the same class. 
Huang \et \cite{huangjiang2019automatic} introduced a new AMC method using grid constellation matrices (GCMs) with a contrastive fully convolutional network (CFCN). GCMs, derived from low-complexity preprocessing of received signals, represent constellation diagrams. They employed a contrastive loss function to train the CFCN, enhancing differentiation between modulation types for more distinct representations. This approach led to improved classification performance, robustness against model mismatches, and shorter training times compared to recent methods.

\subsubsection{Multi-task learning} 
Multi-task learning aims to find optimal weight parameters for multiple tasks such as SNR estimation and AMC. Moreover, numerous works demonstrate that combining other related tasks can help improve the performance of WSR. 
To enable robust AMC under variable SNR conditions, Xie \et \cite{xie2019deep} proposed a two-step approach comprising SNR estimation using M2M4 statistics followed by multi-label deep learning classification. Their results demonstrated performance approaching that of deep learning models trained with fixed SNR data. 
Wang \et \cite{wang2021multi} later presented a novel multi-task learning (MTL) strategy for joint AMC under more realistic channel impairments including white non-Gaussian noise and synchronization errors. Multiple CNNs with shared parameters were trained on data under diverse SNRs using MTL. By extracting generalized features across noisy conditions, the MTL approach improved robustness and generalization versus conventional techniques.

\subsubsection{Transfer learning} 
Transfer learning facilitates the adaptation of pre-trained models to new data distributions, overcoming the common assumption in deep learning that training and test data are drawn from the same distribution. In practical deployments, varying sampling rates and other factors induce distribution shifts over time, resulting in distinct source and target domains with different underlying frequencies. Moreover, curating extensive labeled datasets for every target domain is often infeasible. Transfer learning addresses these challenges by transferring knowledge from a source domain with abundant labeled data to a target domain where labels may be scarce. This enables models trained on one data distribution to generalize to another distribution. Consequently, transfer learning is an essential technique for adapting deep neural networks to the inevitable distribution shifts arising in real-world applications. 
Transfer learning has shown promise for improving AMC performance with limited labeled data. Bu \et \cite{bu2020adversarial} developed an adversarial transfer learning architecture (ATLA) incorporating adversarial training to reduce domain shift between source and target AMC tasks. By additionally transferring knowledge from abundant source domain labels, ATLA achieved competitive accuracy with far less training data than supervised learning. 
In related work, Wang \et \cite{wang2020transfer} proposed a semi-supervised transfer learning approach for AMC in MIMO systems. Their deep reconstruction and classification network (DRCN) first leverages unlabeled data to train an autoencoder, then transfers encoder knowledge to initialize a CNN classifier. Experiments demonstrated superior performance over CNNs trained from scratch with scarce labels, matching the accuracy of CNNs trained on massive labeled data.

\subsubsection{Other learning methods}
Moreover, other kinds of learning methods have also been utilized for boosting the classification accuracy of AMC, including curriculum learning, incremental learning, and \emph{etc}. 
For instance, Zhang \et \cite{zhang2019automaticcurriculum} proposed two networks, namely MentorNet and StudentNet, to overcome the overfitting problem in AMC. 
The MentorNet employs curriculum learning to supervise the training of StudentNet and it addresses the issue of overfitting in StudentNet. 
Experimental results reveal that the proposed curriculum learning can help StudentNet possess great performance. 
Traditional incremental learning (IL) algorithms are not suitable for IoT due to storage limitations and performance degradation without historical data. In \cite{liu2021class}, a new channel separation-enabled IL (CSIL) scheme is proposed for NDI, which avoids conflicts between devices' fingerprints and does not require historical data. The proposed framework is evaluated using real data from an IoT application in aviation and has the potential for accurate device identification in various IoT applications.

\subsection{Implementation}

This subsection explores the practical implementation aspects of WSR systems, focusing on techniques and strategies to enhance efficiency, reduce resource demands, and enable deployment in real-world 6G networks. These include the development of lightweight models, model compression techniques, hardware implementation approaches, and other techniques.

\subsubsection{Lightweight Models}

Lightweight models are essential for deploying WSR systems on resource-constrained devices, such as edge nodes and IoT devices in 6G networks, where processing power, memory, and energy are limited. These models aim to preserve high recognition accuracy while reducing computational complexity and resource demands. Architectures like MobileNet and EfficientNet have been pivotal in this effort, tailored for tasks such as AMC and WTC. For example, Fei \et proposed \cite{fei2024mobileamct} MobileAmcT, a lightweight deep learning model combining efficient convolutional and Transformer modules, achieving higher accuracy and lower computational cost for AMC in drone communication systems. Similarly, the authors in \cite{chen2024efficientnet} demonstrated that the lightweight EfficientNet-B3 model outperforms traditional deep learning models in AMC for linear and OFDM systems, achieving higher accuracy, faster training, and better generalization, making it well-suited for mobile and embedded devices.

Further enhancing scalability, techniques such as pruning and parameter reduction refine these models for efficiency. Wang \et proposed \cite{wang2020lightamc} LightAMC, a novel lightweight AMC method that reduces model size and computational cost for IoT and UAV systems by introducing neuron scaling factors and pruning redundant neurons using compressive sensing, with only a slight performance trade-off. Likewise, the authors in \cite{lin2020improved} proposed a filter-level pruning technique using activation maximization for deep learning-based AMC, enhancing deployment on resource-constrained devices with equal or better accuracy.
These advancements ensure lightweight WSR solutions remain high-performing and deployable across 6G’s diverse, resource-limited environments, from industrial automation to smart cities.

\subsubsection{Model Compression}

Model compression techniques play a vital role in optimizing WSR models by reducing their size and computational overhead, making them well-suited for resource-constrained devices in 6G networks. One widely adopted method is quantization \cite{goez2022methodology, goldbarg2024novel}, which lowers the precision of model weights from high-precision 32-bit floating-point representations to more compact 8-bit or even 4-bit integer formats. For example, Göez \et \cite{goez2022methodology} proposes a methodology using the Brevitas and FINN frameworks to analyze quantization in a VGG10-based deep learning model for AMC, identifying an optimal bit-per-layer configuration that reduces model size by 75.8\% with only a 0.06\% accuracy loss. This compression enables the model to run efficiently on low-power edge devices, such as IoT sensors in a smart grid, where rapid inference is critical for real-time monitoring, and energy efficiency is paramount due to battery constraints.

Another powerful technique, knowledge distillation \cite{yang2023knowledge}, further enhances WSR model efficiency by transferring knowledge from a large, high-performing teacher model to a smaller, lightweight student model. This approach has proven particularly valuable for DL-based WSR tasks. For instance, Yang \et \cite{yang2023knowledge} proposed GSCNET, a lightweight DL model for AMC using Ghost modules and depthwise separable convolution, enhanced by knowledge distillation with cross-entropy, KL divergence, and soft label-based losses, achieving high accuracy with reduced computational complexity. Together, these compression methods ensure WSR systems meet the stringent real-time and resource demands of next-generation wireless networks.

\subsubsection{Hardware Implementation}

Hardware implementation is pivotal for deploying WSR systems, utilizing specialized platforms such as Field-Programmable Gate Arrays (FPGAs), Graphics Processing Units (GPUs), and Application-Specific Integrated Circuits (ASICs) to meet the rigorous demands of 6G networks. These platforms provide high parallelism and energy efficiency, essential for handling the computational intensity of DL-based WSR models. The process of implementing a DL-based modulation recognizer, begins with designing a deep neural network using languages such as Python, C, or MATLAB, followed by training it to optimal performance with simulation software. Subsequently, the trained model is translated into hardware description languages such as VHDL or Verilog and simulated using tools such as ModelSim. The design is then downloaded to a hardware circuit for debugging, enabling real-time signal modulation recognition.

For example, the work \cite{kumar2022automatic} proposed a quantized CNN-based AMC scheme for complex-valued radio signals, implemented on FPGA with low-precision weights and iterative pruning, achieving 1.4\% higher accuracy and 60\% reduced hardware use compared to the baseline, with 527k classifications per second and 7.5 $\mu s$ latency. Similarly, Zhao \et \cite{zhao2024ultra} presented a low-cost, accurate, and fast AMC algorithm optimized for FPGA implementation, achieving high accuracy (56\% at 0dB, 100\% above 6dB) with over 94\% reductions in computing demands and power consumption, 82\% less resource use, and 9.74x faster processing compared to state-of-the-art FPGA designs. 
Instead of using FPGAs, the authors in \cite{jung2023chip} proposed a hybrid STFT-CNN model for AMC that achieves 79\% accuracy at 0dB SNR while requiring significantly less hardware resources than traditional CNN approaches, demonstrated by a 28nm CMOS implementation that offers dramatic improvements in power consumption, area, bandwidth, and memory usage. Woo \et \cite{woo2024efficient} presented a dedicated hardware accelerator for RF signal modulation recognition that utilizes ternary weight quantization in a low-complexity DNN model, proposes a merged layer architecture to maximize efficiency, demonstrates significant improvements in bandwidth and hardware cost reduction through ASIC-based analysis, and verifies the complete system's functionality on an FPGA platform.

\subsubsection{Other Techniques}

Beyond lightweight models, model compression, and hardware implementation, several additional techniques contribute to the effective deployment of WSR systems. These may include strategies such as efficient data preprocessing (e.g., optimized signal transformations such as FFT or wavelet-based methods), pipelined architectures for accelerated inference, and integration with network protocols to support dynamic spectrum management. These approaches address diverse implementation challenges, ensuring WSR systems remain adaptable to the complex and evolving requirements of 6G networks.

\subsection{Summary of Intelligent Methods}

Intelligent WSR in 6G networks harnesses advanced DL methods across models, data, learning strategies, and implementation techniques to achieve robust, efficient, and scalable signal recognition. From a model perspective, architectures such as DNNs, DBNs, CNNs, RNNs (including LSTM and GRU), transformers, and hybrid/heterogeneous designs enhance WSR tasks such as AMC, WTC, and RFFI, with innovations such as MCformer \cite{hamidi2021mcformer} and ORACLE \cite{sankhe2019oracle} achieving high accuracy (e.g., 99\% for RFFI). Data-driven approaches leverage image-based inputs (e.g., constellation diagrams, eye diagrams) and sequence-based data (e.g., IQ, AP, FFT sequences), often combined for improved performance, as seen in \cite{wang2019data}. Learning methods, including contrastive, multi-task, and transfer learning, optimize feature extraction and generalization, with techniques such as ATLA \cite{bu2020adversarial} reducing training data needs. Implementation strategies focus on lightweight models (e.g., MobileAmcT \cite{fei2024mobileamct}), model compression (e.g., quantization \cite{goez2022methodology}), hardware acceleration (e.g., FPGA-based CNNs \cite{kumar2022automatic}), and additional techniques such as pipelined architectures, ensuring WSR systems meet 6G’s real-time, resource-constrained demands across diverse applications.

To assess the feasibility of WSR methods in real-time 6G applications, Table \ref{tab:wsr_complexity} compares the computational complexity of key approaches: LB, FB, ML, and DL. Complexity is quantified using Big-O notation, reflecting operations like hypothesis testing or convolution, critical for meeting 6G’s 1 ms latency goal. Suitability is evaluated based on processing speed and resource demands, with advantages (e.g., LB’s optimality) and disadvantages (e.g., DL’s resource intensity) highlighted. This comparison aids researchers in selecting methods for latency-sensitive scenarios like autonomous vehicles or dynamic spectrum sharing.

\begin{table*}[h]
\centering
\caption{Computational Complexity Comparison of WSR Methods for Real-Time Applications}
\label{tab:wsr_complexity}
\begin{tabular}{|p{2cm}|p{2cm}|p{2cm}|p{2cm}|p{2cm}|p{2cm}|p{2cm}|}
\hline
\textbf{Method} & \textbf{Specific Method} & \textbf{Complexity} & \textbf{Key Operations} & \textbf{Suitability for Real-Time 6G} & \textbf{Advantages} & \textbf{Disadvantages} \\
\hline
\textbf{Likelihood-based (LB)}
  & Max. Likelihood (MLC) & $O(N^2)$ & Hypothesis testing & Moderate: High latency at large $N$ & Optimal if known & High complexity \\
\cline{2-7}
  & Gen. Likelihood (GLRT) & $O(N^2 \log N)$ & Max. over unknowns & Low: Complex estimation & Robust to unknowns & Bias in nested cases \\
\hline
\textbf{Feature-based (FB)}
  & Spectral Features & $O(N)$ & FFT or PSD & High: Low latency & Simple, near-optimal & Feature-critical \\
\cline{2-7}
  & Stat. Moments & $O(N)$ & Sum over samples & High: Efficient & Low complexity & Less adaptive \\
\hline
\textbf{Machine Learning (ML)} 
  & SVM & $O(N^2)$ (train), $O(N)$ (infer) & Kernel computation & Moderate: Fast inference & Feature-flexible & Training overhead \\
\hline
\textbf{Intelligent Methods}
  & CNN & $O(N^3)$ & Convolution & Low: High latency & High accuracy & Resource-heavy \\
\cline{2-7}
  & Lightweight CNN & $O(N^2)$ & Reduced layers & High: Edge-optimized & Balanced speed & Reduced capacity \\
\cline{2-7}
  & RNN & $O(N^2)$ & Sequential proc. & Moderate: Temporal slow & Time dynamics & Sequential delay \\
\hline
\end{tabular}
\end{table*}

\section{Evaluation Framework: Datasets, Metrics, And Standards}
\label{sec:data_evaluation}

In this section, we first introduce the popular datasets for WSR, Then, the evaluation metrics for WSR are presented. 

\subsection{Datasets}
Datasets are critical for training, validating, and testing ML and DL models for WSR. While some studies utilize custom simulated data, benchmark radio signal datasets introduced in seminal papers have gained broad adoption. The recent popular AMC datasets include RadioML2016 \cite{o2016radio}, RadioML2018\cite{OShea2018}, MIMOSigRef-SD \cite{ghasemzadeh2021spatial}, HISARMOD \cite{tekbiyik2020robust}, and RML22 \cite{sathyanarayanan2023rml22}. 
TechRec \cite{fontaine2019towards}, WiFi \cite{estes2021classifying}, and \cite{kapetanovic2020classifying} are the representative datasets for WTC. 
For RFFI, WiSig \cite{hanna2022wisig} contains 10 million packets captured from 174 off-the-shelf WiFi transmitters and 41 USRP receivers over 4 captures spanning a month. 
Shen \et \cite{shen2022towards,shen2023towards} published the datasets for LoRa devices. 
Table \ref{tab:datasets} summarizes key characteristics of these popular public radio ML datasets in terms of modulations, SNR range, size, and channel models. 

\begin{table*}[h]
\centering
\caption{Summary of Dataset for Wireless Signal Recognition.}
\begin{center}
\begin{tabular}{m{1cm}<{\centering}m{2cm}<{\centering}m{6cm}<{\centering}m{1cm}<{\centering}m{1cm}<{\centering}m{1cm}<{\centering}m{3cm}<{\centering}}
\toprule
Task & Dataset & Types & SNR range & Size & Sample length & Characteristics\\
\midrule
AMC 
&
RadioML 2016 \cite{OShea2016}
&
\textbf{11 classes}: BPSK, QPSK, 8PSK, 16QAM, 64QAM, GFSK, CPFSK, 4PAM, WB-FM, AM-SSB, and AM-DSB
&
[-20:2:18]
&
$220$k
&
$128$
&
CFO, sample rate offset, AWGN, and fading
\\ \midrule
AMC 
&
RadioML 2018 \cite{OShea2018}
& 
\textbf{24 Classes}: OOK, 4ASK, BPSK, QPSK, 8PSK, 16QAM, AM-SSB-SC, AM-DSB-SC, FM, GMSK, OQPSK, OOK, 4ASK, 8ASK, BPSK, QPSK, 8PSK, 16PSK, 32PSK, 16APSK, 32APSK, 64APSK, 128APSK, 16QAM, 32QAM, 64QAM, 128QAM, 256QAM, AM-SSB-WC, AM-SSB-SC, AM-DSB-WC, AM-DSB-SC, FM, GMSK, OQPSK
&
$[-20,+30]$
&
$2.5$M
&
$1024$
&
CFO, symbol rate offset, multipath fading, and thermal noise
\\ \midrule
AMC 
&
MIMOSigRef-SD \cite{ghasemzadeh2021spatial}
&
M-QAM, MIL-STD-188-110 B/C standard-specific QAM, M-PSK, M-APSK, DVB-S2/S2X/SH standard-specific APSK, and M-PAM&
$[-20,+18]$
&
$780$k
&
$128$
&
several emulated environments
\\ \midrule
AMC 
&
HisarMod2019 \cite{tekbiyik2020robust}
&
BPSK, QPSK, 8PSK, 16PSK, 32PSK, 64PSK, 4QAM, 8QAM, 16QAM, 32QAM, 64QAM, 128QAM, 256QAM, 2FSK, 4FSK, 8FSK, 16FSK, 4PAM, 8PAM, 16PAM, AM-DSB, AM-DSB-SC, AM-USB, AM-LSB, FM, PM
&
$[-20,+18]$
&
$780$k
&
$1024$
&
5 different wireless communication channels \\ \midrule
AMC 
&
RML22 \cite{sathyanarayanan2023rml22}
&
BPSK, QPSK, 8PSK, 16QAM, 64QAM, PAM4, WBFM, CPFSK, GFSK, AM-DSB&
$[-20,+20]$
&
$420$k
&
$128$
&
-
\\ \midrule
AMC 
&
HKDD \cite{zheng2023towards}
& 
BPSK, QPSK, 8PSK, OQPSK, 16QAM, 32QAM, 64QAM, 4PAM, 8PAM, 2FSK, 4FSK, 8FSK.
&
$[-20,+20]$
&
$252$k
&
$512$
&
-
\\ \midrule
WTC 
&
TechRec \cite{fontaine2019towards}
&
LTE, Wi-Fi and DVB-T 
&
$[-20,+18]$
&
$160$k
&
$128$
&
-
\\ \midrule
WTC 
&
WiFi \cite{estes2021classifying}
&
802.11ax, 802.11ac, 802.11n, 802.11ax with 802.11ac, and 802.11ax with 802.11n
&
$[-20,+18]$
&
$160$k
&
$128$
&
-
\\ \midrule
WTC 
&
WLAN packet dataset \cite{kapetanovic2020classifying}
&
802.11b/g/n
&
-
&
$360,000$
&
$256$
&
-
\\ \midrule
RFFI 
&
WiSig \cite{hanna2022wisig}
&
174 off-the-shelf WiFi transmitters and 41 USRP receivers
&
$[0, 25]$
&
10 million
&
-
&
-
\\ \midrule
RFFI 
&
\cite{shen2022towards,shen2023towards}
&
40 LoRa devices
&
-
&
$20,000$
&
-
&
-
\\
\bottomrule
\end{tabular}
\label{tab:datasets}
\end{center}
\end{table*}

\subsection{Evaluation Metrics}
To measure the quality of a machine learning model, it is necessary to give a test set and use the model to test the test set. 
Each sample is predicted, and the evaluation score is calculated according to the prediction result. 

\subsubsection{Classification Metrics}
WSR, aiming at classifying the true label of the signal type of the input signal, is a classification task. 
For classification tasks, the common evaluation metrics are accuracy, precision, recall and $F_1$ measure.

Given a test set $\mathcal{T}=\{(x_1,y_1),\cdots,(x_N,y_N)\}$ and the label $y_n\in {1,\cdots,C}$, the learned model $f(x;\theta)$ is used to predict the samples in the test set and the predictions are $\{\hat{y}_1,\cdots,\hat{y}_N\}$. 

\paragraph{Commom metrics for WSR}
Common metrics for WSR include accuracy, confusion matrix, precision, recall, and $F_1$ measure. 
\textbf{Accuracy} is defined by
\begin{equation}
  Acc = \frac{1}{N}\sum_{n=1}^N I(y_n=\hat{y}_n),
\end{equation}
where $I(*)$ is the indicator function. 
Accuracy is the average of the overall performance of all categories. Precision and recall, estimation for the performance of each category, are two metrics widely used in the fields of information retrieval and statistical classification and are also heavily used in the evaluation of machine learning. For a class $c$, the results can be divided into four categories
\begin{enumerate}
  \item \textbf{TP, True Positive}, An example has a true class of $c$ and the model correctly predicts it as the category $c$. The number of such samples is recorded as $TP=\sum_{n=1}^N i(y_n=\hat{y}_n=c)$.
  \item \textbf{FN, False Negative}, An example has a true class of $c$ and the model falsely predicts it as other categories. The number of such samples is recorded as $FN=\sum_{n=1}^N i(y_n=\hat{y}_n\ne c)$.
  \item \textbf{FP, False Positive}, An example has a true class of other classes and the model falsely predicts it as the category $c$. The number of such samples is recorded as $FP=\sum_{n=1}^N i(y_n\ne c \wedge \hat{y}_n=c)$.
  \item \textbf{TN, True Negative}, An example has a true class of other classes and the model predicts it as other categories. The number of such samples is recorded as $TN$ and it can be ignored for class $c$.
\end{enumerate}
The four categories can be presented by a confusion matrix as shown in Table \ref{tab:consusionmatrix}.
\begin{table}[h]
\centering
\caption{Confusion Matrix}
\begin{center}
\begin{tabular}{m{1cm}<{\centering}m{1cm}<{\centering}m{1cm}<{\centering}m{1cm}<{\centering}}
\toprule
&  & \multicolumn{2}{c}{Predictions} \\
\cline{3-4}
                    &          & $\hat{y}=c$ & $\hat{y}\ne c$ \\
\midrule
\multirow{2}*{True} & $y=c$    & $TP$      & $FN$          \\
\cline{2-4}
                    & $y\ne c$ & $FP$      & $TN$          \\
\bottomrule
\end{tabular}
\label{tab:consusionmatrix}
\end{center}
\end{table}
\textbf{Precision}, \textbf{Recall}, and \textbf{$F_1$ Measure} can be defined by the definition above. 
\textbf{Precision} for class $c$ is defined as the ratio of the correct prediction among all samples predicted to be class $c$,
\begin{equation}
  P = \frac{TP}{TP+FP}.
\end{equation}

\textbf{Recall} is the proportion of correct predictions for all samples with true label class $c$, expressed as
\begin{equation}
  R = \frac{TP}{TP+FN}.
\end{equation}

\textbf{$F_1$} Measure is a comprehensive indicator, which is the harmonic mean of precision and recall
\begin{equation}
  F_1 = \frac{2\times P\times R}{P+R}.
\end{equation}

\begin{figure}[t]
  \centering
  \includegraphics[width=0.65\linewidth]{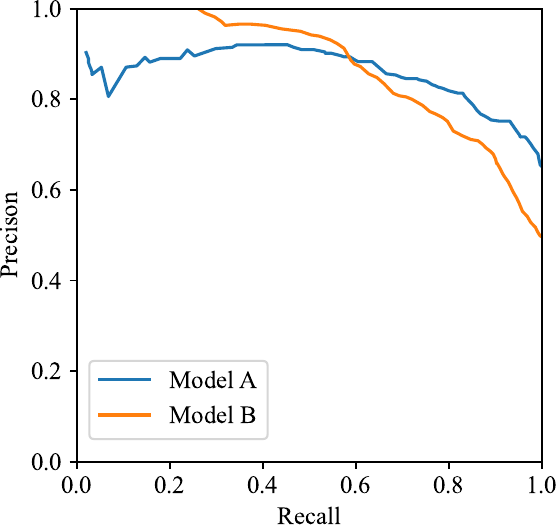}
  \caption{Precision-Recall (PR) curve.}
  \label{fig:pr}
\end{figure}

The final output of the classification model is often a probability value, which is needed to convert the probability value into a specific category. 
For the binary classification, a threshold is often used to judge the output as a positive class or a negative one. 
The above evaluation indicators (Accuracy, Precision, Recall) are all for a specific threshold, so when different models take different thresholds, the Precision-Recall (PR) curve is introduced. 
As shown in Fig. \ref{fig:pr}, the horizontal axis of the PR curve is recall, and the vertical axis is precision. 
For a model, a point on its PR curve represents that, under a certain threshold, the model determines the result greater than the threshold as a positive sample, and the result less than the threshold as a negative sample, and the returned result corresponds to a pair recall and precision as a coordinate on the PR coordinate system. The entire PR curve is generated by shifting the threshold from high to low.
The closer the P-R curve is to the upper right corner $(1,1)$, the better the model. In real scenarios, it is necessary to comprehensively judge the quality of different models according to different decision-making requirements (discussed in the following chapters).

Another indicator for comprehensive evaluation of the model is the receiver operating characteristic (ROC) curve. The ROC curve originated in the military field and was widely used in the medical field. The abscissa of the ROC curve is the false positive rate (False Positive Rate, FPR); the ordinate is the true positive rate (True Positive Rate, TPR). The calculation methods of FPR and TPR are
\begin{equation}
  FPR=\frac{FP}{FP+TN}
\end{equation}
\begin{equation}
  TPR=\frac{TP}{TP+FN}
\end{equation}

The area under the ROC curve, referred to as AUC, is mathematically defined as the integral of the ROC curve. By definition, the value of AUC cannot exceed 1 as the ROC curve lies on or below the line $y=x$. Since ROC curves typically dominate the line $y=x$, AUC values generally range from 0.5 to 1. The AUC value is employed as an evaluation metric because in numerous instances the ROC curve fails to unambiguously indicate the superior classifier, and as a magnitude, classifiers with larger AUC values are preferred. 
Criteria for judging the quality of a classifier (prediction model) from AUC
\begin{enumerate}
  \item $AUC = 1$, it is a perfect classifier, when using this prediction model, there is at least one threshold to get a perfect prediction. In the vast majority of prediction cases, there is no perfect classifier.
  \item $0.5 < AUC < 1$, better than random guessing. This classifier (model) can have predictive value if the threshold is properly set.
  \item $AUC = 0.5$, the following machine guesses the same (for example: losing a copper plate), and the model has no predictive value.
  \item $AUC < 0.5$, worse than random guessing; but better than random guessing as long as it always works against predictions.
\end{enumerate}

\begin{figure}
  \centering
  \subfigure[]{
  \includegraphics[width=0.45\linewidth]{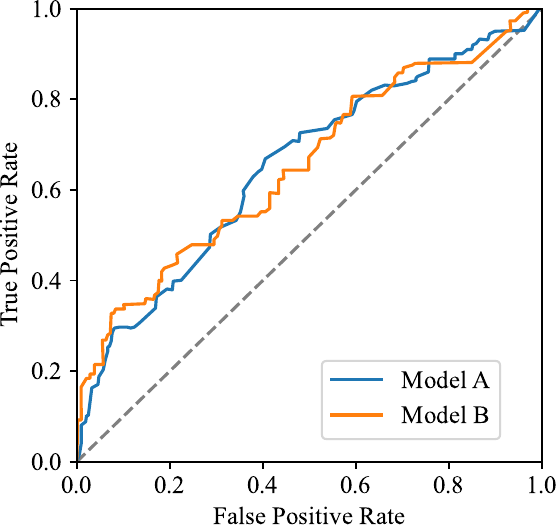} 
  }
  \subfigure[]{
  \includegraphics[width=0.45\linewidth]{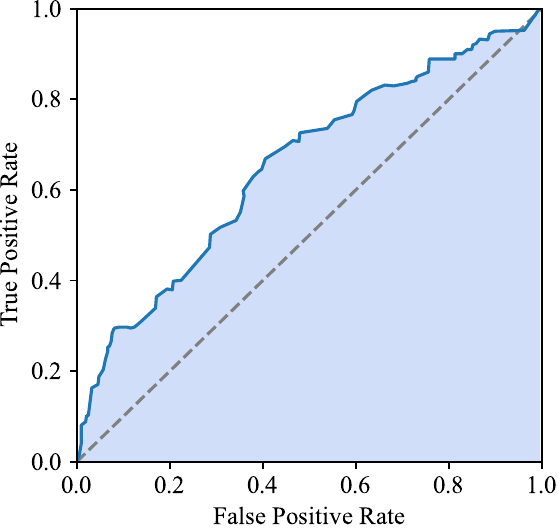} 
  }
  \DeclareGraphicsExtensions.
  \caption{Illustration of (a) the receiver operating characteristic (ROC) curve and (b) the area under ROC curve (AUC).}
  \label{fig:roc}
\end{figure}

\paragraph{Metrics for RFFI} 
False acceptance rate (FAR), false rejection rate (FRR), and equal error rate (EER) are the common metrics for RFFI. 
FAR is the probability that the system incorrectly accepts an unauthorized user, while FRR is the probability that the system incorrectly rejects an authorized user. 
The EER is the point where the FAR and FRR are equal.

\subsubsection{Model Complexity}

\begin{table}
\centering
\caption{Metrics of Model Complexity}
\label{tab:complexity}
\begin{tabular}{m{3cm}<{\centering}m{5cm}<{\centering}}
\toprule
Metrics & Summary \\
\midrule
Number of Parameters & The number of parameters in the model. \\ 
Model Size & The physical storage space required to save the model. \\
FLOPs & The number of floating-point operations per second. \\
Training Time & The time required to train the model. \\
Inference Time & The time required to infer the model. \\
\bottomrule
\end{tabular}
\end{table}

The model complexity of intelligent models mainly includes space complexity and time complexity. 
Space complexity is mainly reflected in the number of parameters and the model size. 
Time complexity is mainly reflected in the number of floating-point operations per second (FLOPs), training time, and inference time. 
Table \ref{tab:complexity} summarizes the metrics of model complexity.

\subsection{Open-Source Tools, Software Libraries, and Platforms}
Table \ref{tab:tools} presents a comprehensive compilation of open-source tools, software libraries, and platforms tailored for wireless signal recognition. Tools such as GNU Radio and Wireshark enable real-time signal processing and protocol analysis, while libraries such as Scapy and Liquid DSP offer programmatic flexibility for custom recognition systems. Platforms such as OpenWiFi and ROS with SDR integrations cater to advanced applications, including Wi-Fi analysis and robotics. These resources, supported by diverse communities, provide a robust foundation for researchers, developers, and enthusiasts to explore and implement wireless signal recognition across various platforms and use cases.

\begin{table*}[h!]
\centering
\caption{Open-Source Tools, Software Libraries, and Platforms for Wireless Signal Recognition}
\label{tab:tools}
\begin{tabular}{m{1cm}<{\centering}m{3cm}<{\centering}m{3cm}<{\centering}m{3cm}<{\centering}m{3cm}<{\centering}m{2.5cm}<{\centering}}
\toprule
\textbf{Category} & \textbf{Name} & \textbf{Description} & \textbf{Key Features} & \textbf{Use Case} & \textbf{Platform Support} \\
\midrule
\multirow{4}{*}{Tools} 
    & GNU Radio & Open-source toolkit for software-defined radio (SDR) with a graphical interface and modular framework & Modulation/demodulation, filtering, spectrum analysis; supports various SDR hardware & Experimenting with wireless signal recognition (e.g., modulation types, protocol decoding) & Linux, Windows, macOS \\
    \cline{2-6}
    & Kismet & Open-source wireless network detector, sniffer, and IDS focused on 802.11 (Wi-Fi), expandable to other protocols & Captures raw packets, identifies signal sources, supports plugins & Wi-Fi signal recognition and tracking & Primarily Linux, some macOS/Windows \\
    \cline{2-6}
    & Wireshark & Well-known open-source packet analyzer supporting wireless protocols & Protocol decoding, filtering, wireless traffic analysis & Analyzing wireless traffic for signal patterns & Linux, Windows, macOS \\
    \cline{2-6}
    & Sigrok & Open-source signal analysis suite for logic analyzers, oscilloscopes, and SDRs & Protocol decoding (e.g., RFID, Zigbee), extensible & Recognizing simple wireless signals from low-power devices & Linux, Windows, macOS \\
\hline
\multirow{4}{*}{Libraries} 
    & Scapy & Python-based packet manipulation library supporting wireless protocols & Packet crafting, sniffing, analysis; supports 802.11 frames & Programmatic wireless signal recognition & Cross-platform (Python) \\
    \cline{2-6}
    & Liquid DSP & Efficient C library for digital signal processing, designed for SDR & Modulation/demodulation, filtering, signal detection, real-time optimized & Building custom wireless signal recognition systems & Cross-platform (C) \\
    \cline{2-6}
    & PySDR & Python-based SDR and signal processing library and educational resource & Simplified signal analysis, visualization, basic recognition tasks & Learning and prototyping wireless signal recognition & Cross-platform (Python) \\
    \cline{2-6}
    & OpenCV (RF extension) & Computer vision library, adaptable for RF signal pattern recognition & Machine learning and image processing for classifying signal spectrograms or waveforms & Advanced signal recognition using visual data & Cross-platform (C++, Python) \\
\hline
\multirow{3}{*}{Platforms} 
    & OpenWiFi & Open-source Wi-Fi implementation on FPGA with Linux integration & Full-stack SDR Wi-Fi solution, supports mac80211 & Deep analysis and recognition of Wi-Fi signals & FPGA + Linux \\
    \cline{2-6}
    & Flashlight ASR & Facebook AI’s open-source speech recognition library, extensible to RF signal analysis & Machine learning framework for pattern recognition, adaptable to wireless signal classification & Training models to recognize complex wireless signal patterns & Cross-platform (C++, Python) \\
    \cline{2-6}
    & ROS (SDR plugins) & Open-source robotics framework with community-developed SDR integrations & Signal processing and recognition for wireless communication in robotic applications & Recognizing wireless signals in autonomous systems or IoT & Primarily Linux \\
\bottomrule
\end{tabular}
\end{table*}

\subsection{Standards}
Standardization efforts for WSR are advancing through contributions from organizations such as the IEEE, ITU, and 3GPP, focusing on interoperability and performance in CR and next-generation networks. The IEEE 802.22 standard supports spectrum sensing \cite{gronsund2014system}, while ITU-R recommendations (e.g., ITU-R SM series \cite{series2018recommendation}) provide guidelines for signal identification. 3GPP's AI integration in 5G and beyond hints at future AMC frameworks \cite{zhu20213gpp}. Research benchmarks such as RadioML further unify evaluation criteria. These efforts collectively shape reliable and efficient signal recognition systems.

\section{Challenges And Future Directions} 
\label{sec:challenges_issues}

While substantial research has been conducted on WSR, this remains an active area with many open challenges and opportunities for future work due to the complexity, dynamics, and security of wireless environments. This section presents a range of potential open issues and future research trends. 

\begin{figure*}[t]
	\centering
	  \includegraphics[width=0.95\linewidth]{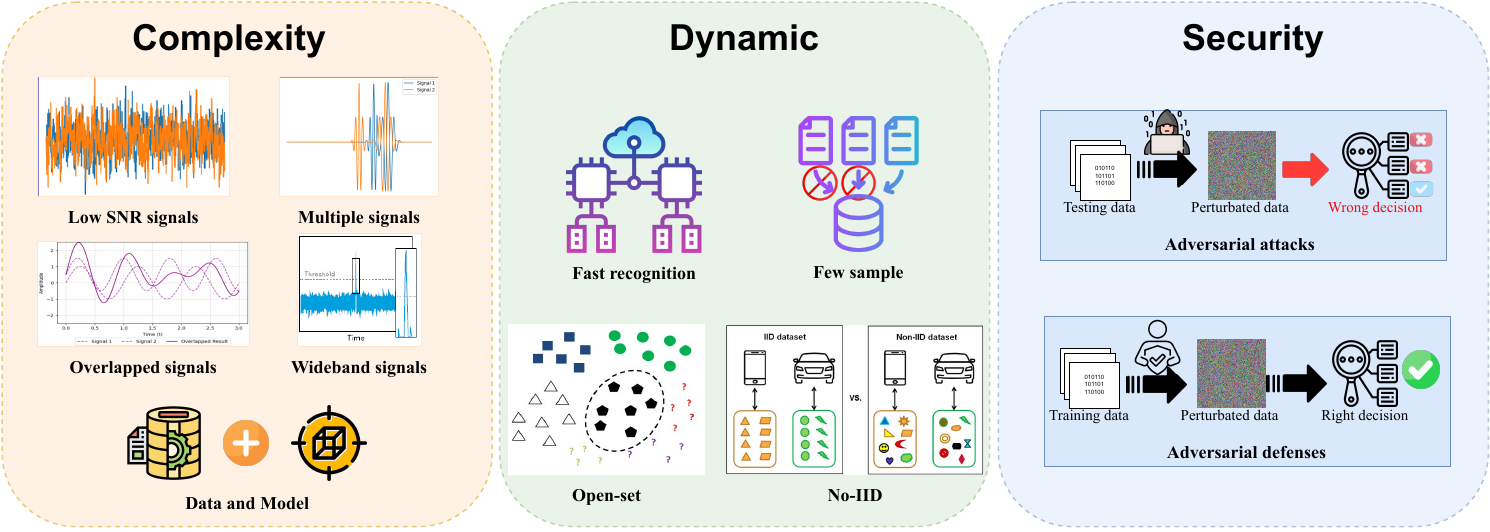}
	  \caption{\color{blue}Challenges of wireless signal recognition and future directions from the view of complexity, dynamic, and security.}
	  \label{fig:challenge}
\end{figure*}

\subsection{Complex Environments}

\subsubsection{Low SNR Conditions}
Real-world wireless environments present multiple sources of noise and channel distortions that pose significant challenges for reliable WSR. The presence of noise can degrade the quality of received signals, making it difficult to extract discriminative features, especially under low SNR conditions. While deep learning methods have demonstrated state-of-the-art performance on WSR benchmarks, their accuracy substantially deteriorates at low SNRs compared to high SNRs. Overcoming the vulnerability of deep models to noisy signals remains an open problem. 
An \et \cite{an2023robust} proposed a novel noise-robust deep learning architecture for AMC termed threshold autoencoder denoiser CNN (TADCNN). The approach integrates a threshold autoencoder denoiser module prior to a CNN classifier. The denoiser component leverages reconstruction-based denoising to improve the quality of the input signals by reducing noise, resulting in a 70\% improvement in classification accuracy at low SNRs and an average accuracy of 66.64\% on the RML2016.10A dataset, outperforming the state-of-the-art AMC model by 6\% to 18\%. 
The authors in \cite{harper2021snr} proposed a two-stage process that employs a deep convolutional SNR estimation model to improve classification performance at decreasing SNRs. The first stage estimates the SNRs and the second stage uses modulation classifiers that have been tuned on signals with similar SNRs to improve classification accuracy.

\subsubsection{Multiple Signals} 
The proliferation of wireless services has led to increasing scarcity in available spectrum resources, motivating a shift toward co-channel deployment where multiple signals share the same bandwidth. A prime example is non-orthogonal multiple access (NOMA) which has been widely adopted in 5G systems to improve spectral efficiency through power-domain multiplexing of user signals. However, receiving and classifying the overlapping transmissions poses major challenges. Conventional signal processing techniques cannot effectively filter and separate the mixed signals in time or frequency. Therefore, developing accurate WSR methods for co-channel signals is crucial but remains an open problem. Recent works have explored applying deep learning to tackle this task, but performance is still far from that of single-signal classification. 
The authors in \cite{yin2019cochannel} proposed a CNN-based AMC method for co-channel multi-signals, which can classify 31 mixed signals from five modulation schemes. The authors evaluated the robustness of the proposed method across a wide range of SNR levels from 0dB to 15dB. Experimental results demonstrated that the classification performance of the proposed approach is resilient to varying noise levels, with consistent accuracy as the SNR changes. 
Hou \et \cite{hou2022multi} developed a novel frequency-domain technique using a fast Fourier transform, energy detection, and a convolutional neural network to accurately detect and classify overlapping signals, demonstrating high accuracy and robustness for Internet of Things applications.
In the 6G context, multiple signals pose additional complexity due to integrated sensing and communication (ISAC), where WSR must simultaneously identify communication signals and sense environmental data (e.g., radar reflections) from the same spectrum [ref-ISAC]. For example, distinguishing a QPSK communication signal from an overlapping radar pulse in a dense urban scenario requires advanced separation techniques beyond current CNN capabilities, highlighting a critical research gap for 6G’s dual-purpose spectrum utilization.

\subsubsection{Overlapped Signals} 
The increasing density of wireless networks has led to a growing number of overlapping signals, which poses a significant challenge for WSR. Overlapping signals can be caused by co-channel interference, multipath propagation, and the presence of multiple transmitters nearby. Conventional signal processing techniques are unable to effectively filter and separate the mixed signals in time or frequency. Therefore, developing accurate WSR methods for overlapping signals is crucial but remains an open problem. Recent works have explored applying deep learning to tackle this task, but performance is still far from that of single-signal classification. 
Ren \et \cite{ren2021automatic} presented a Multi-Domain Squeeze-and-Excitation ResNeXt (SE-ResNeXt) method for AMC of overlapping radar signals, enhancing accuracy and performance in complex electromagnetic environments, especially at low SNRs. 
The authors in \cite{zhang2023reference} introduced a deep learning approach combining ConvNeXt and atrous self-attention transformer for overlapped signals AMC, offering higher accuracy with a simpler and more efficient training process and reduced computational and memory requirements.

\subsubsection{Wideband Signals} 
The increasing demand for high data rates and low latency in wireless communication systems has led to the widespread adoption of wideband signals. Wideband signals are characterized by large bandwidths and high data rates, which pose significant challenges for WSR. Conventional signal processing techniques are unable to effectively capture the complex time-frequency characteristics of wideband signals. Therefore, developing accurate WSR methods for wideband signals is crucial but remains an open problem. Recent works have explored applying deep learning to tackle this task, but performance is still far from that of narrowband signal classification. 
\cite{vagollari2023end} proposed a deep learning-based framework for managing radio spectrum, using signal detection, localization, and modulation classification through wideband spectrogram analysis, and demonstrates high precision, recall, and localization accuracy in signal detection with advanced training strategies.

\subsubsection{Model and Data}
All data-driven WSR schemes require a large amount of training samples, which are difficult to obtain in practical communication scenarios. Additionally, purely data-driven schemes cannot satisfy classification performance requirements under dynamically changing communication scenarios. In particular, the classification performance of these methods is very poor at low signal-to-noise ratios. Therefore, it is necessary to merge model-driven and data-driven methods. Combining traditional model-based approaches with learned representations from data could utilize both physical insights and statistical information to achieve robust classification performance under varying channel conditions. Integrating prior knowledge into deep learning models may also reduce data requirements and improve generalization with limited training samples. The synergy of model-driven and data-driven methods has the potential to address the challenges of WSR in practical scenarios. 
The authors in \cite{huang2020identification} proposed a cyclic correntropy vector-based AMC method using a long short-term memory densely connected network. The extracted cyclic correntropy vector features fed into the data-driven long short-term memory densely connected network, which combines a long short-term memory network and dense network, showed improved performance over other recent schemes in simulations. An additive cosine loss utilized during training of the long short-term memory densely connected network was shown to maximize inter-class feature differences while minimizing intra-class feature variations, further improving classification performance.
Marey \et \cite{marey2021code} proposed a novel AMC algorithm for multiuser uplink single-carrier frequency division multiple access systems that utilize channel decoder soft information to improve classification performance. The proposed algorithm, designed using a space-alternating generalized expectation-maximization approach, was shown in simulations to outperform traditional algorithms with reduced processing time. 
The work \cite{ding2022data} presented a novel data-and-knowledge dual-driven AMC scheme based on radio frequency machine learning by exploiting attribute features of different modulations. Simulation results demonstrated that the proposed scheme, which utilizes a visual model to extract visual features and an attribute learning model to derive attribute semantic representations converted to the visual space by a transformation model, achieved better performance than benchmark schemes in classification accuracy, especially at low signal-to-noise ratios, while also reducing confusion between high-order modulations.
\cite{zhang2023data} proposed a data-and-knowledge dual-driven radio frequency fingerprint identification (RFFI) scheme using a multiscale attention convolutional network (AttMsCN), demonstrating improved identification accuracy and convergence speed, especially in low signal-to-noise ratio environments. 
The authors in \cite{li2024kgamc} presented a novel knowledge graph-driven AMC (KGAMC) scheme for 6G wireless networks, which enhances classification performance, particularly at low signal-to-noise ratios, by integrating domain knowledge into network training and feature aggregation, thereby improving interpretability and reducing signal confusion.

\subsection{Dynamic Environments}

\subsubsection{Fast Recognition}

Intelligent communications with high reliability and low latency are key characteristics for beyond fifth-generation (5G) wireless communication networks. However, traditional WSR schemes based on convolution may not satisfy the classification performance and low computational cost requirements of beyond 5G wireless communication networks. The powerful feature extraction ability of deep learning often relies on deeper network models to achieve deep extraction and recognition of signal features, which may not be suitable beyond 5G networks that require low latency and cost. Therefore, new WSR methods that can achieve high performance with low computational complexity are needed for beyond 5G wireless communication networks. 
Zhang \et \cite{zhang2021automatic} a residual neural network (ResNet) based WSR scheme utilizing involution in place of convolution. Involution incorporates a self-attention mechanism to enhance the model's discrimination capability and expressiveness. Simulation results demonstrate the effectiveness of the proposed involution-based ResNet for WSR. The proposed scheme was shown to have better classification performance and faster recognition speed compared to convolutional neural networks.
The authors in \cite{fu2021lightweight} designed a lightweight network using separable CNN (S-CNN). Separable convolution layers replaced standard convolution layers and most fully connected layers were removed. Model aggregation was achieved by a central device (CD) aggregating edge device (ED) model weights and by multiple EDs training ED models. The S-CNN, with separable convolutions and reduced fully connected layers, reduced model parameters and computational complexity compared to conventional CNNs.
Looking ahead, a promising direction for 6G is integrating WSR with large language models (LLMs) to enable adaptive spectrum management \cite{liu2024llm4cp}. By leveraging LLMs’ natural language understanding, WSR systems could interpret real-time spectrum policies or user demands (e.g., from network logs), dynamically adjusting recognition strategies—such as switching between lightweight CNNs and deeper models—within milliseconds. This fusion could address 6G’s 1 ms latency goal while optimizing resource allocation in fluctuating environments, offering a novel paradigm beyond current DL approaches.

\subsubsection{Few Samples}

While DL-based methods have surpassed traditional handcrafted feature-based methods in classification performance in recent years, they require significantly larger numbers of labeled samples for training. When the number of labeled samples is insufficient, the classification performance of DL-based methods declines sharply. With diversifying signal acquisition methods and rapid storage technology development, obtaining a large number of unlabeled signals is simple. However, obtaining the same amount of labeled signals is very difficult because data labeling requires substantial manpower, material resources, and time. Additionally, a large number of signals exist in the electromagnetic spectrum in real environments. It is impractical to label all collected signals due to time requirements for rapidly changing scenarios. Therefore, studying few-shot WSR with insufficient labeled signals is particularly important. A practical WSR system requires high performance with a limited number of labeled samples. 
The authors in \cite{li2020automaticwcl,shi2020few,zhou2021amcrn,zhang2022gan} investigated AMC with fewer samples to achieve higher classification accuracy. 
The work \cite{ghasemzadeh2022gs} presents a highly efficient AMC architecture using stacked quasi-recurrent neural network (S-QRNN) layers for feature extraction and minimalist recurrent pooling to aggregate features over time. The proposed model demonstrates higher trainability, classification accuracy and efficiency compared to state-of-the-art classifiers, making it well-suited for resource-constrained IoT devices. 
Shao \et \cite{shao2020convolutional} studied a few-shot learning algorithm to identify interference signals with limited training samples. The proposed method achieved promising classification performance for various typical jamming types. The authors demonstrated that their approach can effectively identify interference signals with only a small number of labeled training samples, overcoming the data requirements of traditional deep learning-based modulation classifiers.
Liang \et \cite{liang2022few} investigated RFFI with fewer samples to achieve higher classification accuracy.

\subsubsection{Open Set}

Current deep learning approaches used for WSR make a closed-set assumption, where all test categories appear during training. However, this is unrealistic in practice. In real-world scenarios, deep learning faces the challenge of unknown categories not present in the training data (unknown unknown classes, UUCs), while only known categories (known known classes, KKCs) are used during training. In fact, UUCs are often misclassified as one of the KKCs with high probability. Therefore, the goal of open set recognition (OSR) is to correctly identify unknown classes while also classifying known classes accurately \cite{geng2020recent}. OSR aims to address the practical scenario where the test set contains not only known classes seen during training but also unknown classes not present in the training data. 
In \cite{chen2023openset}, an open-set automatic modulation recognition scheme is proposed combining feature representation and classification. A triplet loss function from metric learning is used by the representation network to form distinct clusters for N known modulation classes. The degree of membership is then calculated via extreme value theory by modeling the distance between known training data to its corresponding clustering center, followed by $N$ binary classifiers. Comprehensive experiments on public datasets confirm that the proposed scheme outperforms state-of-the-art methods in terms of balanced accuracy and openness.
Shebert  \et \cite{shebert2021openset} proposed a CNN-based open set classifier able to detect signals not from known classes by thresholding the output sigmoid activation. The closed set classifier achieves 94.5\% accuracy for known signals with SNRs $>$ 0 dB, but cannot detect unknown class signals. The open set classifier retains 86\% accuracy for known signals, but can detect 95.5\% of unknown class signals with SNRs $>$ 0 dB.
The work \cite{xu2020openset} presented a novel method for the OSR problem and its application to wireless interference signal recognition. The proposed method modifies and combines intra-class splitting and adversarial sample generation to construct precise boundary samples. Experiments on image and wireless interference signal datasets demonstrate the effectiveness of the proposed open-set recognition method.
The authors in \cite{shebert2023open} proposed an open-set hybrid classifier, which combines deep learning and expert feature classifiers to leverage the reliability and explainability of expert feature classifiers and the lower computational complexity of deep learning classifiers. 
The authors in \cite{zhao2024meta} proposed a meta-learning-based few-shot open-set recognition method for AMC, namely, Meta Supervised Contrastive Learning (MSCL). It combines the strengths of supervised contrastive learning and meta-learning to effectively amplify inter-class distinctions and reinforce intra-class compactness, exhibiting superior performance in both few-shot and open-set AMC. 

\subsubsection{Non I.I.D. Data} 
The training data distribution and test data distribution should be the same in the traditional machine learning model. However, in the real world, the data distribution of the training set is different from that of the test set, which is called the non-IID data distribution. The non-IID data distribution is a common problem in WSR for the following reasons. 
First, the dynamic nature of wireless channels and the mobility of wireless devices lead to non-IID data distributions. 
Second, the non-IID data distribution is also caused by the heterogeneity of wireless signals, such as different modulation types, different signal-to-noise ratios, and different channel conditions. 
Third, the existing interference and jamming signals in the wireless environment also lead to non-IID data distributions. 
Thus, it is necessary to study the non-IID data distribution in WSR. 
For example, HisarMod2019 \cite{tekbiyik2020robust} is a non-IID dataset for AMC, which contains 26 modulation types and 20 SNR levels from 5 different channel models.

\subsection{Open Environments}
Open wireless environments are characterized by the presence of adversarial attacks. Adversarial attacks are a major threat to the security and reliability of WSR systems. Thus, adversarial defense is a critical requirement for WSR systems. 

\subsubsection{Adversarial attacks} 

The authors in \cite{lin2020adversarial} investigated how well-designed adversarial perturbations can significantly reduce the accuracy of convolutional neural networks (CNNs) in modulation recognition tasks, showing that even subtle changes undetectable to humans can lead to a 50\% drop in accuracy, and emphasizes the need for improving CNN resilience against such attacks. 
Similarly, the work \cite{bao2021threat} examined how adversarial attacks affect deep neural network-based device identification in IoT, demonstrating that increased perturbation and iteration step size degrade identification accuracy, and introduced combined evaluation indicators to enhance robustness in IoT systems. 
Liu \et \cite{liu2023robust} proposed a generation adversarial perturbations problem formulation considering both perfect and imperfect channel state information (CSI), with a spoofing attack algorithm exploiting the S-procedure to address the non-convex robust optimization under imperfect CSI, achieving superior spoofing performance compared to benchmarks.

\subsubsection{Adversarial defenses} 

\cite{chen2024learn} proposed an Adversarial Multi-Distillation (AMD) framework for robust training of deep learning-based automatic modulation recognition (AMR) models, where two teacher models transfer classification and defense knowledge respectively to a student model through knowledge distillation, significantly improving the model's robustness against adversarial attacks while maintaining high accuracy and enabling lightweight robust decision making.
To against adversarial examples in AMC, Zhang \et \cite{zhang2021countermeasures} introduced a countermeasure that uses neural rejection, combined with label smoothing and Gaussian noise injection, to detect and reject adversarial examples with high accuracy. The results show that this approach effectively protects DL-based AMC systems from adversarial attacks.
Similarly, the authors in \cite{yi2021gradient} considered a data-driven subsampling setting for the Carlini-Wagner attack. 
The paper \cite{sahay2020frequency} introduced a novel receiver architecture employing DL models that exhibit resilience to transferable adversarial interference. 
Evaluations revealed that utilizing frequency-domain features instead of time-domain features significantly enhanced model robustness against transferable adversarial attacks, resulting in classification performance improvements of over $30\%$ for RNNs and over $50\%$ for CNNs.

\section{Conclusion} \label{sec:conclusion}

This survey has comprehensively surveyed the advancements in wireless signal recognition (WSR) from the perspective of applications, main tasks, methods, datasets and evaluation, and challenges and open issues. 
First, we have introduced the applications of WSR from civilian and military aspects. 
Then, WSR algorithms are categorized into model-based and intelligent methods, where model-based methods can be classified into likelihood-based, feature-based methods, and machine-learning methods and intelligent methods are introduced from the view of model, data, learning and others. 
Public datasets of WSR and evaluation metrics are also presented. 
Finally, we have discussed the challenges and open issues of WSR. 
The comprehensive survey can be a good reference for researchers to understand the current status of WSR and future research directions.

\bibliographystyle{IEEEtran}
\bibliography{wsr_papers}

\end{document}